\documentclass[aps, prd, nofootinbib, amsmath, preprintnumbers, superscriptaddress,secnumarabic, notitlepage,twocolumn]{revtex4-2} 
\usepackage{amssymb,amsmath,mathrsfs,graphicx,latexsym,amsthm,slashed,eso-pic,hyperref}
\usepackage{comment}
\usepackage{bbold}
\usepackage{aas_macros}

\newcommand{\beq}{\begin{equation}} \newcommand{\eeq}{\end{equation}}
\newcommand{\bea}{\begin{eqnarray}} \newcommand{\eea}{\end{eqnarray}}

\newcommand{\be}{\begin{equation}}
\newcommand{\ee}{\end{equation}}
\newcommand{\nn}{\nonumber}

\definecolor{orange}{rgb}{1,0.5,0}

\def\mpuel#1{{\color{blue}MP: #1}}

\def\resp#1{{\color{violet} #1}}
\usepackage{slashed}
\usepackage{appendix}
\usepackage{siunitx}

\newcommand{\SK}{S_{i}}
\newcommand{\SKg}{S_{\phi}}
\newcommand{\SKnu}{S_{\nu_s}}

\begin{document}

\begin{flushright}
KANAZAWA-23-09    
\end{flushright}

%dvipdfm
\title{Phantom fluid cosmology:
Impact of a phantom hidden sector\\ on cosmological observables}% and direct detection
%and dark matter searches}
\author{James M.\ Cline}
\affiliation{McGill University, Department of Physics, 3600 University Street, Montr\'eal, QC H3A 2T8, Canada}
\author{Matteo Puel}
\affiliation{McGill University, Department of Physics, 3600 University Street, Montr\'eal, QC H3A 2T8, Canada}
\author{Takashi Toma}
\affiliation{Institute of Liberal Arts and Science, Kanazawa University, Kanazawa, Ishikawa 920-1192 Japan}
\affiliation{Institute for Theoretical Physics, Kanazawa University, Kanazawa, Ishikawa 920-1192 Japan}
\author{Qiu Shi Wang}
\affiliation{McGill University, Department of Physics, 3600 University Street, Montr\'eal, QC H3A 2T8, Canada}

\date{\today}

\begin{abstract}
Phantom scalar theories are widely considered in cosmology, but rarely at the quantum level, where they give rise to negative-energy ghost particles.  These cause decay of the vacuum
into gravitons and photons, violating observational gamma-ray limits unless the ghosts are effective degrees of freedom with a cutoff $\Lambda$ at the few-MeV scale. We update the constraints on this scale, finding that $\Lambda\lesssim 19\,$MeV.
We further explore the possible coupling of ghosts to a light, possibly massless, hidden sector particle, such as a sterile neutrino. Vacuum decays can then cause the dark matter density of the universe to grow at late times.  The combined phantom plus dark matter fluid has an effective equation of state $w<-1$, and functions as a new source of dark energy.
We derive constraints from cosmological observables on the rate of vacuum decay into such a phantom fluid.
We find a mild preference for the ghost model over the standard cosmological one, and a modest amelioration of the Hubble and $S_8$ tensions.

\end{abstract}

%\preprint{}

\maketitle

\tableofcontents

\section{Introduction}

Ghosts, or phantoms, are scalar fields with a wrong-sign kinetic term.
They have been widely studied in cosmology, as a source of 
matter with equation of state $w<-1$, as was suggested in early determinations of the dark energy equation of state
\cite{Hannestad:2002ur,Melchiorri:2002ux,Lima:2003dd,Bamba:2012cp}.  If such an object exists in nature, it must have not only classical behavior, but also quantum mechanical:
negative energy particles which are the quanta of the ghost field.~\footnote{They are unlike the Faddeev-Popov ghosts for
fixing gauge symmetries in quantum field theory, 
which carry positive energy but with negative probability.}

Even if such exotic particles have no direct couplings to the standard model (SM), they cannot avoid being gravitationally coupled.  As a result, through virtual graviton exchange, the vacuum is unstable to decay into pairs of ghosts plus positive-energy particles \cite{Carroll:2003st}, and the rate for such processes is unbounded unless there is an ultraviolet cutoff $\Lambda$ on the phase space
of the ghosts: in other words, any consistent theory of ghosts must be a low-energy approximation that breaks down at momenta larger than $\Lambda$.  Non-observation of excess gamma rays from vacuum decay leads to a bound that was roughly estimated as $\Lambda\lesssim 3\,$MeV \cite{Cline:2003gs}.

As pointed out in Ref.\ \cite{Cline:2003gs}, the low-energy effective theory of ghosts must be Lorentz-violating, since $\Lambda$ is a bound on the magnitude of the 3-momenta of the ghosts, which implies a preferred
reference frame.  Explicit, UV-complete models that give such a momentum cutoff were constructed in Refs.\ \cite{Holdom:2004yx,Rubakov:2006pn,Libanov:2007mq}.  In the present work, we will not be concerned with the microphysics underlying such ghost models, but rather with the 
late-time cosmological implications of the low-energy effective theory.

Another possible signature of phantom theories could be decay of the vacuum into ghosts plus dark matter (DM).  If both types of particles had the same dispersion relation, apart from the overall sign, there would be no net cosmological effect, at the homogeneous level, due to the cancellation of stress-energies of the two components.  Here we will consider the case of massless ghosts, which can have a nonvanishing effect, since the negative-energy density of the ghosts redshifts faster than that of the created dark matter.  The result is a net creation of DM at late times.  We will show that for purely gravitationally coupled ghosts, this process is too slow to give detectable results.  But it is consistent to couple the ghosts and the DM more strongly,
for example by exchange of a heavy gauge boson, allowing for significant late-time DM creation.

In the present work, we examine the effect of such late-time DM creation on the cosmic microwave background (CMB) and structure formation.  Even if the contributions of the two fluids cancel exactly at the homogeneous level, in the case of massless ghosts and massless sterile neutrinos, their perturbations do not cancel. This leads to constraints on the rate of DM
plus phantom fluid creation, and consequently on the energy scale of new physics linking phantom particles to dark matter.

Unlike most previous studies, we are interested here in the quantum production of the phantom particles, rather than a classical field condensate.  To keep these issues cleanly separated, we focus on the case of massless phantoms, whose potential vanishes.  It is therefore consistent to neglect any phantom vacuum expectation value in the present work. 

We start by revisiting the upper bound on $\Lambda$ from diffuse gamma rays in Section \ref{LambdaSect}, and make a fully accurate determination. 
In Section \ref{sect:model} we 
describe the particle physics model for a phantom sector coupled to dark matter, and
solve for the cosmological component densities created by decay of the vacuum. There we also derive preliminary constraints based on type Ia supernovae.  In Section \ref{sect:cmb}, the consequences for the CMB and large scale structure are studied using a full Monte Carlo search of the parameter space, resulting in an upper bound on the vacuum decay rate and corresponding new-physics energy scales.   We investigate the potential of the model for providing a new source of dark energy, and for addressing various cosmological tensions.
Conclusions are given in Section \ref{sec:conc}.

\section{Constraint on $\Lambda$}
\label{LambdaSect}
In Ref.\ \cite{Cline:2003gs}, a rough estimate was made of the 
rate for vacuum decay into ghosts and photons.  Here we compute
the rate more quantitatively.  The computation is identical to that
for gravitational scattering of ghosts into photons,
$\phi\phi\to\gamma\gamma$, except that the initial state particles
are treated as final state particles, and integrated over with
momenta restricted by $|\vec p_i| \le \Lambda$.  The matrix element
has been computed in Refs.\ \cite{Holstein:2016fxh,Bernal:2018qlk}, and is given by~\footnote{We use the interaction Lagrangian $\mathcal{L}_{\rm int} = h_{\mu \nu} T^{\mu \nu} / (2\, m_P)$, where $h_{\mu \nu}$ is the graviton field and $T^{\mu\nu}$ is the total energy-momentum tensor, as done in Ref.~\cite{Bernal:2018qlk}. This choice differs from the Lagrangian used in Ref.~\cite{Holstein:2016fxh} by a factor of $1/2$.}
\be
	%\sum|{\cal M}|^2 = {2\over m_P^4}\left(t u\over s\right)^2\,,
    \sum_{\text{spin}}|{\cal M}|^2 = {1\over 8 m_P^4}\left(t u\over s\right)^2\,,
\ee
where the Mandelstam invariants are defined in the usual way,
and we can pretend that the ghost energies are positive since the
momentum-conserving delta function is $\delta(E_1 + E_2 - E_3-E_4)$ where $E_{1,2}$ are ghost energies and $E_{3,4}$ are photon energies.
Here $m_P=2.43\times 10^{18}$\,GeV is the reduced Planck mass. 
%\jc{We are considering real ghosts here I presume.  We will get some small difference for complex ghosts (bound smaller by $2^{1/7}$ factor, do you agree)?}
%\mpuel{Yes you are definitely right that we are considering ghosts as real scalars. For complex ghosts. This translates into a factor $(1/2)^{1/9}$ in the bound of $\Lambda$. I will write it below.}

\begin{figure}[t]
\begin{center}
 \includegraphics[scale=0.29]{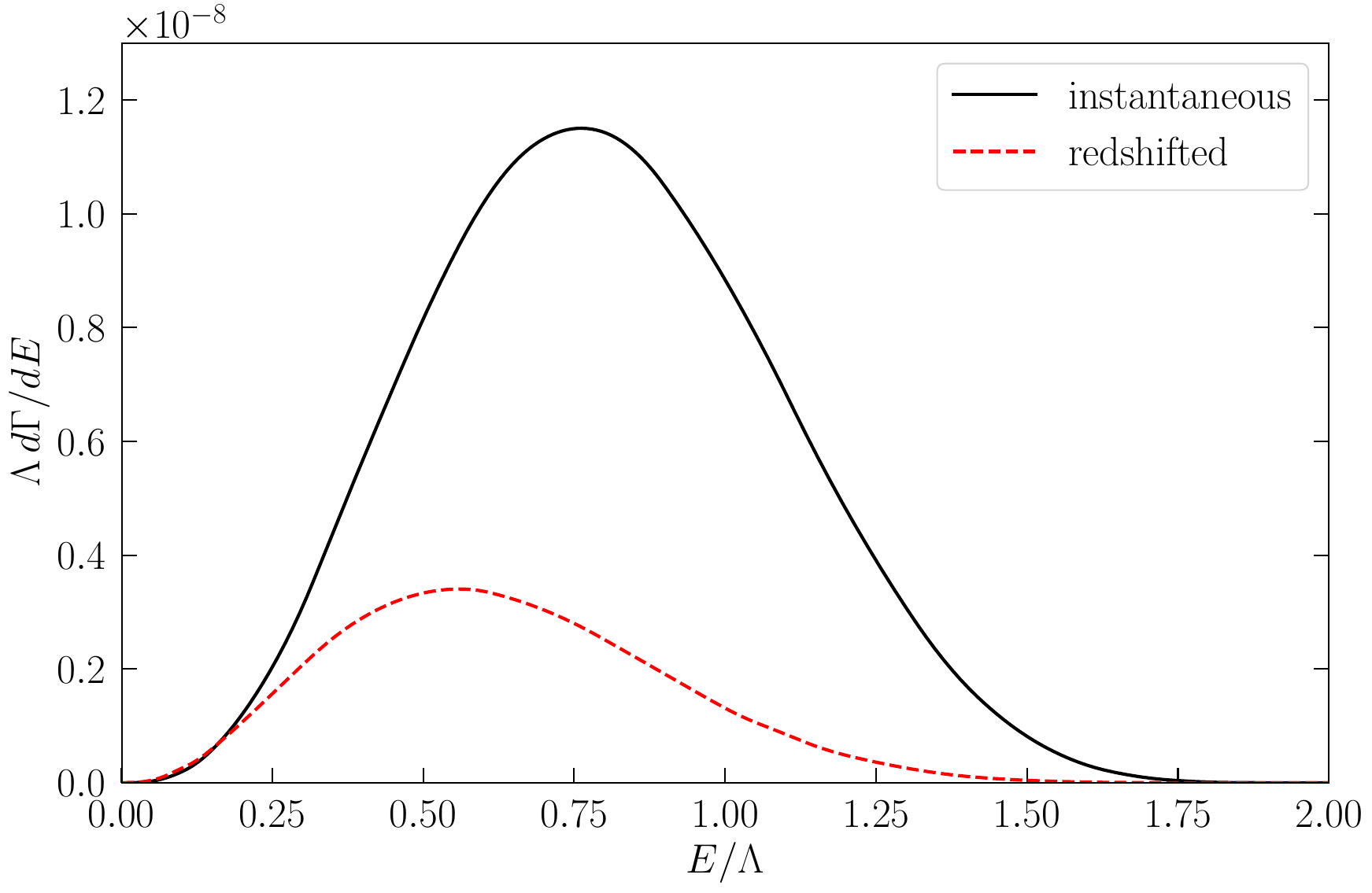}
 \caption{Solid: Instantaneous photon spectrum ($\Lambda\times$ decay rate per unit volume and energy) from vacuum $\to \phi\phi \gamma\gamma$
 (real scalar ghosts plus photons) decay, with units of $\Lambda^8/m_P^4$.  Dashed: Observed flux accounting for cosmological redshift, with units of $\Lambda^7/(4\pi H_0 m_P^4)$, as given by Eq.~\eqref{eq:redshifted_photon_spectrum}.}
 \label{fig:gam-spect}
\end{center} 
\end{figure}

Carrying out the phase space integrals as described in appendix
\ref{appA}, we first compute the instantaneous spectrum of gamma rays $d\Gamma/dE$.  The result is shown as the black solid curve in Fig.\ \ref{fig:gam-spect}.  However the observed flux is a line-of-sight integral, similar
to that for decaying dark matter \cite{Slatyer:2021qgc}, which can be expressed as an integral over the redshift of the corresponding emission distance,
\be
\label{eq:redshifted_photon_spectrum}
    {dJ\over dE} = {1\over 4\pi}\int_1^\infty dz\, {1\over H(z)(1+z)^3}\, {d\Gamma\over dE}((1+z)E)\,,
\ee
where the Hubble rate is $H(z) \cong H_0\sqrt{\Omega_m(1+z)^3 + \Omega_\Lambda}$, ignoring the small contribution from radiation. 
Here $\Omega_m \simeq 0.32$ and $\Omega_{\Lambda} \simeq 0.68$ are the total matter and dark energy abundance, respectively.
The result is shown as the red dashed curve in Fig.\ \ref{fig:gam-spect}, in units of $\Lambda^7 / (4 \pi H_0 m_P^4)$. 
This must be compared to the observed diffuse electromagnetic spectrum, which in the region of interest was measured by the COMPTEL experiment~\cite{2000AIPC..510..467W} to be approximately $dJ/dE \cong 5.3\times 10^{-3}\,(E / E_0)^{-2.4}\,{\rm MeV^{-1}\,cm^{-2}\,s^{-1}\,sr^{-1}}$ with $E_0 = 1\,\,\text{MeV}$.  
By demanding that the predicted flux in Eq.~\eqref{eq:redshifted_photon_spectrum} not exceed the observed one, we derive a $95\%$ Confidence Level (C.L.) limit of
\be
	%\Lambda \lesssim 14\,{\rm MeV}\,.
    \Lambda \lesssim 18.8\,{\rm MeV}\,.
	\label{Lambda_bound}
\ee
which is somewhat weaker than the rough estimate originally derived in Ref.\ \cite{Cline:2003gs}. Here we assumed the ghost particles are real scalars; the bound in Eq.~\eqref{Lambda_bound} should be reduced by a factor $(1/2)^{1/9.4} \approx 0.93$ for complex scalar ghosts,
giving $\Lambda\lesssim 17.5$\,MeV.

We have also estimated the constraint arising from vacuum decay to ghosts and $e^+e^-$ pairs; using the matrix elements computed in
Ref.\ \cite{Bernal:2018qlk} (see Eq.\ (\ref{eq:M2_gravity}) below), we find a decay rate per unit volume $\Gamma_d = \Lambda^8/(368640\,\pi^5\,m_P^4)$ in the approximation $m_e\ll\Lambda$. The present density of $e^\pm$ is of order
$n_e=\Gamma_d/H_0$, giving the rate $\Gamma_s \sim n_e\sigma$ for a CMB
photon to scatter on $e^\pm$, with $\sigma \sim 8\pi\alpha^2/(3\Lambda^2)$.
The probability to scatter is of order $\Gamma_s/H_0$, which should be less than $\sim 10^{-4}$ \cite{Fixsen:1996nj}.  This
leads to a bound of $\Lambda\lesssim 350\,$MeV, significantly weaker than Eq.\  (\ref{Lambda_bound}).

\section{Ghost/dark matter models}
\label{sect:model}
We next consider the decay of the vacuum into massless ghosts $\phi$ and massive hidden sector particles $\nu_s$, which for definiteness is considered to be a sterile Dirac neutrino.  (The main results are not expected to be sensitive to different possible choices, such as scalar dark matter.)
Such a process can increase the energy density of the Universe at late times, due to the faster redshifting of the ghost versus the massive particle contribution.

\subsection{Gravitational coupling}

The matrix element for the gravitational scattering between Dirac fermions with mass $m_{\nu_s}$ and scalar particles is~\cite{Bernal:2018qlk}
\be
\label{eq:M2_gravity}
    %|\mathcal{M}|^2 = \frac{1}{32\,m_P^4} \frac{-t (s + t) (s + 2 t)^2}{s^2}
    |\mathcal{M}_g|^2 = -\frac{(t - m_{\nu_s}^2) (s + t - m_{\nu_s}^2) (s + 2 t - m_{\nu_s}^2)^2}{32\,m_P^4\, s^2}\,,
\ee
where the Mandelstam invariants $s, t, u$ have their usual definitions, after changing $p_{1,2}\to -p_{1,2}$.
The vacuum decay rate, per unit volume, is given by the phase-space integral
\bea
\label{eq:decay_rate}
    \Gamma &=& (2\pi)^4 \int \prod_{i = 1}^4 {d^{\,3}p_i\over (2\pi)^3\,2 E_i}\,\delta(E_1 + E_2 - E_3 - E_4)\nn\\
    &\times &\delta^{(3)}({\vec p}_1 + {\vec p}_2 - {\vec p}_3 - {\vec p}_4)\,|{\cal M}_g|^2,
\eea
where we have taken $p_{1,2}\to -p_{1,2}$,
so that all energies are positive.  The ghost momenta $p_1$ and $p_2$ should be integrated up to the cutoff $\Lambda$.  
In appendix~\ref{appA} we carry out the numerical integration of the vacuum decay rate per unit volume.
By numerically fitting, the result can be approximated as~\footnote{In reality $\Gamma_g$ falls to zero when $m_{\nu_s} \to \Lambda$ since the phase space vanishes, but the small discrepancy with 
Eq.\ (\ref{eq:gammag}) is numerically insignificant for our purposes.\label{fnl} }
\be
\label{eq:gammag}
\Gamma_g \cong 4.4 \times 10^{-9}\,\frac{\Lambda^8}{m_P^4}\,\Big[1 - \big(\tfrac{m_{\nu_s}}{\Lambda}\big)^2\Big] \,\exp\!{\big[\!-\!5.3\big(\tfrac{m_{\nu_s}}{\Lambda}\big)^{4.2}\big]},
\ee
assuming the ghosts $\phi$ are real scalar particles,
and that $m_{\nu_s} \le \Lambda$ by energy conservation. 

One can estimate the expected energy density in $\nu_s$ today following Ref.~\cite{Cline:2003gs},
\be
\rho_{\nu_s} (t_0) \sim \Lambda\,n_{\nu_s} (t_0) \sim \Lambda\,\Gamma_{g}\,t_0\,,
\ee
where $n_{\nu_s}(t)$ is the number density of sterile neutrinos and $t_0 \sim 1 / H_0$ is the age of the Universe.   Taking the maximum value 
$\Lambda \sim 19\,$MeV and dividing by the critical density today $\rho_{\rm crit}$, we find that
$\Omega_{\nu_s} = \rho_{\nu_s}/\rho_{\rm crit} \lesssim 10^{-9}$, which is far below the sensitivity of observational constraints.  It follows that the vacuum decay rate must be $\sim 9$ orders of magnitude larger than that arising from purely gravitational interactions to have an observable cosmological impact.

\subsection{Scalar or vector coupling}

\begin{figure}[t]
\begin{center}
\includegraphics[width=8.5cm]{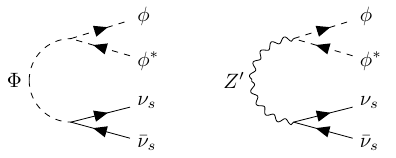}
\caption{Diagrams of vacuum decay mediated by a scalar $\Phi$ or a vector boson $Z^\prime$.}
\label{fig:1}
\end{center} 
\end{figure}

Since the gravitational interaction of $\nu_s$ is too weak to have appreciable cosmological effects, we introduce a heavy mediator coupling $\phi$ to $\nu_s$, which can be either a scalar $\Phi$ or a vector $Z'$. Integrating out the mediator results in an effective interaction of the form
\be
   {\Lambda\over M^2_s}|\phi|^2\, \bar\nu_s\nu_s\hbox{\quad or \quad }
    {i\over M_v^2}(\phi^*\!\!\stackrel{\leftrightarrow}{\partial}_\mu\!\phi)\,\bar\nu_s\gamma^\mu\nu_s,
    \label{effint}
\ee
where $M_s$ and $M_v$ are the new-physics mass scales, to be constrained.~\footnote{The choice of parametrization for the scalar coupling strength is motivated by assuming that the dimensionful coupling of the scalar to ghosts will not exceed the cutoff $\Lambda$.}
For simplicity, we take the ghost to be a complex scalar in both cases, and $\nu_s$ to be a Dirac fermion.  This choice insures there is no gauge anomaly for the vector mediator case. 
Figure~\ref{fig:1} shows the Feynman diagrams of the vacuum decay processes.

Labeling the ghost momenta as $p_{1,2}$ and the $\nu_s$ momenta as $p_{3,4}$, we obtain the following new-physics matrix elements for the vacuum decay
\be
\label{eq:M2}
|{\cal M}_n|^2 = \left\{\arraycolsep=1.4pt\def\arraystretch{2.2}
\begin{array}{ll}
   \frac{2\Lambda^2}{M_s^4}\left[s - 4 m_{\nu_s}^2\right],& \hbox{scalar}\\
   %\,[p_3\cdot p_4 - m_{\nu_s}^2], & \hbox{scalar}\\
    \frac{2}{M_v^4}\left[ s^2 - (t-u)^2\right],& \hbox{vector}
%    \frac{1}{M_v^4}\,[2q\cdot p_3\, q\cdot p_4 -
%    q^2(p_3\cdot p_4 + m_{\nu_s}^2)],& \hbox{vector}
    \end{array}\right.,
\ee
where again
the Mandelstam invariants $s, t, u$ have their
usual definitions, after negating the 4-momenta of the
ghosts so that kinematically they look like initial-state particles with positive energy.
%, despite the ghosts being in the final state.
%The vacuum decay rate, per unit volume, is given by the phase space integral
%\bea
%\label{eq:decay_rate}
%    \Gamma_n &=& (2\pi)^4 \int \prod_{i = 1}^4 {d^{\,3}p_i\over (2\pi)^3\,2 E_i}\delta(E_1 + E_2 - E_3 - E_4)\nn\\
%    &\times &\delta^{(3)}({\vec p}_1 + {\vec p}_2 - {\vec p}_3 - {\vec p}_4)\,|{\cal M}|^2
%\eea
%where we have taken $p_{1,2}\to -p_{1,2}$ (notice that $|{\cal M}|^2$ is invariant) so that all energies are positive.  The ghost momenta $p_1$ and $p_2$ should be integrated up to the cutoff $\Lambda$.
%In appendix~\ref{appA} we carry out the numerical integration of the vacuum decay rate per unit volume for both mediators. The results can be analytically approximated by
Integrating Eq.~\eqref{eq:decay_rate} numerically  we find that the decay rate for both mediators can be  approximately fit by the formulas~\footnote{See footnote \ref{fnl}.}
\be
\label{eq:Gamma}
\Gamma_n \cong 1.1 \times 10^{-5}\,\Lambda^8\,\left\{\arraycolsep=1.4pt\def\arraystretch{2.2}
\begin{array}{ll}
   \frac{1}{M_s^4}\,\exp{\big[- 6.7\, \big(\frac{m_{\nu_s}}{\Lambda}\big)^{2.1} \big]}, & \hbox{scalar}\\
   \frac{1}{M_v^4}\,\exp{\big[- 5.7\, \big(\frac{m_{\nu_s}}{\Lambda}\big)^{4.2} \big]},& \hbox{vector}
    \end{array}\right.,
\ee
(see appendix~\ref{appA}).

\subsection{Phantom sector spectra}
Once the massless ghosts $\phi$ and the massive sterile neutrinos $\nu_s$ are produced from the vacuum decay, their cosmological evolution is described by the Boltzmann equation in an expanding universe,
\be
\label{eq:Botlzmann_eq_f}
\frac{\partial f_{i}^0}{\partial t} - H p_i \frac{\partial f_{i}^0}{\partial p_i} = \SK\,,
\ee
where $p_i = |\vec{p}_i|$,  $t$ is the physical time, $H$ is the Hubble parameter, $f_{i}^0 (t, p_i)$ is the (dimensionless) phase-space distribution function and 
%$S_i \equiv \big(\frac{d f_i}{dt} \big)_{\mathcal{C}}^{(0)}$ 
$S_i$ is the collision or source term for the species $i = \phi, \nu_s$, both at the homogeneous background level. The latter quantity is defined by~\cite{Kolb:1990vq}~\footnote{The collision term for both species is positive since it describes the probability of collision. For massless ghosts, which carry negative energies, one has generally that $E_{\phi} = - p_{\phi} \leq 0$ with $p_{\phi} \geq 0$, and hence $f_{\phi}^0 \geq 0$, $n_{\phi} \geq 0$, $\rho_{\phi} \leq 0$, $P_{\phi} \leq 0$ and $w_{\phi} \geq 0$, from Eqs.~\eqref{eq:nus_density} and~\eqref{eq:rho_P_omega_def}. The negative pressure indicates that ghosts anti-gravitate.}
%\footnote{In principle the collision term should contain all possible interactions that affect the phase-space distributions. The interaction given by Eq.~\eqref{effint} allows also the scattering process $\phi \phi \leftrightarrow \nu_s \nu_s$ with ghosts in the initial state. However, this contribution is always subdominant to vacuum deacy, as discussed in appendix~\ref{appD}.}
\bea
\label{eq:dfidt_0}
\SK &=& \frac{(2\pi)^4}{2 E_i} \int \prod_{j \neq i}^4 {d^{\,3}p_j\over 2 E_j (2\pi)^3}\delta(E_1 + E_2 - E_3 - E_4)\nn\\
    &\times &\delta^{(3)}({\vec p}_1 + {\vec p}_2 - {\vec p}_3 - {\vec p}_4)\,|{\cal M}_n|^2 \,\mathcal{F}_0\,.
\eea
where we used the same labeling as in Eq.~\eqref{eq:decay_rate} (i.e., ghost energies are replaced by their absolute values), % with the $+$ sign for $i = \nu_s$ and the $-$ sign for $i = \phi$, 
and
\bea
\label{eq:F0}
\mathcal{F}_0 &=& [1 + f_{1}^0 (\vec{p}_1)]\,[1 + f_{2}^0 (\vec{p}_2)]\, [1 - f_{3}^0 (\vec{p}_3)]\, [1 - f_{4}^0 (\vec{p}_4)] \nn\\
&-& f_{1}^0 (\vec{p}_1)\, f_{2}^0 (\vec{p}_2)\, f_{3}^0 (\vec{p}_3)\, f_{4}^0 (\vec{p}_4)\,.
\eea
If the occupation number of the produced particles from the vacuum decay is small, namely $f_{i}^0 \ll 1$ (we will validate this assumption \textit{a posteriori}), then $\mathcal{F}_0 \simeq 1$ and the collision term in Eq.~\eqref{eq:dfidt_0} becomes independent of the phase-space distributions. This allows us to integrate Eq.~\eqref{eq:Botlzmann_eq_f} directly, obtaining~\citep{Bae:2017dpt,Dvorkin:2019zdi}
\bea
\label{eq:fi0_gen_sol}
f_{i}^0 (t, p_i) &=& \int_{t_{\rm in}}^{t} d t'\, \SK\Big( \tfrac{a(t)}{a(t')}\, p_i,\, t'\Big) \nn \\
&=& \int_{a_{\rm in}}^{a(t)} \frac{d a'}{a' H(a')}\,\SK \Big( \tfrac{a(t)}{a'}\, p_i,\, a'\Big)
\eea
where $t_{\rm in}$, or interchangeably the scale factor $a_{\rm in}$, is the initial time of integration, and the factor $a(t)/a(t')$ accounts for the redshifting of the momenta between $t'$ and $t$.  
We take $a_{\rm in} \simeq 10^{-8}$ to be well deep inside the radiation era, when the vacuum decay is completely negligible; our results are insensitive to the precise value of $a_{\rm in}$. We will compute the collision terms first.

The source term for the ghosts is relatively easy to evaluate, since the integrals over final-state particles have exactly the same form as for the cross section $\phi\phi\to\nu_s\nu_s$ of positive-energy massless $\phi$ particles.  
One can as usual perform this calculation in a Lorentz-invariant manner to obtain the cross section as a function of the Mandelstam variable $s = (p_1^\mu+p_2^\mu)^2 = 2 p_1 p_2 (1-\cos\theta)$, where $\theta$ is the angle between $\vec p_1$ and $\vec p_2$,
\be
\label{eq:dfphidt_0}
    \SKg (p_1) = %- 
    {1\over  2 (2\pi)^2 \,p_1} \int d \cos\theta \int_0^\Lambda{dp_2\, p_2}\, s\, \sigma(s)\,,
\ee
with
\be
\label{eq:scattering_cross_section}
    \sigma(s) = \left\{\arraycolsep=1.4pt\def\arraystretch{2.2}
    \begin{array}{ll}       \frac{\Lambda^2}{8 \pi\,M_s^4} \Big(1 - 4 \frac{m_{\nu_s}^2}{s}\Big)^{3/2} \,,& \hbox{scalar}\\ \frac{s+2 m_{\nu_s}^2}{12\pi\,M_v^{4}}\,\sqrt{1 - 4 \frac{m_{\nu_s}^2}{s}} \,,& \hbox{vector}
    \end{array}\right.\,.
\ee
The neutrino spectrum is slightly more complicated to calculate because the phase space of the ghosts is restricted by the momentum cutoff $\Lambda$; hence one cannot write a completely analogous expression involving the cross section for $\nu_s\nu_s\to \phi \phi$.  Instead, 
\bea
\label{eq:dfnusdt_0}
\SKnu (p_3) &=& \frac{1}{16 (2\pi)^4} {1\over E_3}\int dc_{12}\,dc_{13}\,d\varphi \nn\\
&\times& \int_0^\Lambda dp_2\,{p_1^2\, |{\cal M}|^2\,\Theta(\Lambda-p_1)\over E_3-p_3 c_{23}}\,,
\eea
where $c_{ij}=\hat p_i\!\cdot\!\hat p_j$,
and $p_1 = p_2 \,(E_3-c_{23}p_3)/(p_2(1-c_{12})-(E_3-c_{13} p_3))$.  The integrals in Eqs.~\eqref{eq:dfphidt_0} and~\eqref{eq:dfnusdt_0} can be evaluated numerically and they correspond to the instantaneous spectrum $d\Gamma_n / dp_i$ once multiplied by the factor $4\pi\,p_i^2 / (2\pi)^3$.

\begin{figure*}[t]
\begin{center}
\centerline{ %\includegraphics[scale=0.4]{mnu-spectra}\hfil
%\includegraphics[scale=0.4]{ghost-spectra}}
%\caption{Left: Neutrino momentum spectra, with arbitrary normalization, for several choices of $m_\nu/\Lambda$.  Right: redshifted ghost spectra, preserving relative normalization between the different choices of $m_\nu$, due to phase space suppression. }
\includegraphics[scale=0.3]{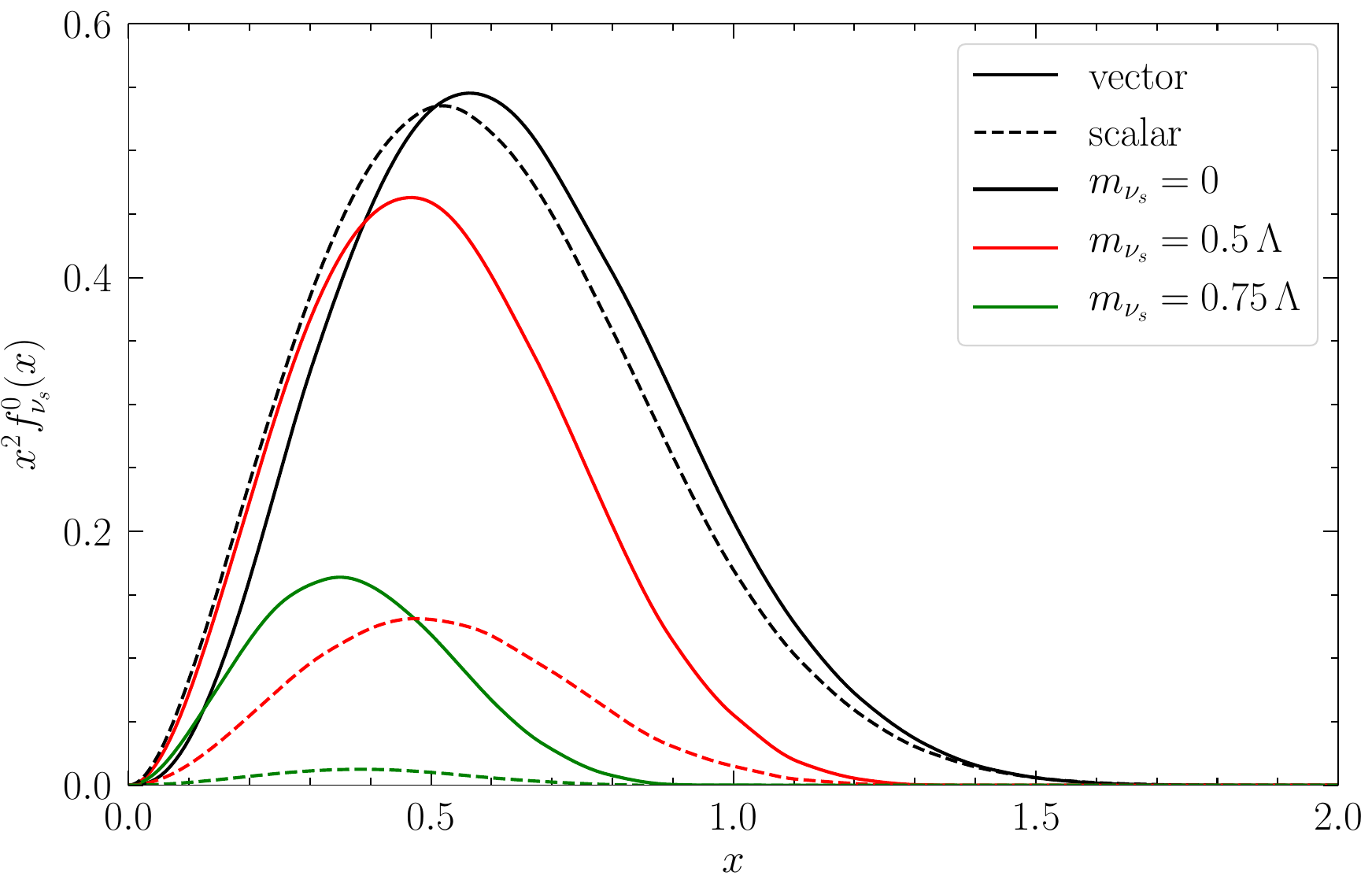}\hfil
\includegraphics[scale=0.3]{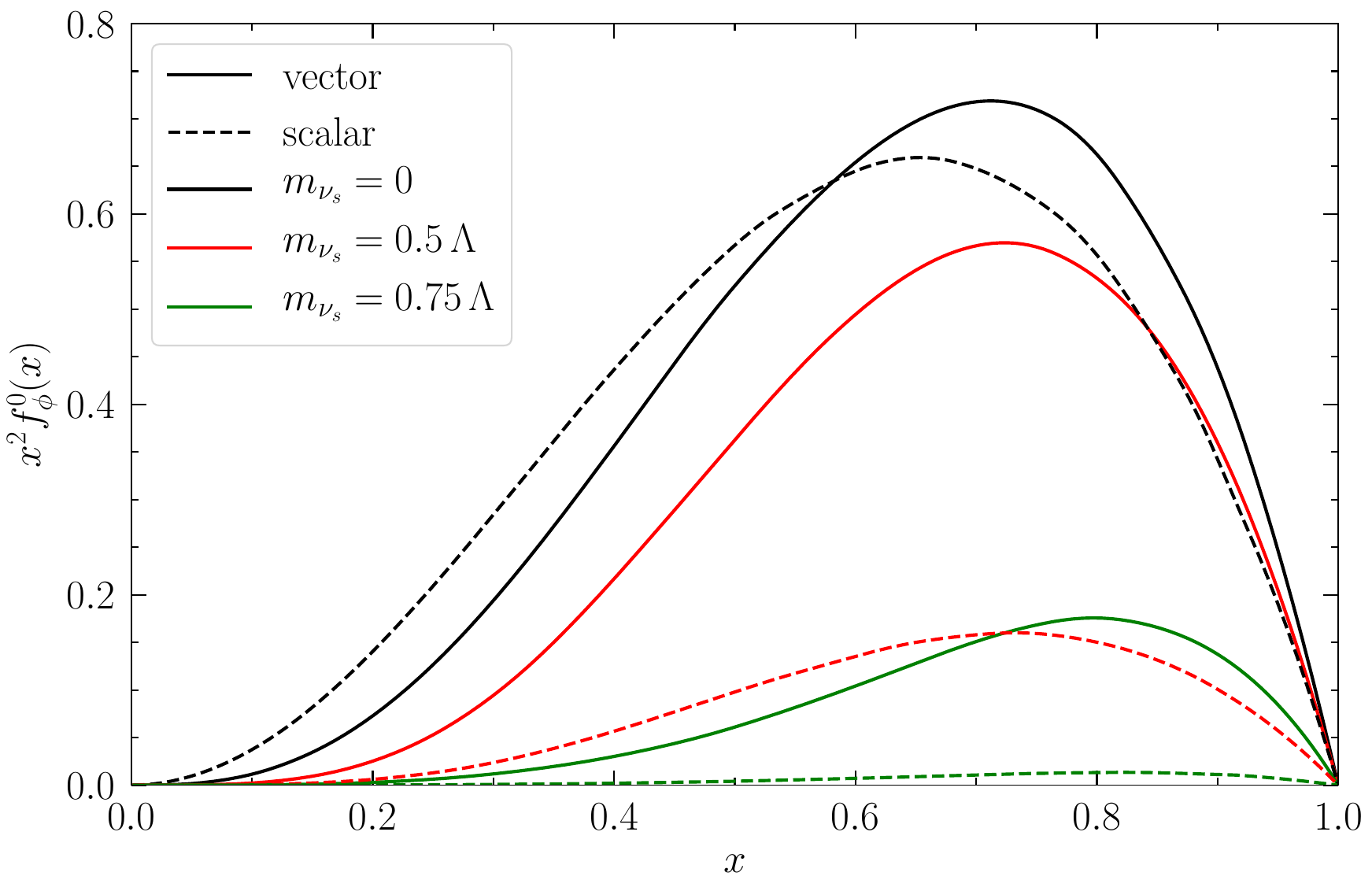}}
\caption{\textit{Left}: Present-day sterile neutrino momentum spectrum, divided by 
$\Lambda^5 / (128 \pi^6 M_i^4 H_0)$, with $M_i=M_s$ (scalar mediator) or $M_v$ (vector mediator), for several choices of $m_{\nu_s}/\Lambda$. Here $x = p / \Lambda$ and we used $\Omega_m = 0.32$. Solid (dashed) curves are for the vector (scalar) mediator model.  \textit{Right}: Momentum distribution for ghosts.}
 \label{fig:spectra}
\end{center} 
\end{figure*}

We can use the computed collision terms to derive the phase-space distributions at the present time, which are given by Eq.~\eqref{eq:fi0_gen_sol} with $t = t_0$ or equivalently $a = a_0 = 1$ and where
\be
\label{eq:Friedmann}
H (a) = H_0 \sqrt{ \frac{\Omega_{m}}{a^3} + \frac{\Omega_r}{a^4} + \Omega_{\Lambda} + \Omega_g (a)}\,.
\ee
Here, $\Omega_m$, $\Omega_r$ and $\Omega_{\Lambda}$ are the relative abundances of total matter (baryonic plus cold dark matter), radiation (photons plus massless SM neutrinos) and dark energy described by a cosmological constant, respectively, and the contribution from the ghost plus $\nu_s$ densities is $\Omega_{g}(a) = (\rho_{\phi} (a) + \rho_{\nu_s} (a))/{\rho_{\rm crit}}$ with the critical density defined as usual, $\rho_{\rm crit} = 3 H_0^2 / (8 \pi G)$.
In practice, one can neglect $\Omega_r$ since the ghost fluid is produced mainly at late times.  In the following, we will assume for simplicity a flat universe, 
which is also the expectation after a period of early-time acceleration dubbed as inflation~\cite{Guth:1980zm}.
%, so that 
In this way, we have
$\Omega_{\Lambda} + \Omega_m + 
\Omega_g = 1$, where we have defined $\Omega_g \equiv \Omega_g (a_0)$ for simplicity.
%In the following, we will take $\Omega_m = 0.32$ and $\Omega_{\Lambda} + \Omega_m + \Omega_g^{0} = 1$, with $\Omega_g^0 \equiv \Omega_g (a_0)$, unless stated otherwise.

\begin{comment}
As a first approximation, we can neglect the back-reaction of ghosts and sterile neutrinos in the Hubble expansion rate, which is generally a good assumption as we will show later.
Neglecting the small contribution from early times during radiation domination, the Friedmann equation reduces to
\be
\label{eq:H_a_approx}
H (a) \approx H_0 \sqrt{ \frac{\Omega_{m}}{a^3} + (1 - \Omega_m) }
\ee
where we replaced $\Omega_{\Lambda}$ with $1 - \Omega_m$ to satisfy the requirement that $H(a_0) = H_0$ today. We will take $\Omega_m = 0.32$ in the rest of the paper if not otherwise stated.
\end{comment}

The distribution functions in~\eqref{eq:fi0_gen_sol} are normalized such that the particle number densities are given by~\footnote{In Eq.~\eqref{eq:nus_density} and from now on, we include the numerical factor $4\pi / (2\pi)^3$ in the definition of the phase-space distribution $f_{i}^0$, given by Eq.~\eqref{eq:fi0_gen_sol}.}
\bea
\label{eq:nus_density}
n_i (t) &=& % \frac{4 \pi}{(2\pi)^3} 
\int dp_i\,p_i^2 f_{i}^0 (t, p_i) \nn\\
&=&% \frac{\Lambda}{2 \pi^2}
\Lambda^3\int dx\,x^2 f_{i}^0 (t, x)\,.
\eea
%where $x^2 f_{i}^0 (t, x)$ includes a factor of $ \Lambda^5 M_i^{-4}/(128 \pi^6 H_0)$, and $M_i= M_s$ or $M_v$ depending on whether the mediator is a scalar or a vector respectively.
The shapes of the spectra at the present epoch for the vector and scalar mediator models are shown in Fig.~\ref{fig:spectra}. 
It is straightforward to show that the redshifted spectra $f_{i}^0 (t_0, p_i)$ computed via Eq.~\eqref{eq:fi0_gen_sol} are identical to those derived through Eq.~\eqref{eq:redshifted_photon_spectrum}.

It can be convenient to have an explicit formula for $f^0_{\nu_s}$ in the case where $m_{\nu_s} = 0$
and vector mediator,  
\be
    p^2 f^0_{\nu_s}(p) \,dp = A\, \Lambda^4 H_0^{-1} x^2 e^{-\pi x^2}\, dx\,,
    \label{eq:nusdensity}
\ee
where $x=p/\Lambda$ and $A = 7.7\times 10^{-4}(\Lambda/M_i)^4$.   Although this is a numerical fit, the coefficient $\pi$ in the exponent turns out to be accurate to three digits.~\footnote{For other choices of $m_{\nu_s}$ we find similar fits by replacing $x^2 e^{-\pi x^2}\to
(2/3) x^{1.59}e^{-4\, x^{2.28}}$ for $m_{\nu_s}/\Lambda = 0.5$ and  $\to 0.349\,x^{1.57}e^{-8.255\,x^{2.39}}$ for $m_{\nu_s}/\Lambda = 0.75$\,.}

\begin{figure*}[t]
\begin{center}
\centerline{ \includegraphics[scale=0.3]{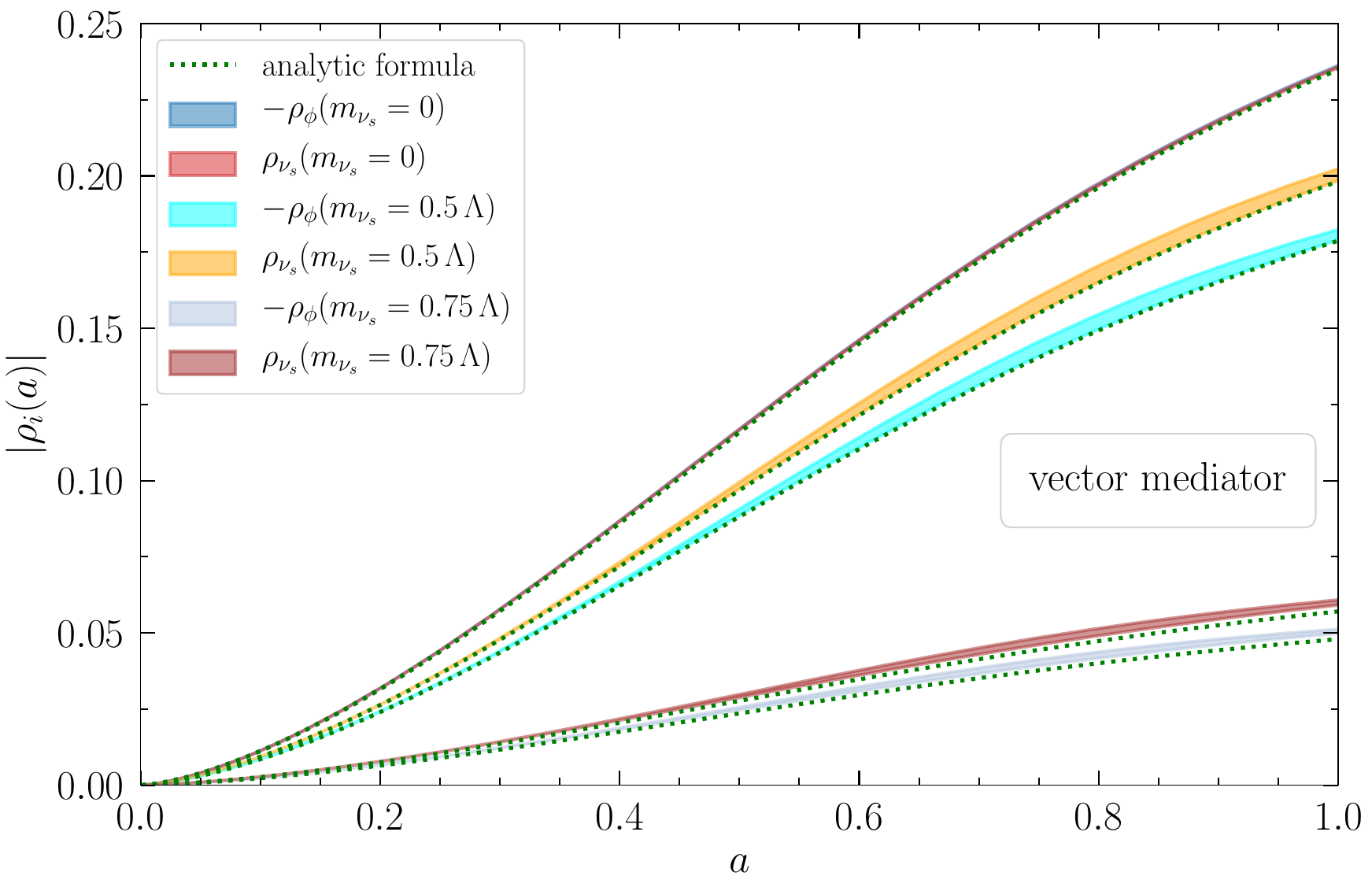}\hfil
\includegraphics[scale=0.3]{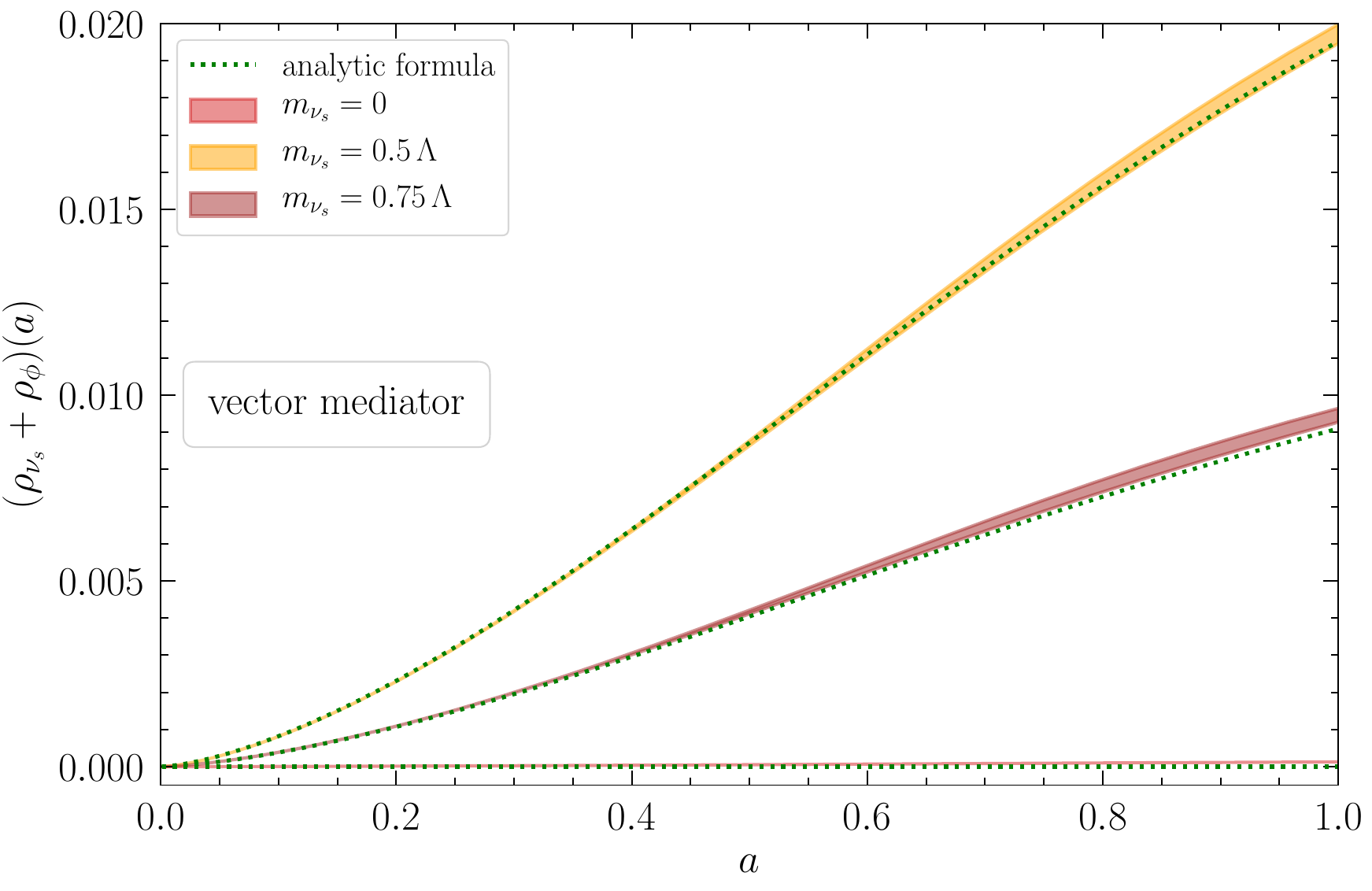}
}
\caption{Scale-factor evolution of the absolute value of the energy densities for ghosts $\phi$ (which have
$\rho_{\phi} < 0$) and sterile neutrinos $\nu_s$ 
(\textit{Left}), and of their sum (\textit{Right})
for the vector mediator 
case.
%(\textit{Left}) and scalar mediator case (\textit{Right}). 
The units of $\rho_{i} (a)$ are $\Lambda^9 / (128\pi^6 M_j^4 H_0)$ ($j = v, s$). 
Solid lines are the result of numerically solving Eq.~\eqref{eq:rho_P_omega_def}. The color shaded band represents the uncertainty on the back-reaction from $\Omega_g (a)$ in the Friedmann equation~\eqref{eq:Friedmann} for the values of $\Gamma_{\rho}$ allowed by Type Ia supernova (see Section~\ref{sec:TypeIaSNe}).
The green dotted lines are given by the analytic formulas in Eq.~\eqref{eq:energy_densities_approx}, derived from the Boltzmann equations in~\eqref{eq:Boltzmann_eq_t} when back-reaction is neglected.
For both plots, we used $\Omega_m = 0.32$ and $\Omega_{\Lambda}$ is determined by the flatness condition $\Omega_{\Lambda} = 1 - \Omega_m - \Omega_g$.
}
\label{fig:densities}
\end{center} 
\end{figure*}

\begin{figure*}[t]
\begin{center}
\centerline{ \includegraphics[scale=0.3]{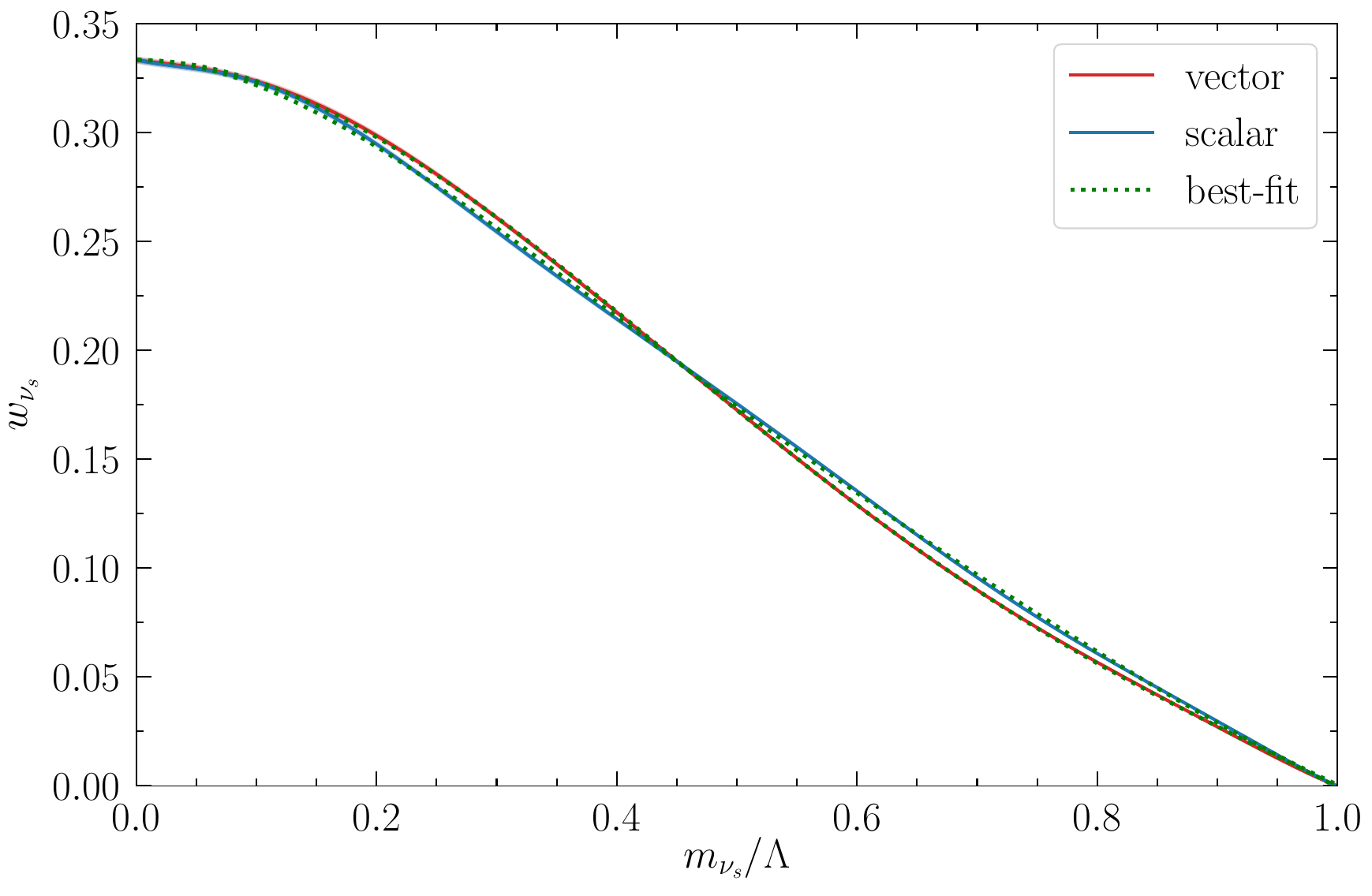}\hfil
\includegraphics[scale=0.3]{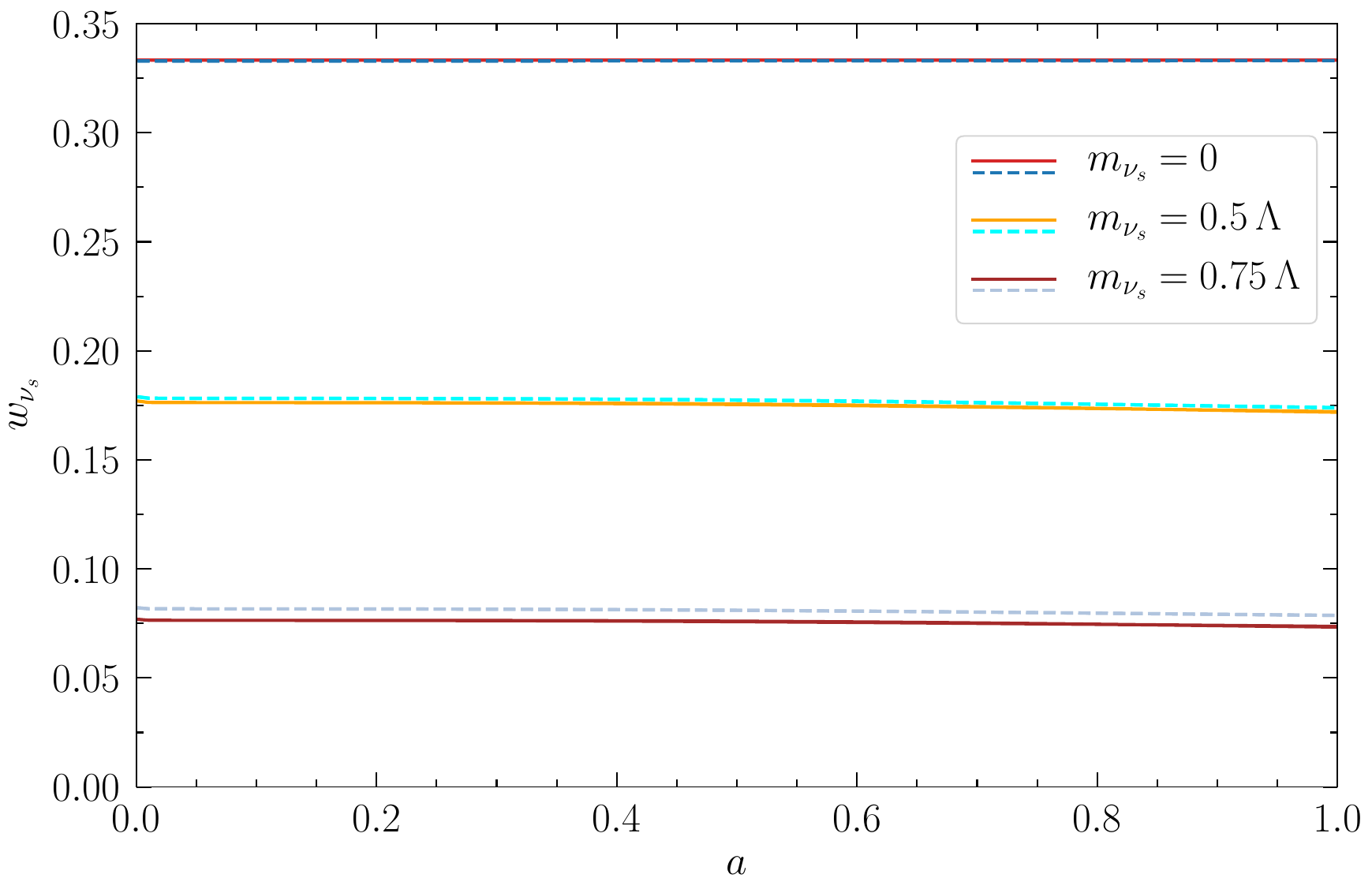}}
\caption{\textit{Left}: Present-day equation of state of sterile neutrinos $w_{\nu_s}$ as a function of the sterile-neutrino mass to cutoff ratio. The numerical results are shown with solid lines, while green dotted curves correspond to the numerical-fit functions 
(\ref{eq:eos_nus_m}). \textit{Right}: Scale-factor evolution of $w_{\nu_s}$ for different choices of $m_{\nu_s}$. The solid curves are for the vector mediator model, whereas the dashed curves are for the scalar model. The band thicknesses and 
 $\Omega_m $ and $\Omega_{\Lambda}$ are as in
 Fig.\ \ref{fig:densities}.
}
 \label{fig:eos_vs_a_m}
\end{center} 
\end{figure*}

\subsection{Phantom fluid evolution}
Treating the ghosts and sterile neutrinos as cosmological fluids, one can compute their energy density, pressure and equation of state (EOS) at the background level as
\bea
\label{eq:rho_P_omega_def}
    \rho_i (t) &=& %\frac{4 \pi}{(2\pi)^3} 
    \int dp_i\,p_i^2\,E_i\, f_{i}^0 (t, p_i)\,, \nn\\
    P_i (t) &=& %\frac{4 \pi}{(2\pi)^3} 
    \int dp_i\,p_i^2\,\frac{p_i^2}{3 E_i}\, f_{i}^0 (t, p_i)\,, \nn\\
    w_i (t) &=& \frac{P_i (t)}{\rho_i (t)}\,,
\eea
where $f_{i}^0 (t, p_i)$ is given by Eq.~\eqref{eq:fi0_gen_sol}, and $E_\phi \leq 0$. 
Figure~\ref{fig:densities} shows the scale-factor evolution of the energy densities of $\phi$ and $\nu_s$ for different values of the sterile neutrino mass $m_{\nu_s}$,
in addition to their sum. 
As expected, in the case of massless $\nu_s$ there is no net contribution of the two new species to the total energy density of the Universe, while for $m_{\nu_s}>0$, there is a clear difference between $|\rho_i (a)|$ for the two components.

The energy densities shown in Fig.~\ref{fig:densities} can be derived in a simpler way, directly from the Boltzmann equation~\eqref{eq:Botlzmann_eq_f}, after weighing by the energy $E_i$ and integrating over the particle three-momentum. This gives
\bea
\label{eq:Boltzmann_eq_t}
	{d\rho_{\phi}\over dt} + 4H \rho_{\phi} &=& -\Gamma_\rho,\nn\\
	{d\rho_{\nu_s}\over dt} + 3H (1 + w_{\nu_s}) \rho_{\nu_s} &=& +\Gamma_\rho,
\eea
where $\Gamma_{\rho}$ is the rate of change in the particle energy density due to the vacuum decay, computed in appendix~\ref{appB} (see Eq.~\eqref{eq:Gamma_rho}) which gives approximately
\be
\label{eq:Gamma_rho_approx}
\Gamma_{\rho} \cong
\Gamma_n\,\Lambda\,\times 
\left\{
\begin{array}{ll} 0.76,& \hbox{scalar}\\ 0.82,& \hbox{vector}
\end{array}\right. \equiv c_m\, \Gamma_n\,\Lambda\,,
\ee
in terms of $\Gamma_n$ from Eq.~\eqref{eq:Gamma}.

In Eq.~\eqref{eq:Boltzmann_eq_t}, the equation of state for sterile neutrinos $w_{\nu_s}$ depends upon $m_{\nu_s}$, and it could
{\it a priori} also be a function of time.  $w_{\nu_s}$ ranges between zero for nonrelativistic particles when $m_{\nu_s} \sim \Lambda$ to $1/3$ for relativistic ones, when $m_{\nu_s} \sim 0$. This is confirmed by Fig.~\ref{fig:eos_vs_a_m} (left), which shows the present-day $w_{\nu_s}$ computed via Eq.~\eqref{eq:rho_P_omega_def} as a function of $m_{\nu_s}$. 
We find that a good fitting formula is
\be
\label{eq:eos_nus_m}
    %w_{\nu_s} (\tilde{x}) \simeq 0.33 + 0.04 \tilde{x} - 1.29 \tilde{x}^2 + 1.40 \tilde{x}^3 - 0.48 \tilde{x}^4
    w_{\nu_s} (\tilde{x}) \simeq 
    \left\{%\arraycolsep=1.0pt\def\arraystretch{2.2}
    \begin{array}{ll} 1/3 - 1.33\, \tilde{x}^2 + 1.92\, \tilde{x}^3 - 1.07\, \tilde{x}^4 + 0.16\, \tilde{x}^6, \\%&
%    \hbox{\qquad\quad\qquad\qquad\qquad\qquad scalar mediator};\\ 
1/3 - 1.05\, \tilde{x}^2 + 0.82\, \tilde{x}^3 + 0.01\, \tilde{x}^4 - 0.12\, \tilde{x}^6, \\%& 
%    \hbox{\qquad\quad\qquad\qquad\qquad\qquad vector mediator}.
    \end{array}\right.
\ee
respectively for the vector and scalar mediator,
where $\tilde{x} \equiv m_{\nu_s} / \Lambda \in [0, 1]$. 

The $\nu_s$ equation of state turns out to be time independent, to a good approximation, as displayed in Fig.~\ref{fig:eos_vs_a_m} (right).
It can be shown that the time independence is exact whenever the scale factor has a simple power law behavior, $a(t) = t^p$, because in that case the integrals in Eq.\ (\ref{eq:rho_P_omega_def}) factorize to the form $\rho_i = F(t)\hat\rho_i(p)$ and $P_i = F(t)\hat P_i(p)$
such that the time dependence cancels in their ratio.  Since most of the ghost fluid production takes place while the Universe is approximately matter dominated, this analytic behavior is realized by the full numerical solution.

\begin{comment}
This behavior may be physically understood by the continuous out-of-equilibrium  creation of ghosts and sterile neutrinos at a constant rate in time (since the source term in Eq.~\eqref{eq:Boltzmann_eq_t} is constant), which induces the same time dependence for the energy density and pressure, since the redshifting effect acts equally on both quantities.
\resp{We checked that the inclusion of back-reaction of the new species in the Friedmann equation~\eqref{eq:Friedmann} does not spoil such a conclusion for the values of $\Gamma_{\rho}$ allowed by cosmological data, as discussed next.}
\end{comment}

It is convenient to recast the
Boltzmann equations \eqref{eq:Boltzmann_eq_t}
in dimensionless form, with $y=a^3$ playing the role of time,
\bea
\label{eq:Boltzmann_y}
    {d\Omega_{\phi}\over dy} &+& {4\over 3y}\Omega_{\phi} = -{
    \epsilon_\rho\over y \hat H(y) }\,,\nn\\
    {d\Omega_{\nu_s}\over dy} &+& {1+w_{\nu_s}\over y}\Omega_{\nu_s} = {\epsilon_\rho\over y \hat H(y) }\,,\nn\\
    \hat H(y) &=& \sqrt{\Omega_{\Lambda} + \Omega_m/y + \Omega_g(y)}\,,
\eea
and
 $\epsilon_\rho \equiv \Gamma_\rho/(3 H_0 \rho_{\rm crit})$.
Evidently, in the limit of massless $\nu_s$, $w_{\nu_s} = 1/3$ and the net contribution $\Omega_{g} (y) =\Omega_{\phi} (y) + \Omega_{\nu_s} (y)$ remains exactly zero, as expected.

Taking massless ghosts and massive neutrinos rather than vice versa is an arbitrary model-building choice.  The system of equations for the alternative choice is trivially related, by exchanging the roles of $\Omega_\phi$ and $\Omega_{\nu_s}$ in Eq.\ (\ref{eq:Boltzmann_y}), and relabeling $w_{\nu_s}\to w_\phi$.  The first choice results in $\Omega_g > 0$ while the second gives $\Omega_g < 0$.  We will consider the general case in the following, which results in 
a sign choice $\pm$ for the solutions given below.

By solving the system (\ref{eq:Boltzmann_y}) numerically, one finds that the net production $\Omega_{g} (y)$ is approximately
linear in $\epsilon_{\rho}$, for $\epsilon_\rho \lesssim \mathcal{O}(10)$.  For larger $\epsilon_\rho$, linearity breaks down because the back-reaction of
the ghost plus $\nu_s$ fluid appears in $\hat H$ and reduces the production; however  such large values are
ruled out by observations of late-time cosmology, namely supernovae, as we will show.
Hence we focus on the linear regime.  Useful approximate fits to the numerical solutions are given by 
\bea
    \label{eq:Ofits}
    \Omega_{\nu_s/\phi} (y) &=& \pm\epsilon_\rho\,A_0\, y^{A_1 - A_2 y - A_3 y(1-y)}\,,\\
    \Omega_{g}(y)&\cong& \pm \epsilon_\rho\,
    f(w_{\nu_s})
    \left(1 + 0.109\,\ln y\right)^{3.09}\,,\nn\\
   f(w_{\nu_s}) &\cong& 0.163 -0.594\,w_{\nu_s} +0.321\,w_{\nu_s}^2\nn
\eea
where the coefficients $A_i$ are linear functions of the equation-of-state parameter $w$, shown in Table \ref{tab:fits}, and the 
factor  $(1 + 0.109\,
\ln y)^{3.09}$ should be interpreted as zero at small $y$
where it would become negative.
The choice of sign corresponds to taking massless ghosts and massive
$\nu_s$ ($+$) or vice versa ($-$).
For small $m_{\nu_s}$, one can read from Eq.\ (\ref{eq:eos_nus_m}) that
$(1/3-w_{\nu_s}) \cong C_m(m_{\nu_s}/\Lambda)^2$ with 
$C_m \cong 4/3$ for the scalar mediator and $C_m \cong 1$
for the vector mediator.

\begin{table}[t]\centering
\setlength\tabcolsep{5pt}
\def\arraystretch{1.5}
\begin{tabular}{| c |}
\hline
$A_0 \approx 0.7969 - 0.486\,  w$\\
$A_1 \approx 0.4128 - 0.017\,  w$\\
$A_2 \approx 0.0587 - 0.003\,  w$\\
$A_3 \approx 0.4296 + 0.076\,  w$\\
\hline
\end{tabular}
\caption{Dependence of the coefficients $A_i$ in Eq.\ (\ref{eq:Ofits}) on the equation-of-state parameter $w$.  For massive neutrinos, $w = w_{\nu_s} < 1/3$, for massless ghosts
$w = 1/3$, and vice versa for massless neutrinos and massive ghosts.}
\label{tab:fits}
\end{table}

\begin{figure*}[t]
\begin{center}
\includegraphics[scale=0.35]{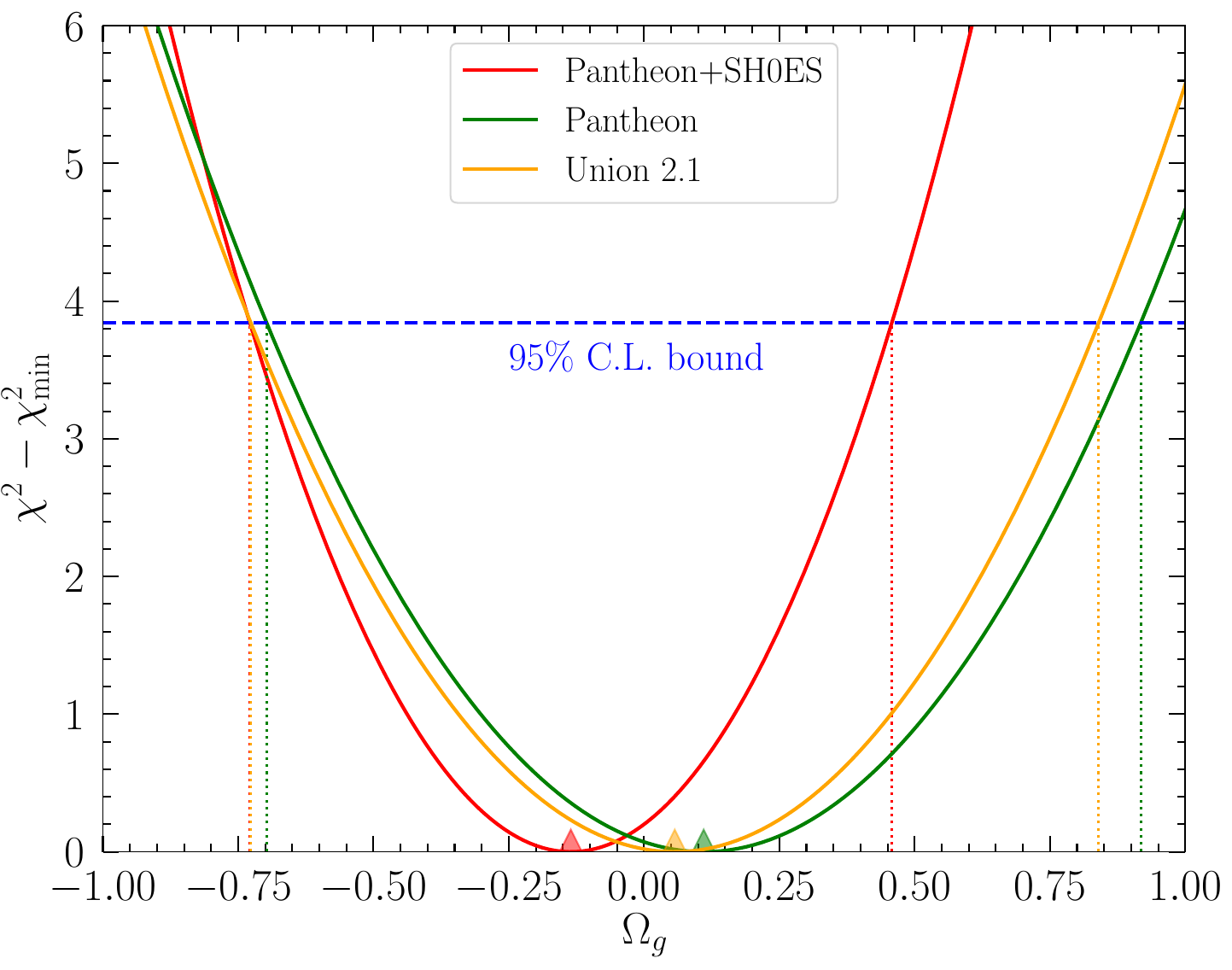}
\includegraphics[scale=0.345]{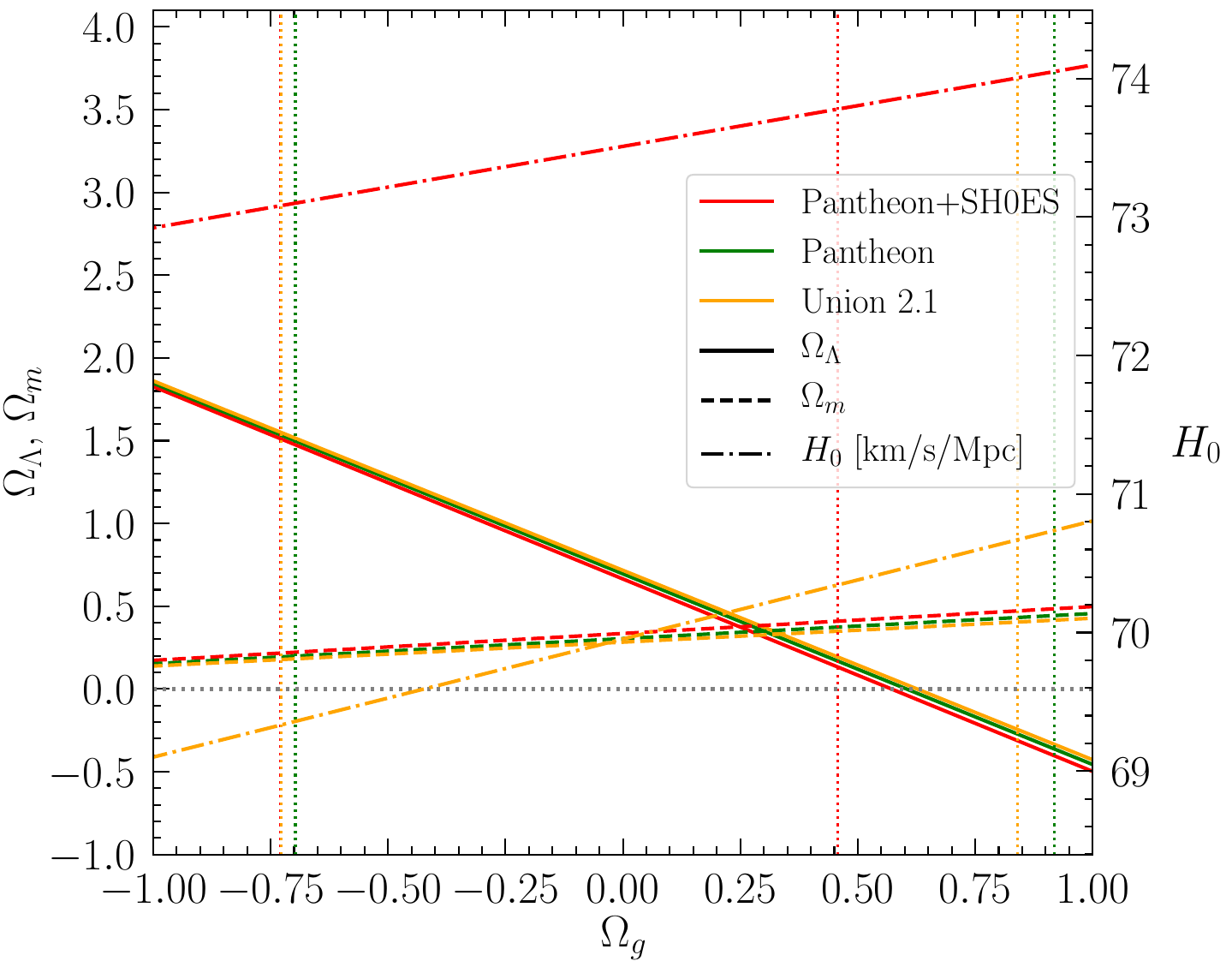}
\caption{\textit{Left}: $95\%$ C.L.\ allowed regions for the ghost plus sterile neutrino contribution to the present composition of the Universe, for the three supernovae data sets considered. 
The triangles indicate the best-fit values for $\Omega_g$.
\textit{Right}: Most likely values of $\Omega_\Lambda$ (solid), $\Omega_m$ (dashed) and $H_0$ (dot-dashed) as a function of 
$\Omega_g$.  The 95\% C.L.\ regions of the latter are indicated as dotted vertical lines.} 
\label{fig:SN_likelihood}
\end{center} 
\end{figure*}

\begin{figure*}[t]
\begin{center}
\includegraphics[scale=0.332]{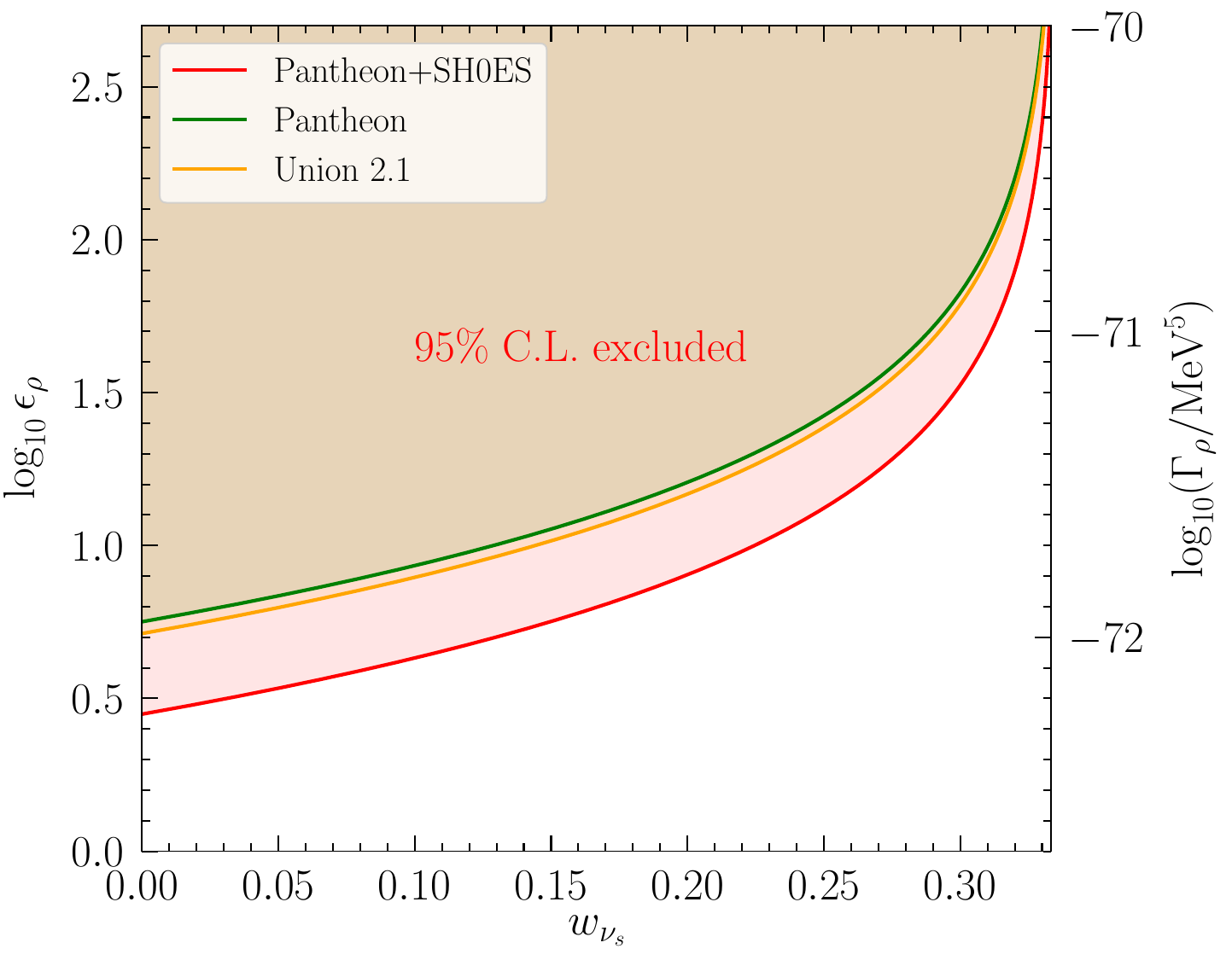}
\includegraphics[scale=0.326]{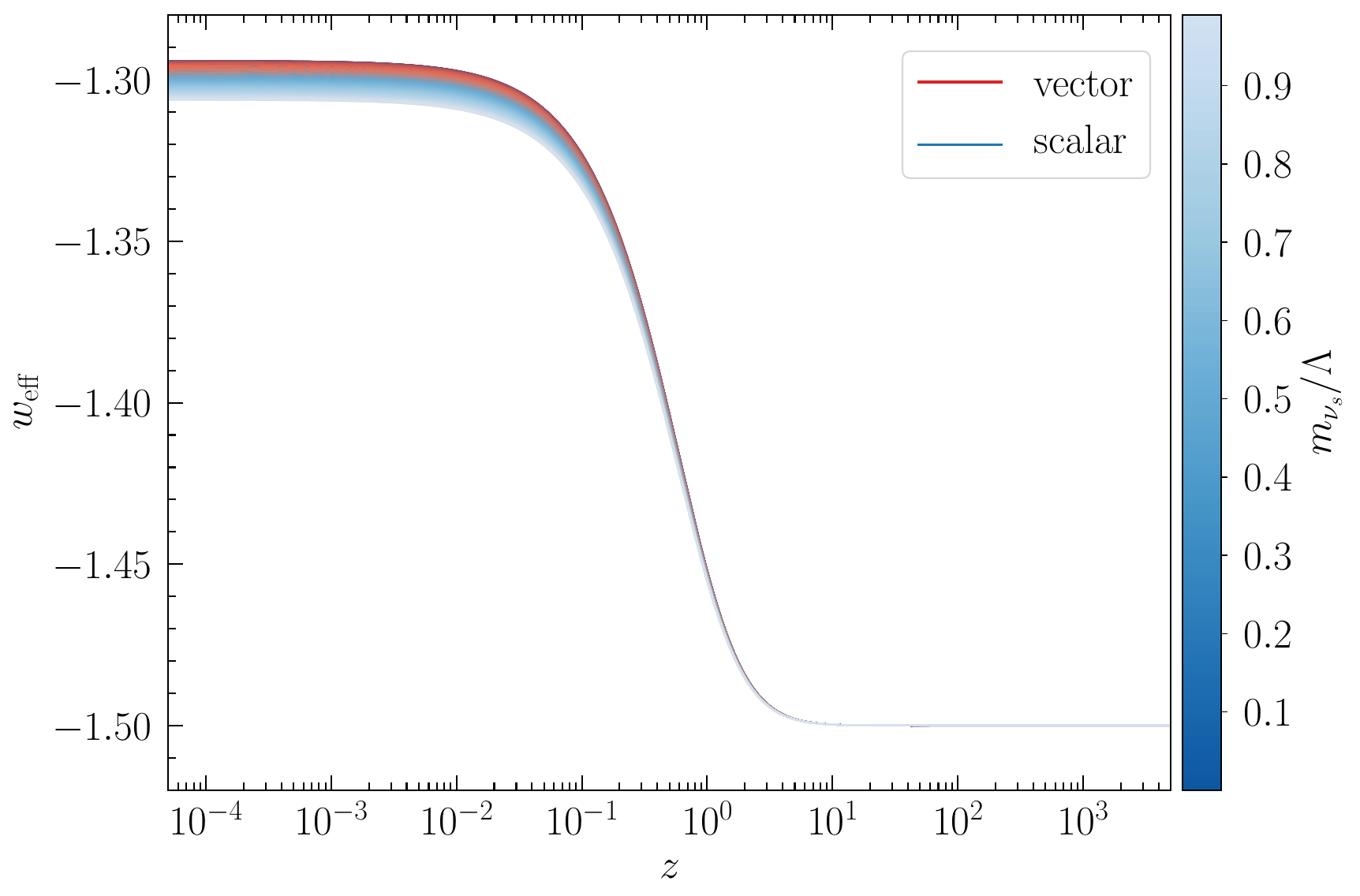}
\caption{\textit{Left}:
Mapping of Fig.~\ref{fig:SN_likelihood} to the parameters $\epsilon_{\rho} = \Gamma_\rho/(3H_0\rho_{\rm crit})$ or $\Gamma_{\rho}$ as a function of the equation-of-state $w_{\nu_s}$ for sterile neutrinos. 
\textit{Right}: Redshift evolution of the effective equation-of-state parameter $w_{\rm eff}$ for the $\phi+\nu_s$ fluid. The band colors distinguish the vector and scalar mediator models respectively. $\Omega_m$ and $\Omega_{\Lambda}$ are fixed as in previous figures ($w_{\rm eff}$ depends only weakly on these assumptions).
} 
\label{fig:SN_bounds}
\end{center} 
\end{figure*}

\subsection{Constraints from type Ia Supernovae} \label{sec:TypeIaSNe}
Because the net fluid of ghosts plus sterile neutrinos is produced at late times, a sensitive probe of its effect on cosmology is the use of supernovae as standard candles \cite{Perlmutter_1999,Riess_1999}.
Below, in Section \ref{sect:massive_nu}, we will do a full analysis including other cosmological data sets.  This preliminary study serves to define
the parametric region of interest for the later Monte Carlo exploration.

The luminosity distance $d_L(z)$ of type Ia supernovae  can probe deviations from the standard expansion history of the Universe.   It is given by 
\be
    d_L(z) = (1+z)\, c\, \int_0^z {dz'\over H(z')}
\ee
where $H(z)$ is given by Eq.\ (\ref{eq:Friedmann})
with $a =(1+z)^{-1}$.  It is related to the distance
modulus $\mu(z) = 5\,\log_{10}(d_L(z)/10\,{\rm pc})$,
which is experimentally determined from the observed and modeled absolute luminosities of the supernovae.
Given a covariance matrix $C_{ij}$, the log-likelihood can be computed in the standard way,
\be
   -2 \ln{\cal L} = \chi^2 = \delta\vec\mu^T C^{-1}\delta\vec\mu\,,
\ee
where $\delta\vec\mu$ is the vector of residuals between theory and observation.  One minimizes the $\chi^2$ with respect to the cosmological parameters
to find the best-fit values and confidence intervals. 
This is known in the literature as frequentist profile likelihood, which is complementary  to the Bayesian Markov Chain Monte Carlo (MCMC) since the latter identifies large regions in the parameter space that exhibit a good fit to the data.
In the present case, the parameters over which to minimize the $\chi^2$ are the Hubble rate $H_0$,  the present contribution to the energy density from the ghosts and
sterile neutrinos $\Omega_g$, and that from the total matter contribution, $\Omega_m$. 
There may also be a nuisance parameter,
the correction factor of the absolute magnitude of the supernovae $M$, which is degenerate with $H_0$ for uncalibrated supernovae.
We assume flatness, so that the dark energy contribution is determined as $\Omega_\Lambda = 1 - \Omega_m - \Omega_g$.

This procedure was followed for three different
(but not all mutually exclusive) data sets:
Union 2.1 \cite{2012ApJ...746...85S}, Pantheon \cite{Pan-STARRS1:2017jku}, and Pantheon+SH0ES \cite{Brout:2022vxf}, comprising 580, 1048 and 
1657 supernovae, respectively.~\footnote{The Pantheon+ sample contains 1701 spectroscopically confirmed type Ia supernovae up to redshift $z \simeq 2.3$, but only 1657 of them have $z > 0.01$. We followed Ref.~\cite{Brout:2022vxf} in omitting supernovae with $z < 0.01$ from the analysis because of their large peculiar velocities,  which preclude an accurate determination of the cosmological parameters.}  In an alternative
analysis of Pantheon+, the smaller sample of 1590 SNe is used, omitting the nearby 
SNe whose absolute distances are calibrated using Cepheid variables; this leaves $H_0$ and $M$ undetermined.
The determination of the other parameters gives the same result in both cases.

The results of minimizing the $\chi^2$ for the different data sets are shown in  Fig.\ \ref{fig:SN_likelihood} (left).  There is no preference for
$\Omega_g \ne 0$, but large magnitudes are allowed
at 95\% C.L.,  down to $-0.75$ for the case of 
massive ghosts and up to $0.45$-$0.9$ for massless ghosts, depending upon the data set. 
Fig.\ \ref{fig:SN_likelihood} (right) displays
the preferred values of $\Omega_\Lambda$ and $\Omega_m$ as a function of $\Omega_g$,
along with 
the dependence of the Hubble parameter $H_0$ on $\Omega_g$.  Two values are shown, corresponding to the Union 2.1 and Pantheon+ samples (since Pantheon does not determine $H_0$).  Both are linearly increasing functions
of $\Omega_g$, but are offset from each other by about
$4$ km/s/Mpc.  The Union 2.1 fit gives a minimum value of 
$H_0 \cong 69.3$\,km/s/Mpc at $\Omega_g \cong -0.75$, which is still in tension with the Planck determination
$67.4\pm 0.5$  \cite{Planck:2018vyg} at nearly 4$\sigma$. 
In Section \ref{sect:massive_nu} we will show that this preliminary result agrees with a more careful analysis where all of the standard $\Lambda$CDM parameters are allowed to vary.

%\mpuel{I think the fact that the $H_0$ curves shown in the figure do not reach $\sim 68$ as Planck wants does not mean that we can not explain the Hubble tension for some models, which perhaps are not favored compared to the LCDM best-fit. It just tells us that the model giving the lowest $\chi^2$ can not address the Hubble tension.}

The $95\%$ C.L.\ bounds on $\Omega_g$ in Fig.~\ref{fig:SN_likelihood} restrict the allowed values of the dimensionless parameter $\epsilon_{\rho}$ (defined below Eq.~\eqref{eq:Boltzmann_y}) or its
dimensionful counterpart
$\Gamma_{\rho}$.
Fig.~\ref{fig:SN_bounds} (left)  shows these limits for massless ghosts and massive sterile neutrinos as a function of the $\nu_s$ equation-of-state parameter $w_{\nu_s}$. As already anticipated, values of $\epsilon_{\rho}$ larger than $\mathcal{O}(10)$ are excluded by supernova data, except when $w_{\nu_s} \sim 1/3$. However, for such values of $w_{\nu_s}$ the magnitude and shape of the energy densities of the new species become irrelevant since their contributions cancel, giving a negligible effect in the Friedmann equation~\eqref{eq:Friedmann}. Therefore, the use of the approximate fits in Eq.~\eqref{eq:Ofits}, which assume linearity in $\epsilon_\rho$, 
is justified even for these larger
values of $\epsilon_\rho$.

The minimum value of $\chi^2$ for each data set, compared to the $\chi^2$ value for the standard $\Lambda\rm{CDM}$ parameter values, can be characterized by  $\Delta \chi^2 = \chi^2_{\rm min} - \chi^2_{\Lambda\rm{CDM}}$.  It is given by $-2810.4$ (Pantheon+), $-1.1$ (Pantheon), and $-94.5$ (Union 2.1). The discrepancy for Union and Pantheon+ arises from the dependence on $H_0$ in the $\chi^2_{\rm min}$, which is calibrated using Cepheids.
The superficial appearance that the ghost model gives a better fit than
$\Lambda$CDM is a reflection of the Hubble tension between the supernovae
and the larger cosmological data sets.  This tension is hidden for the Pantheon analysis since $H_0$ is marginalized out. 
We will confirm a mild preference for the ghost model over $\Lambda$CDM in Section \ref{sect:massive_nu}. 

An interesting question is whether $\Omega_g$ can explain all of the dark energy, replacing the cosmological constant.  Figure\ \ref{fig:SN_likelihood} suggests that this is indeed possible, using the Union and Pantheon data sets, by taking $\Omega_g \sim 0.6$.  However it is in tension with the Pantheon+
upper bound $\Omega_g < 0.45$, which requires 
$\Omega_\Lambda \gtrsim 0.2$.
In general, the capability of $\Omega_g$ to function as dark energy can be understood through the effective EOS of the ghost fluid
\be
    w_{\rm eff} = {P_{\phi} + P_{\nu_s} \over
        \rho_{\phi} + \rho_{\nu_s}}\,,
\ee
whose redshift dependence is plotted in Fig.\ 
\ref{fig:SN_bounds} (right).  Notably, it violates the weak energy condition since $w_{\rm eff} < -1.3$ both at early and late times.  The early-time limit $w_{\rm eff} \to -1.5$ can be derived analytically from the exact solution (\ref{eq:energy_densities_approx}) that will be introduced below, in the regime where back-reaction of the ghost fluid is negligible.  
The late-time limit
$w_{\rm eff}\cong -1.3$ depends weakly on the value of $m_{\nu_s}/\Lambda$.  
Coincidentally, such a value of $w_{\rm eff}$ was shown to solve the Hubble tension by shifting the central value of $H_0$ without broadening its uncertainty, although the associated model would be strongly disfavored with respect to the $\Lambda$CDM baseline model~\cite{Vagnozzi:2019ezj}.
It is striking that the
ghost fluid (in conjunction with the produced positive-energy particles) can function as dark energy even in the absence of a classical condensate, or indeed a potential $V(\phi)$, 
which was the original motivation for phantom fields
in cosmology.
The latter conclusion was originally drawn by Ref.~\cite{Holdom:2004yx}, which obtained similar redshift evolution of $w_{\rm eff}$ as we derived in Fig.~\ref{fig:SN_bounds} (right), directly from the conservation of the energy-momentum tensor.

\section{CMB and large scale structure}
\label{sect:cmb}
Even if their combined contributions to the energy density of the Universe are negligible, the creation of massless ghosts and positive-energy particles can leave an imprint on the CMB and matter power spectra. 
In particular, if both species are massless, their energy densities cancel exactly,
but they can nevertheless alter the cosmological evolution and the formation of structure through their perturbations, depending on the production rate.  

As noted above, the expansion history of the Universe is modified if the ghosts and $\nu_s$ are not degenerate in mass.
We estimate that the results for  $m_{\nu_s}>0$ will revert to those of the massless case when $m_{\nu_s}  \lesssim 2 \times 10^{-5}\,\Lambda$. This was found by comparing the effects on the CMB and matter power spectra, including the linear order perturbations, and either
including or neglecting the zeroth-order contributions of the new species.
In the following sections we will study the two qualitatively distinct regimes, depending on the scale of $m_{\nu_s}$.

\subsection{Massless $\nu_s$ case}
\label{subsec:massless_case}
If $\nu_s$ is sufficiently light, its contribution to the total energy density of the Universe cancels that of the presumed massless ghosts and there is no appreciable effect at the background level.
The first nontrivial contribution arises at linear order in the cosmological perturbations.
We derive the relevant equations in Appendix~\ref{appB} and find that the perturbations for $\phi$ and $\nu_s$ do not cancel each other, mainly because of the different phase-space distribution between the two species, shown in Fig.\ \ref{fig:spectra}; see also Eqs.~\eqref{eq:nus_pert_equations_final} and~\eqref{eq:ghost_pert_equations_final}. 

The perturbations were implemented in a modified version of the Boltzmann code \texttt{CAMB}~\cite{Lewis:1999bs}.  To the  six parameters comprising the $\Lambda$CDM model,~\footnote{$\Omega_b h^2$ and $\Omega_c h^2$ are the present-day abundances of baryons and cold DM, respectively, multiplied by the square of the reduced Hubble constant $h$. $100\, \theta_{\rm MC}$ is a \texttt{CosmoMC} parameter describing approximately the position of the first acoustic peak, and is used instead of $H_0$ because it is less correlated with other parameters. $\tau$ is the reionization optical depth, while $A_s$ and $n_s$ are the primordial scalar amplitude and spectral index respectively.}
\beq
\label{eq:six_params}
    \Omega_b h^2,\, \Omega_c h^2,\, 100\, \theta_{\rm MC},\, \tau, \,\ln(10^{10} A_s),\, n_s\,,      
\eeq
we add two new ones:
$\Gamma_{\rho}$, which controls the amplitude of the background energy densities of the dark species, and $\Gamma_{\rho} / \Lambda^4$, which determines the strength of the first-order perturbed collision term.
We assume vanishing initial conditions for the $\phi$ and $\nu_s$ perturbations in the early radiation era ($a \ll 10^{-5}$). %, which seems a reasonable assumption being the background density of the new species completely negligible at that early times (see for instance Fig.~\ref{fig:densities}). 
%%\jc{this might need further explanation} \mpuel{I think this comment was associated to $\ell_{\rm max} = 45$. I will put the information about the maximum hierarchy multipole below Eqs.~\eqref{eq:nus_pert_equations_final} and~\eqref{eq:ghost_pert_equations_final} since it is not relevant here.}

\begin{figure*}[t]
\begin{center}
\includegraphics[scale=0.349]{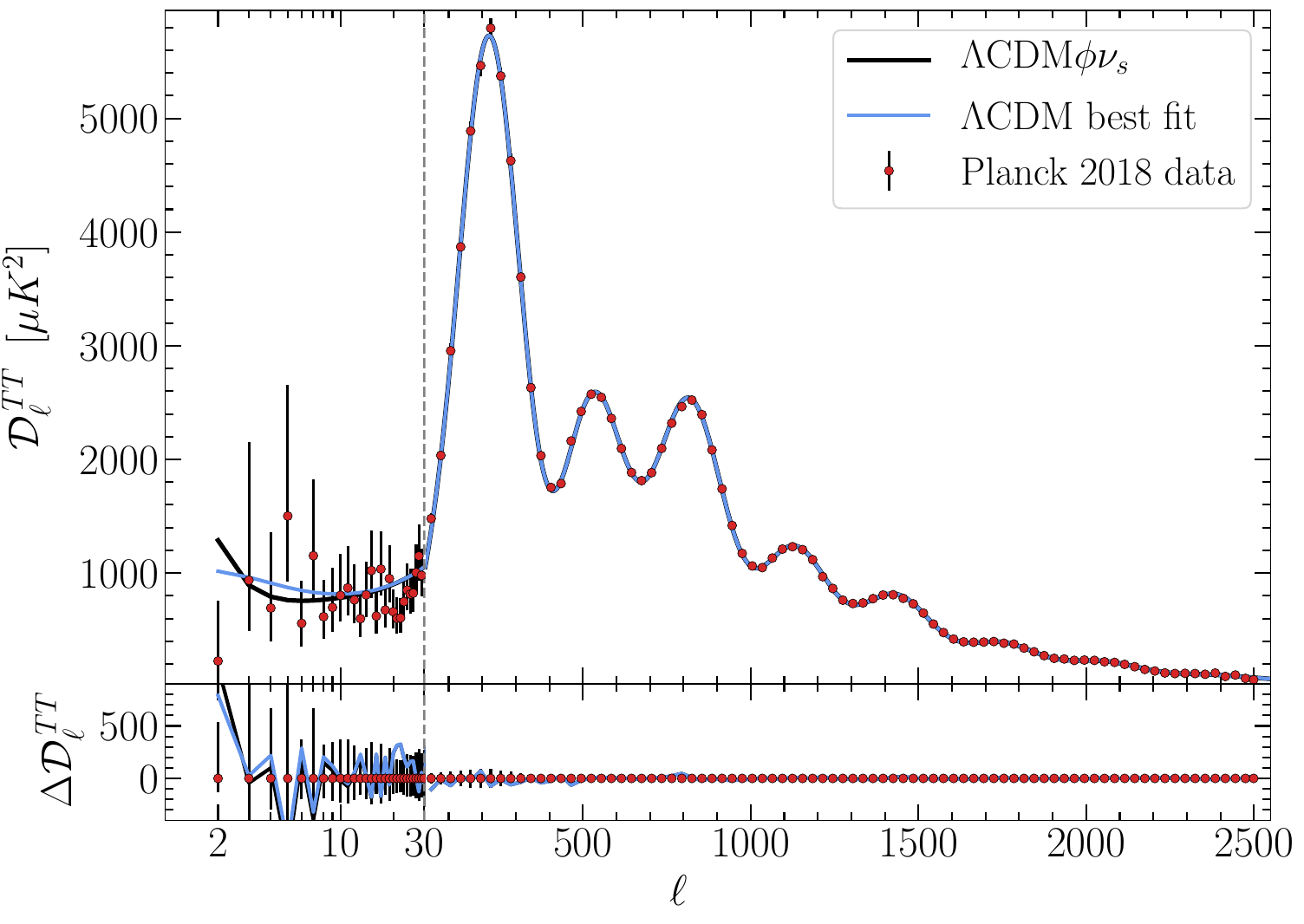}
\includegraphics[scale=0.352]{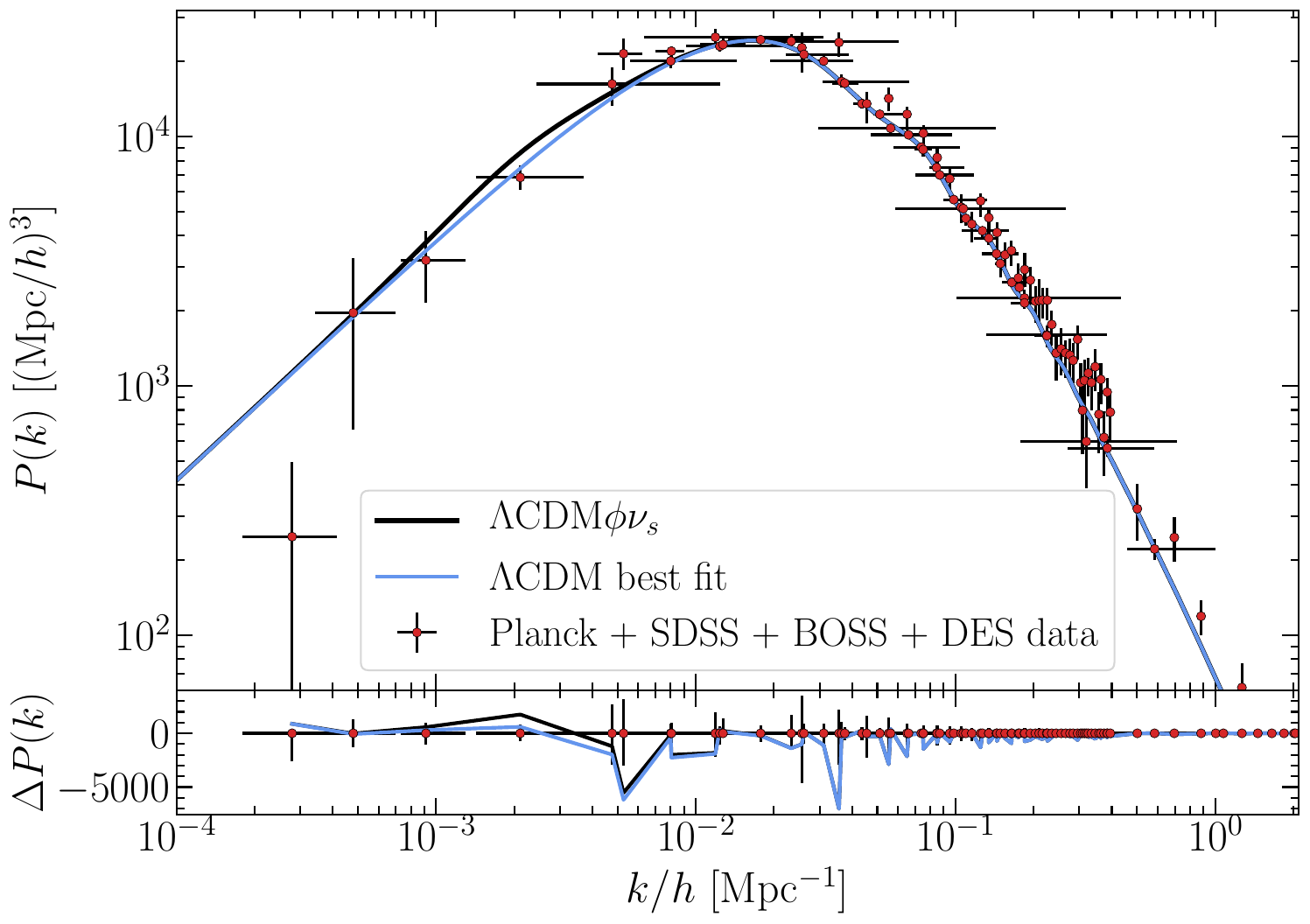}
\caption{For the special case of $m_{\nu_s}=0$.  \textit{Left:} Prediction of the $\Lambda$CDM$\phi\nu_s$ model (black curve) for the CMB temperature autocorrelation spectrum, compared to the $\Lambda$CDM best-fit model (blue curve). The Planck 2018 data points are taken from Ref.~\cite{Planck:2018vyg}.
\textit{Right:} Predictions for the linear matter power spectrum. The data points, used only for display purposes, are given by Ref.~\cite{Chabanier:2019eai}, which contains data from Planck 2018~\cite{Planck:2018nkj}, eBOSS DR14~\cite{eBOSS:2018}, SDSS DR7~\cite{Ata:2017dya} and DES YR1~\cite{DES:2017qwj}. For both plots, 
the vector mediator is assumed, we used $\Gamma_{\rho} \simeq 10^{-69}\,\,\text{MeV}^5$, $\Gamma_{\rho} / \Lambda^4 \simeq 10^{-41}\,\,\text{MeV}$ and fixed the six $\Lambda$CDM parameters and $H_0$ to their Planck 2018 best-fit values. Residuals between the model predictions and the data points are shown at the bottom. }
\label{fig:Dl_TT_m0}
\end{center} 
\end{figure*}

For the background energy densities, the Boltzmann equations  \eqref{eq:Boltzmann_eq_t} can be rewritten in terms of the scale factor $a$ and solved analytically, since the net density of the new species vanishes in the Friedmann equation~\eqref{eq:Friedmann}.
The solutions are
\bea
\label{eq:energy_densities_approx}
    \rho_{\phi} (\bar{x}) &=& - \frac{2\,\Gamma_{\rho}}{11\,H_0 \sqrt{\Omega_\Lambda}} \sqrt{\bar{x}}  \nn\\
    &\times& {}_2 F_1 \bigg(\frac{1}{2}, \frac{11}{6}, \frac{17}{6}, -\bar{x}\bigg)\,, \nn\\
    \rho_{\nu_s} (\bar{x}) &=& \frac{\Gamma_{\rho}}{3\, (3/2 + w_{\nu_s})\, H_0 \sqrt{\Omega_\Lambda}} \sqrt{\bar{x}} \nn \\
    &\times& {}_2 F_1 \bigg(\frac{1}{2}, \frac{3}{2} + w_{\nu_s}, \frac{5}{2} + w_{\nu_s}, -\bar{x}\bigg)\,,
\eea
defining $\bar{x} \equiv a^3\, \Omega_\Lambda/\Omega_m$, where ${}_2 F_1 (a, b, c, z)$ is the hypergeometric function.~\footnote{Eq.~\eqref{eq:energy_densities_approx} are valid whenever back-reaction of the new species on the Hubble-parameter evolution can be neglected, and they apply for any value of $m_{\nu_s}$. One should keep in  mind that $\Gamma_{\rho}$ is generally a function of  $m_{\nu_s}$ (see Eqs.~\eqref{eq:Gamma_rho_approx} and~\eqref{eq:Gamma}), which in turn affects the value of $w_{\nu_s}$ as described by Eq.~\eqref{eq:eos_nus_m}.}

The effects of the two new species on the CMB temperature and linear matter power spectra are shown in Fig.~\ref{fig:Dl_TT_m0}. 
The main modification on the former is an overall decrease of the late Integrated Sachs-Wolfe (ISW) effect, visible at large angular scales
(small $\ell$), which is due to the increase of the gravitational potential along the path of photons between the time of last scattering and today. This makes the photons lose energy as they travel through gravitational potential wells that are growing in time. 

More  precisely, the late ISW contribution to the CMB anisotropy in Fourier space is given by~\cite{Hu:1998kj}
\be
\label{eq:Cl_ISW}
    \mathcal{C}_{\ell}^{\rm ISW} \propto \int dk\,k^2\,\bigg[\int_{\tau_{\ast}}^{\tau_0} d\tau \,(\Psi' - \Phi')\,j_{\ell} (k (\tau_0 - \tau)) \bigg]^2 \,,
\ee
where $\tau_{\ast}$ and $\tau_0$ are the conformal times at last scattering and today, respectively, $\Psi$ and $\Phi$ are the potential and curvature in conformal Newtonian gauge~\cite{Kodama:1984ziu}, the prime indicates the conformal time derivative, and $j_{\ell}$ are the spherical Bessel functions.
The Newtonian curvature $\Phi$ and potential $\Psi$ for the mode $k$ are set by the total density contrast $\delta_{\rm tot}$, velocity $\theta_{\rm tot}$ and shear $\sigma_{\rm tot}$ perturbations~\cite{Ma:1995ey},
\bea
\label{eq:Phi_Psi_potentials}
    \Phi &=& \frac{4 \pi G a^2}{k^2} \bigg[\rho\, \delta + \frac{3 \mathcal{H}}{k^2} (\rho + P)\, \theta \bigg]_{\rm tot}\,, \nn\\
    \Psi &=& - \Phi - \frac{12 \pi G a^2}{k^2} [(\rho + P)\, \sigma]_{\rm tot}\,,
\eea
where $\mathcal{H} = a H (a)$ is the conformal Hubble parameter. The terms in the square brackets in Eq.~\eqref{eq:Phi_Psi_potentials} are gauge invariant and hence they can be evaluated in the frame comoving with the cold DM (CDM), corresponding to synchronous gauge. 
In a flat $\Lambda$CDM universe, the late ISW contribution is zero during matter domination because both $\Psi$ and $\Phi$ remain constant, but as dark energy begins to dominate, the curvature $\Phi$ decays, reducing the magnitude of the potential $\Psi$. This effect is enhanced if (i) there is extra radiation pressure, (ii) the Universe expands faster than in $\Lambda$CDM, or (iii) a nonzero spatial curvature is present~\cite{Lopez:1998jt,Kaplinghat:1999xy,Hu:2001bc}.

\begin{figure*}[t]
\begin{center}
\includegraphics[scale=0.35]{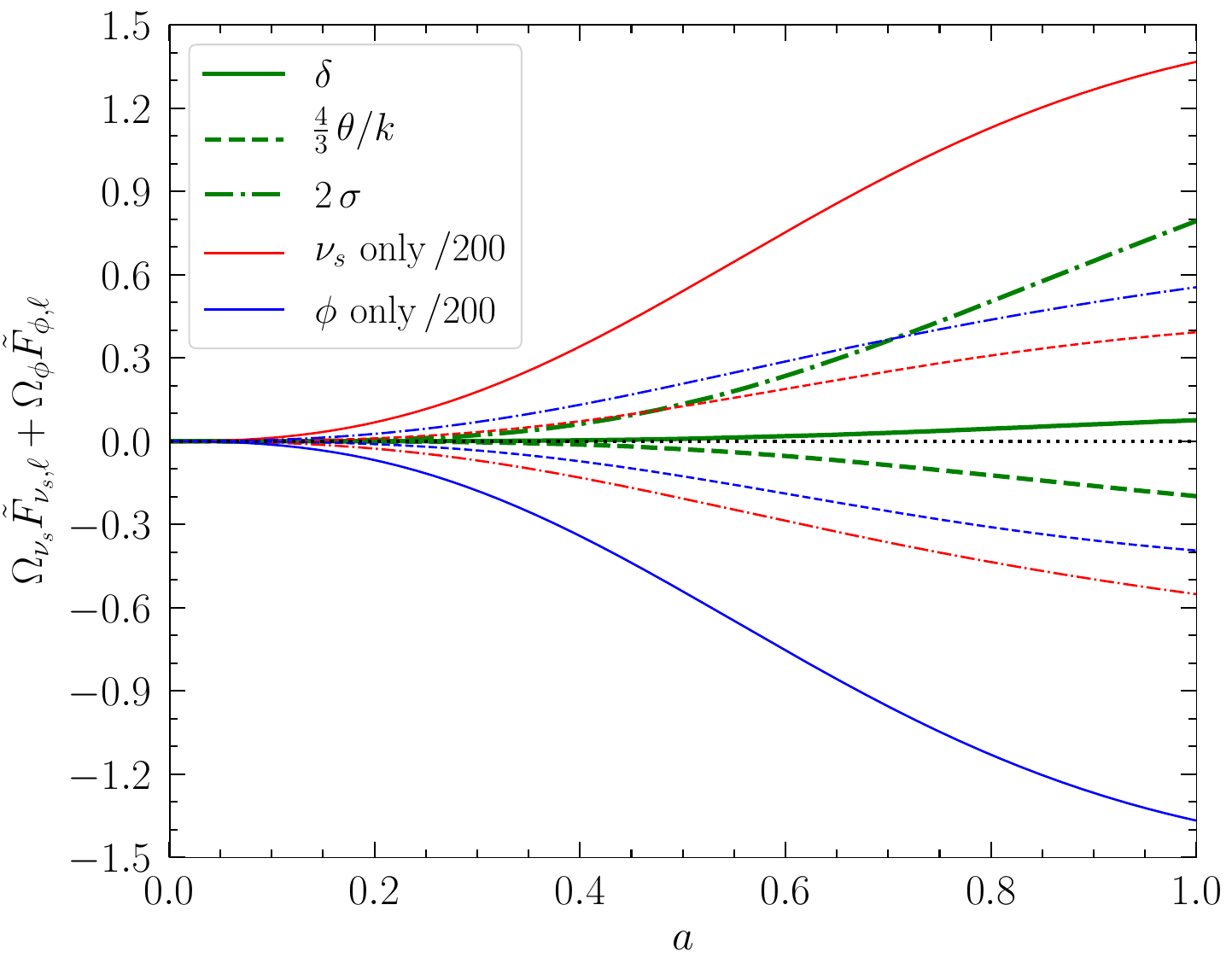}
\includegraphics[scale=0.35]{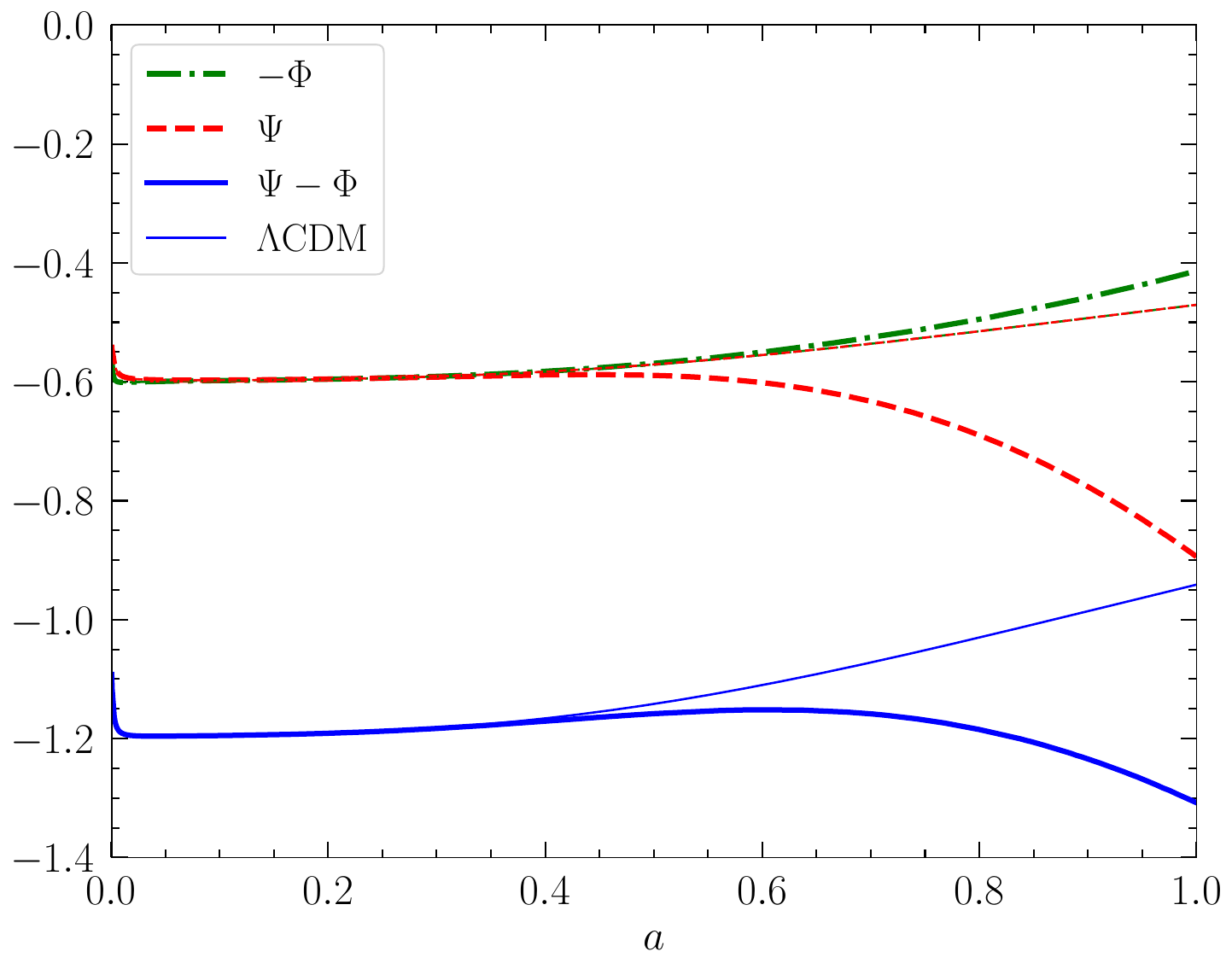}
\caption{\textit{Left:} Scale-factor evolution of the cosmological perturbations at large scales ($k / h = 5 \times 10^{-4} \,\,\text{Mpc}^{-1}$) for sterile neutrinos $\nu_s$ (red) and ghosts $\phi$ (blue), rescaled by $1/200$. The physical perturbations are given by $\Omega_{i} F_{i, \ell}$, where $i=\phi,\,\nu_s$, $\Omega_i (a)$ is the fractional energy density and $F_{i, \ell}$ is the relative perturbation, Eq.~\eqref{eq:Fil_def}: $F_{i, 0} = \delta_i$ (solid line), $F_{i, 1} = (1 + w_i) \theta_i / k$ (dashed line) and $F_{i, 2} = \tfrac{3}{2} (1 + w_i) \sigma_i$ (dot-dashed line); $\delta_i$, $\theta_i$ and $\sigma_i$ are the perturbations for the density contrast, velocity and shear, respectively. For massless $\phi$ and $\nu_s$, $w_i = 1/3$.
The sum $\sum_i\Omega_i F_{i,\ell}$ for $\delta$, $\theta$ and $\sigma$, is shown in green (not rescaled).
\textit{Right:} Scale-factor evolution of the Newtonian curvature $\Phi$ and potential $\Psi$ at the same scale. Thick lines include the contribution from the new species, while thin lines correspond to $\Lambda$CDM. 
$\Gamma_\rho$, $\Lambda$, $H_0$ and mediator are as
in Fig.\ \ref{fig:Dl_TT_m0}.
}
\label{fig:ISW_m0}
\end{center} 
\end{figure*}

In the present case, one can understand the decrease of the late ISW effect directly from Eq.~\eqref{eq:Phi_Psi_potentials}. Figure~\ref{fig:ISW_m0} (left) shows the large-scale evolution of the synchronous-gauge physical perturbations for sterile neutrinos $\nu_s$ and ghosts $\phi$, which can be treated as a single effective fluid since both species are relativistic. 
The ghost nature as a negative-energy particle manifests itself through the opposite sign in the physical perturbations $(\rho_{\phi} F_{\phi, \ell})$ with respect to those for $\nu_s$, having negative density contrast and velocity, but positive shear stress perturbations, which is due to $\rho_{\phi} \lesssim 0$. 
Positive physical shear stress is a generic feature of phantom cosmological species and the main effect is to drive clustering~\cite{Mota:2007sz}. This is exactly the opposite of what happens for normal species like SM neutrinos, where the presence of the shear stress leads to erasure of fluctuations that might initially be present in the fluid, suppressing further growth.

Due to the larger phase space available for $\nu_s$,~\footnote{The momentum of $\nu_s$ can reach $2\Lambda$ for extreme kinematic configurations, while that of the ghost is limited to $\Lambda$.} which increases its first-order perturbed collision term, the relative perturbations for $\nu_s$ are expected to dominate over those for $\phi$ at high hierarchy multipoles (see discussion in section  (\ref{subsec:perts_m0}.\ref{subsec:considerations_perts}) of Appendix~\ref{appB}). 
Since the relative shear $\sigma$ and higher hierarchy multipoles for a single relativistic species are generally negative at large scales, as occurs for SM massless neutrinos and photons, we expect $\sigma_{\phi}$ to be more negative than $\sigma_{\nu_s}$. As a consequence, the physical shear for the effective $\nu_s + \phi$ fluid, defined by  $\rho_{\nu_s} \sigma_{\nu_s} + \rho_{\phi} \sigma_{\phi}$, becomes positive, as borne out in Fig.~\ref{fig:ISW_m0} (left) with the dot-dashed green curve.
The effect of the new species is therefore to decrease the value of the potential $\Psi$ in Eq.~\eqref{eq:Phi_Psi_potentials}, favoring matter clustering. This is shown with the red dashed curve in Fig.~\ref{fig:ISW_m0} (right).

On the other hand, the density contrast $\delta$ and the velocity perturbation $\theta$ for the effective single fluid turn out to be positive and negative, respectively, which makes it harder to predict {\it a priori} the net contribution of the new species to $\Phi$. The evolution of the latter is shown with the green dot-dashed curve in Fig.~\ref{fig:ISW_m0} (right). 
The net effect of the $\nu_s + \phi$ fluid is to reduce $\Psi - \Phi$, which generally leads to smaller $\mathcal{C}_{\ell}^{\rm ISW}$ in Eq.~\eqref{eq:Cl_ISW} than in $\Lambda$CDM.
A similar trend occurs in phantom dark energy models with non-negligible anisotropic stress~\cite{Koivisto:2005mm}. 

For larger values of $\Gamma_{\rho}$ and $\Gamma_{\rho} / \Lambda^4$, at the largest angular scales, the shear perturbation $\sigma$ increases faster than the lower multipole perturbations $\delta$ and $\theta$, driving $(\Psi - \Phi)$ to smaller values on shorter time scales. This results in a larger contribution to the late ISW effect in the CMB temperature power spectrum because of the square in Eq.~\eqref{eq:Cl_ISW}, which is visible in Fig.~\ref{fig:Dl_TT_m0} (left) around $\ell \sim 2 - 3$.  
%\jc{If the effect is only slightly noticeable, is it worth commenting on?} \mpuel{For the new choice of parameters, the effect is visible. It is even more visible (if not diverging) if I increase $\Gamma_{\rho} / \Lambda^4$ for instance. I would like to avoid showing a divergent plot, because this spoils the discussion on the reduction of the late ISW effect made before, which is a feature also present for massive $\nu_s$ in the next section when $w_{\nu_s} \to 1/3$.}
The divergent nature of the perturbations for large values of $\Gamma_{\rho} / \Lambda^4$ or $\Gamma_{\rho}$ is not surprising, since theories with ghosts are generically subject to instabilities.  The growth of the instability is controlled by the ghost momentum cutoff $\Lambda$ and the mediator mass $M_i$, and the analysis carried out below will exclude parameter values leading to excessive growth.~\footnote{
Conceivably, higher order perturbations could cut off the instabilities.  In this work we constrain the effect of the first order perturbations, such that the higher-order contributions, proportional to higher powers of 
$\Gamma_\rho/\Lambda^4\ll 1$, are negligible.}
We note that the effect of weak gravitational lensing on the CMB power spectra ({\it e.g.,} Ref.~\cite{Lewis:2006fu}), in the presence of ghosts and sterile neutrinos, does not deviate appreciably from the $\Lambda$CDM scenario because the Weyl potential $(\Psi - \Phi) / 2$ in the two cases is essentially identical at small angular scales (large $\ell$).

The modifications to the matter power spectrum in our scenario are also negligible for allowed values of $\Gamma_{\rho}$ and $\Gamma_{\rho} / \Lambda^4$, and the main effect is a small rise of the spectrum at large scales, as shown in Fig.~\ref{fig:Dl_TT_m0} (right). 
The small positive contribution to $P(k)$ is due to its dependence on the ratio between the matter density perturbation $\delta \rho_{m}$ and the background matter density $\rho_{m}$. In fact, neither $\nu_s$ or $\phi$ contribute directly to $\delta \rho_{m}$ nor $\rho_{m}$, except through the metric perturbations appearing in the equation governing the evolution of $\delta \rho_{m}$. 
The increase in the matter power spectrum signals that the ghost and and sterile neutrino effective fluid favors a mildly more efficient matter clustering than $\Lambda$CDM at large angular scales, which was already noticed from the CMB power spectrum.

\subsubsection{Method and Results}\label{subsec:results_m0}

\begin{figure}[t]
\begin{center}
\includegraphics[scale=0.35]{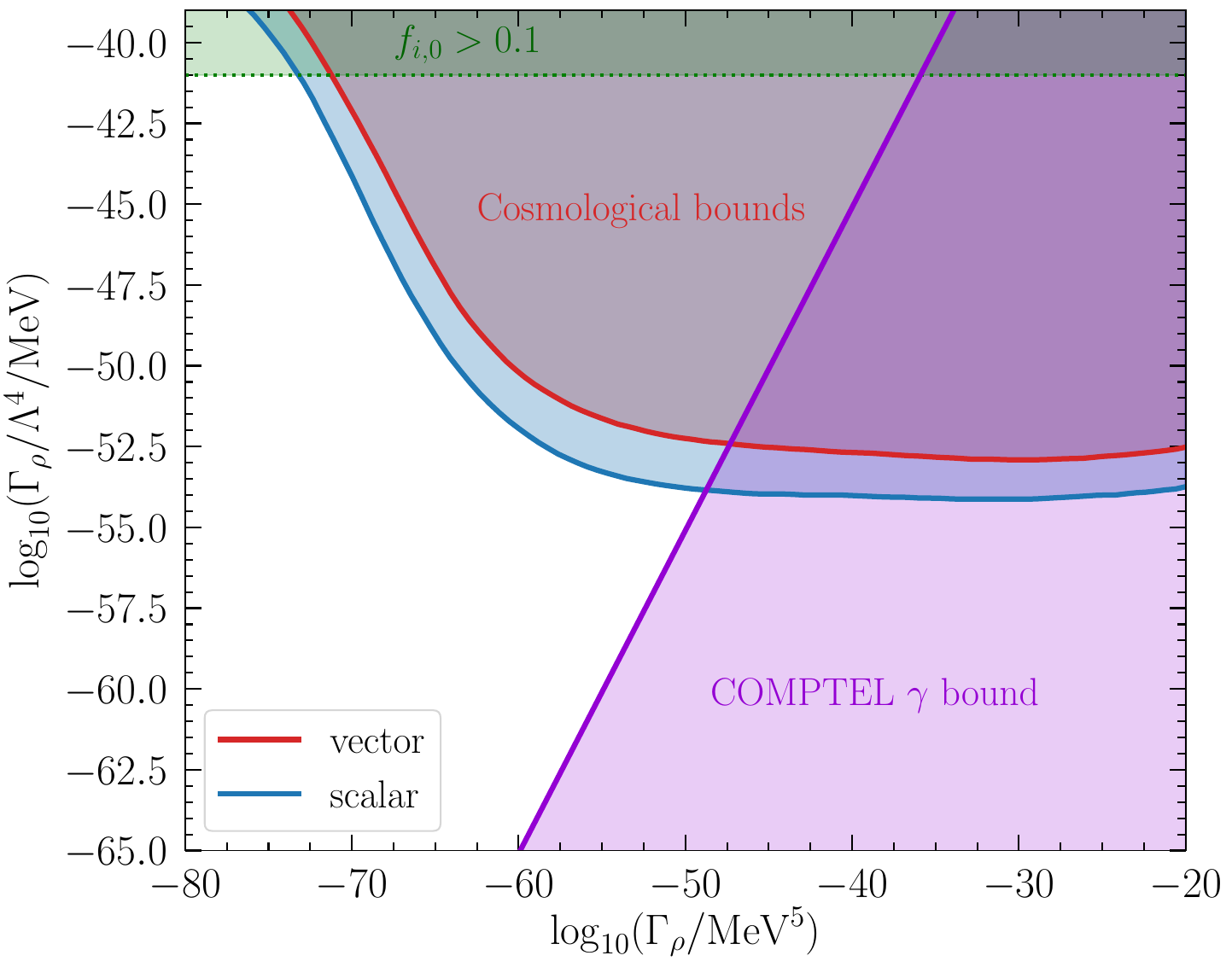}
\caption{Constraints on  $\Gamma_{\rho} / \Lambda^4$ versus $\Gamma_{\rho}$, assuming $m_{\nu_s}=0$. 
Red and blue regions correspond to 
the $95\%$ C.L.
cosmological bounds for the vector and scalar mediator models, respectively.
Violet indicates the gamma-ray excluded region, Eq.~\eqref{Lambda_bound}. 
The green region is inconsistent with the technical assumption $f_{i}^0 \lesssim 0.1$ in Eq.\ (\ref{eq:F0}).
}
\label{fig:bounds_m0_gen}
\end{center} 
\end{figure}

\begin{figure*}[t]
\begin{center}
\includegraphics[scale=0.35]{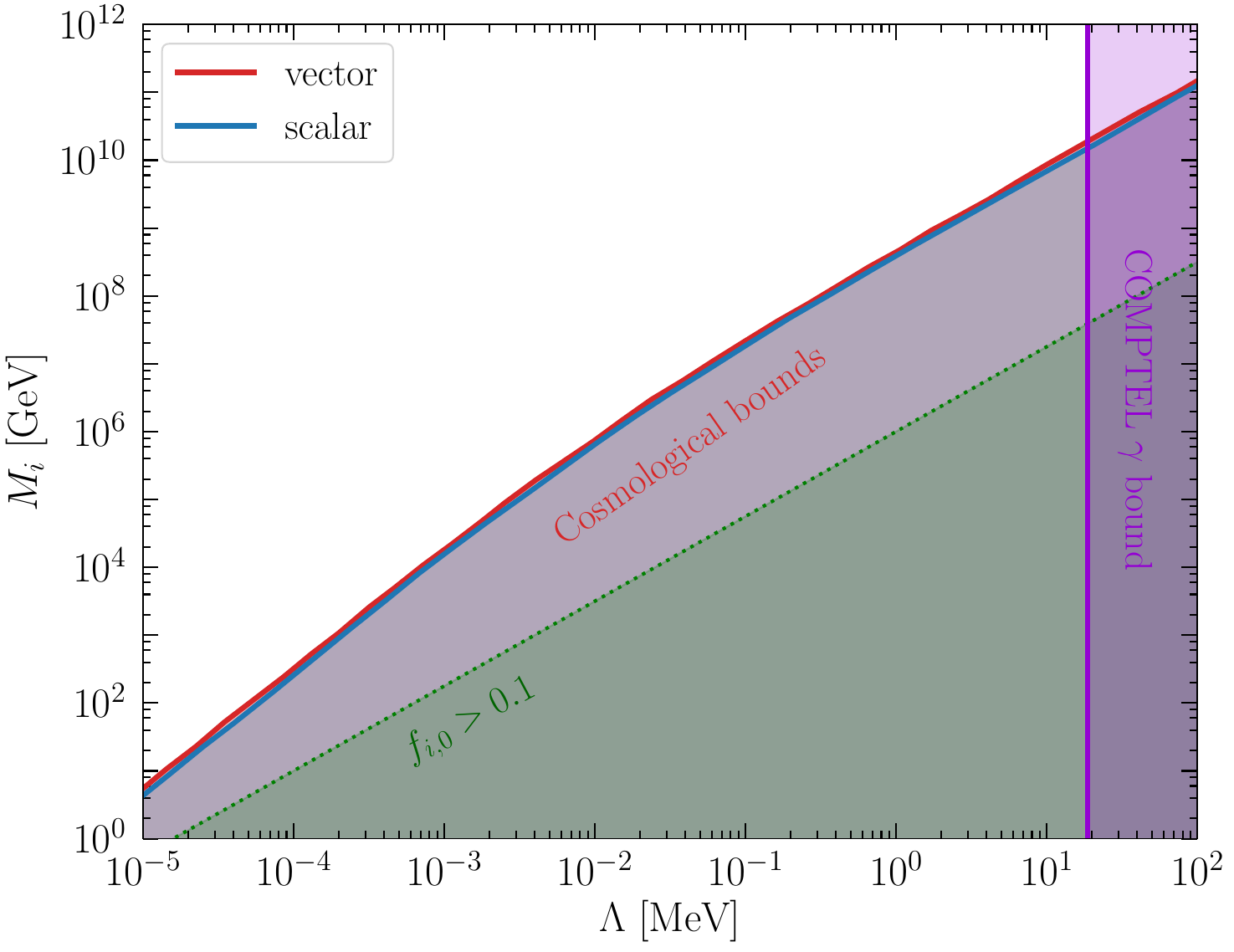}
\includegraphics[scale=0.35]{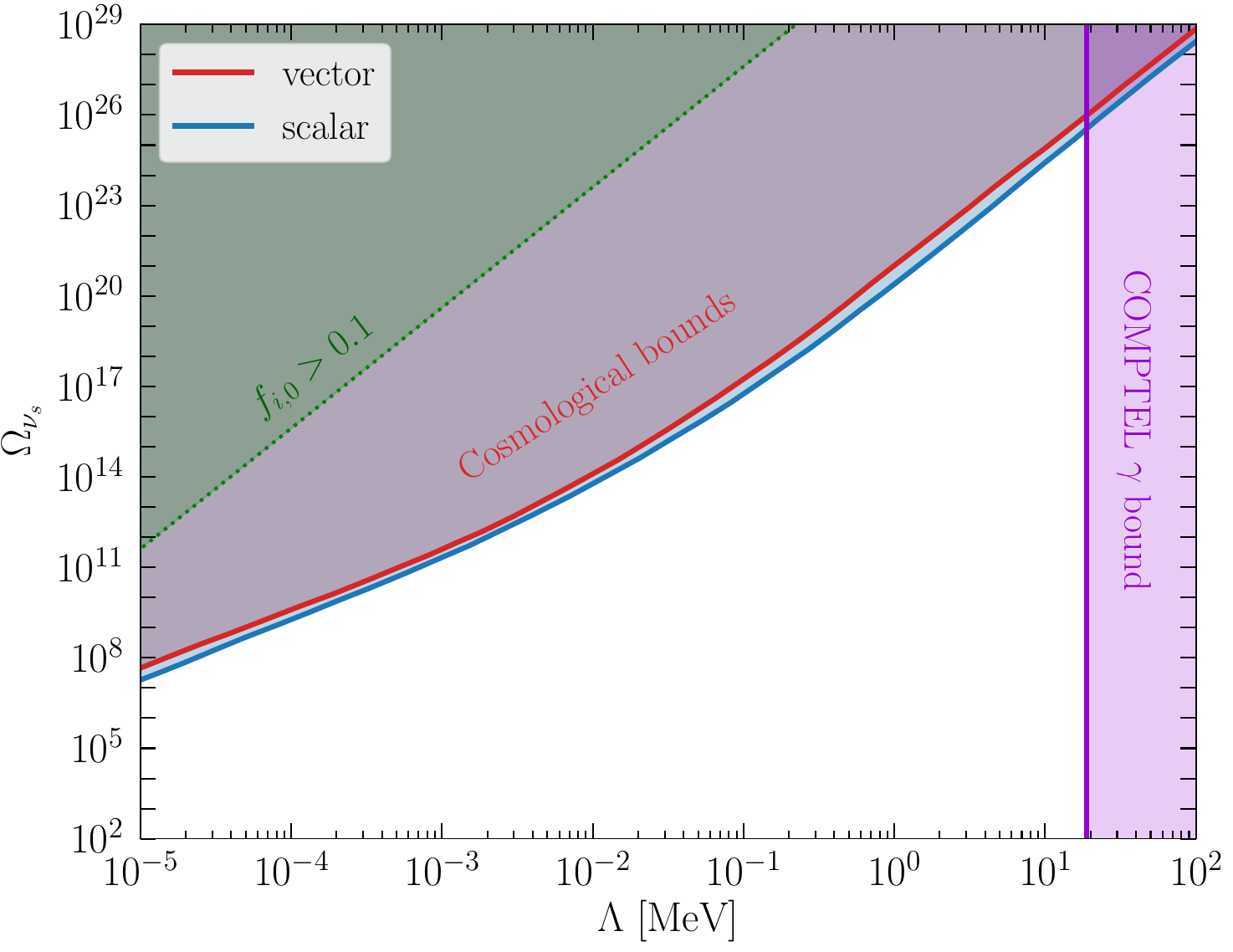}
\caption{For the special case of $m_{\nu_s}=0$.
\textit{Left}: Mapping of excluded regions in Fig.~\ref{fig:bounds_m0_gen} to the microphysics parameters, the mediator mass $M_i$ ($i = v, s$) versus the ghost momentum cutoff $\Lambda$.
\textit{Right}: Conversion of the same constraints to the present-day sterile neutrino abundance $\Omega_{\nu_s}$ as a function of the ghost momentum cutoff $\Lambda$. 
The color coding is the same as in Fig.~\ref{fig:bounds_m0_gen}
}
\label{fig:bounds_m0}
\end{center} 
\end{figure*}

To constrain the new parameters $\Gamma_{\rho}$ and $\Gamma_{\rho} / \Lambda^4$, in conjunction with the $\Lambda$CDM ones, we ran the publicly available 
%Markov Chain Monte Carlo (MCMC) 
MCMC
code \texttt{CosmoMC}~\cite{Lewis:2002ah}. 
Convergence of the MCMC chains was monitored through the Gelman-Rubin parameter $R - 1$~\cite{Gelman:1992zz}, by requiring $R - 1 \lesssim 0.01$. 
The cosmological data sets we considered are:
\begin{itemize}
    \item \textbf{Planck}: the full range of the CMB measurements from Planck 2018, which include the temperature and polarization anisotropies, as well as their cross-correlations~\cite{Planck:2018vyg}; 
    \item \textbf{Lensing}: the 2018 Planck measurements of the CMB lensing potential power spectrum, reconstructed from the CMB temperature four-point function~\cite{Planck:2018lbu};
    \item \textbf{BAO}: the distance measurements of the baryon acoustic oscillations given by the 6dFGS~\cite{Beutler:2011hx}, SDSS-MGS~\cite{Ross:2014qpa}, and BOSS DR12~\cite{BOSS:2016wmc} surveys;
    \item \textbf{DES}: the first year of the Dark Energy Survey galaxy clustering and cosmic shear measurements~\cite{DES:2017qwj,DES:2017myr,DES:2017tss};
    \item \textbf{Pantheon}: distance measurements of (uncalibrated) type Ia supernovae from the Pantheon sample, comprising $1048$ data points in the redshift range $z \in [0.01, 2.3]$~\cite{Pan-STARRS1:2017jku}.
\end{itemize}
We did not include the \texttt{Halofit} modeling of the nonlinear part of the matter power spectrum because additional modifications of the standard nonlinear clustering model might be needed. Since this is beyond the scope of the present work, we consider only linear observables in our analysis.

Figure~\ref{fig:bounds_m0_gen} shows the constraints on $\Gamma_{\rho}$ and $\Gamma_{\rho} / \Lambda^4$, for the vector and scalar mediator models, as derived from the \texttt{CosmoMC} analysis, in addition to the COMPTEL bound of Eq.~\eqref{Lambda_bound}.
Regarding the cosmological limits, one observes two features: the models in agreement with the data lie in the  region of $\Gamma_{\rho} / \Lambda^4 \lesssim 10^{-38}$ MeV, and for sufficiently small $\Gamma_{\rho} / \Lambda^4$, the dependence on $\Gamma_{\rho}$ disappears. We qualitatively explain such features in Appendix~\ref{appB}, since they are closely related to the form of the perturbation equations.
The constraints on the scalar mediator model are slightly stronger than  those of the vector model, due to differences in the perturbation equations for the two interactions, which arise from their energy dependence; see, {\it e.g.,}
Eq.\ (\ref{eq:M2}).

The constraints on $\Gamma_{\rho} / \Lambda^4$ versus $\Gamma_{\rho}$ can be mapped onto the plane of the ghost momentum cutoff $\Lambda$ versus the mediator mass scale $M_i$, using Eq.~\eqref{eq:Gamma_rho_approx}. The resulting limits are shown in  Fig.~\ref{fig:bounds_m0} (left) and they are essentially independent of the mediator  type.
Also shown is the consistency condition for neglecting the Pauli-blocking and Bose-Einstein stimulated-emission factors in the collision term of Eq.~\eqref{eq:dfidt_0}. This is inferred by demanding that $f_{i}^0 \sim 2 \pi^2\, \Gamma_{\rho} / (\Lambda^4 H_0) \lesssim 0.1$, which translates into the bound $\Gamma_{\rho} / \Lambda^4 \lesssim 10^{-41}\,\,\text{MeV}$. 
For massless $\nu_s$, this is equivalent to $M_i \gtrsim 1.0 \times 10^6\,(\Lambda / \rm MeV)^{1.25}$ GeV for both the vector ($i = v$) and scalar ($i = s$) mediator cases.
The $95\%$ C.L. bounds on $M_i$ versus $\Lambda$ can be approximately expressed by
\be
\label{eq:fit_bounds_m0}
    \frac{M_i}{\rm{GeV}} \gtrsim 4.6 \times 10^8 \,\hat{\Lambda}^{1.3 - 0.05 \log_{10}{\hat{\Lambda}}}\,,
    %\log_{10} \bigg(\frac{M_i}{\rm{GeV}}\bigg) \gtrsim a_0 + a_1 x + a_2 x^2 +  a_3 x^3 + a_4 x^4 +  a_5 x^5
\ee
where $\hat{\Lambda} \equiv \Lambda / \text{MeV} \in (10^{-7}, 10^2)$.

\begin{figure*}[t]
\begin{center}
\includegraphics[scale=0.34]{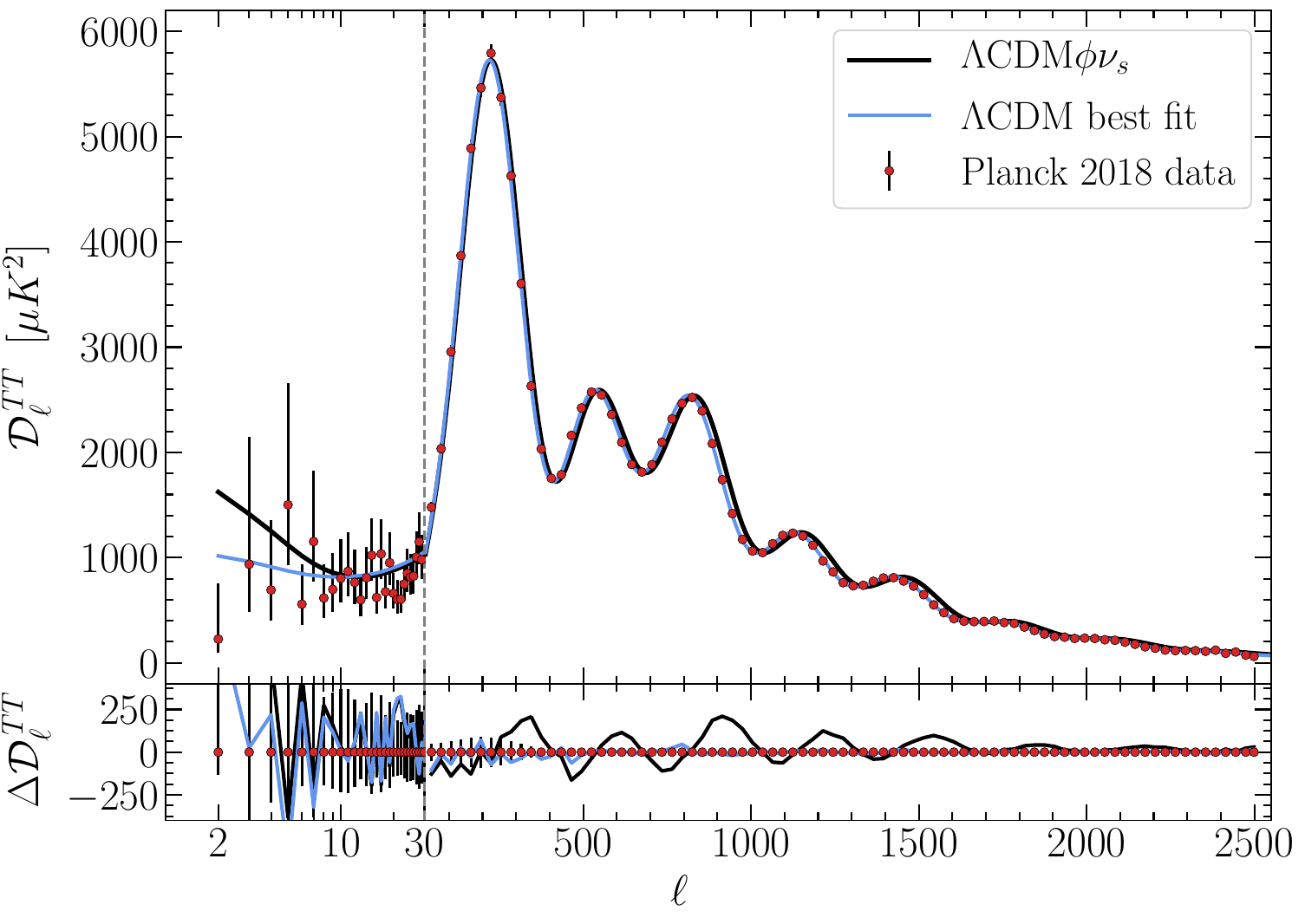}
\includegraphics[scale=0.345]{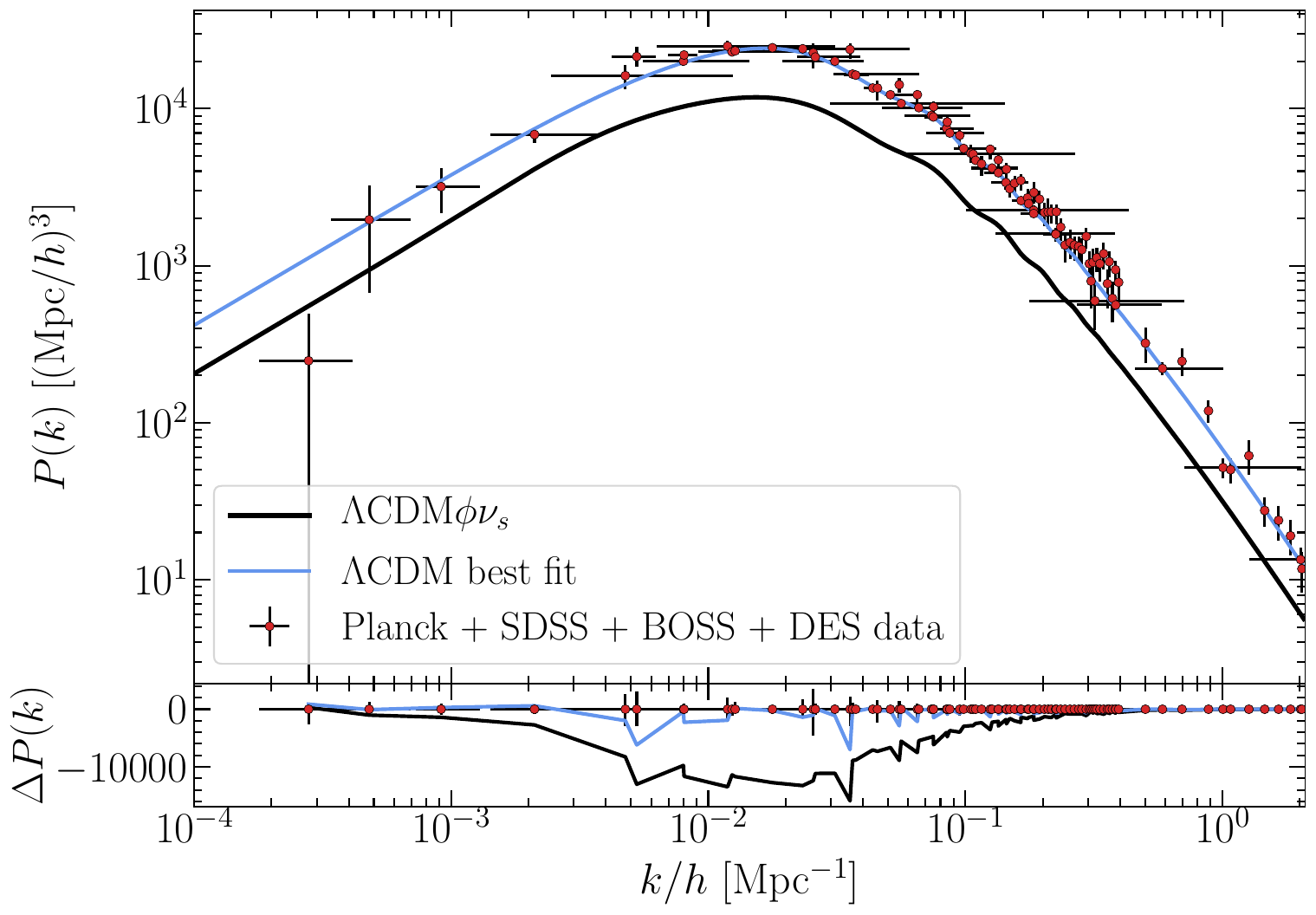}
\caption{Like Fig.\ \ref{fig:Dl_TT_m0}, but for
$m_{\nu_s}>0$.
We assumed $\Omega_g \simeq 0.6$, $w_{\nu_s} \simeq 0.02$, $\Omega_{\Lambda} \simeq 0.1$, and
spatial flatness.}
\label{fig:Dl_TT}
\end{center} 
\end{figure*}

We found that the two new parameters in the massless $\nu_s$ case leave the standard six $\Lambda$CDM parameters almost unchanged compared to their Planck 2018 best-fit values, without ameliorating any current cosmological tensions, such as the $H_0$ and $S_8$ tensions \cite{DiValentino:2021izs,Abdalla:2022yfr}. 
This will be considered in more detail in Section \ref{sect:tensions} below.  The correlations between these parameters are shown in Appendix~\ref{appC}.
Based on the foregoing discussion, the constraints on $\Gamma_{\rho}$ and $\Gamma_{\rho} / \Lambda^4$ are expected to be dominated by the CMB power spectra rather than large-scale-structure observables.
We verified this by excluding BAO, DES, and Pantheon data sets in a separate \texttt{CosmoMC} run, finding that the results in  Fig.~\ref{fig:bounds_m0} (left) were only slightly modified. 

From the bounds in Fig.~\ref{fig:bounds_m0_gen}, one can estimate how large the abundance of massless sterile neutrinos can be today, while remaining consistent with cosmological data. This is shown in Fig.~\ref{fig:bounds_m0} (right), displaying $\Omega_{\nu_s}$ versus the ghost momentum cutoff $\Lambda$ for both mediator models. The surprisingly large values of $\Omega_{\nu_s} \gg 1$ are viable because of their exact cancellation by $\Omega_\phi$, so that the constraints arise only at the level of the perturbations. The maximum values of $\Omega_{\nu_s}$ are reached by saturating the 
gamma-ray constraint, Eq.~\eqref{Lambda_bound}, leading to
\be
    \Omega_{\nu_s}^{95\%\,\,\rm C.L.} \lesssim
    \left\{
    \begin{array}{ll} 9.6 \times 10^{25},& \hbox{scalar\ mediator}\\ 3.4 \times 10^{25},& \hbox{vector\ mediator}
\end{array}\right.\,,
\ee
for $\Lambda\sim 19$\,MeV.
Despite the large numbers, these $\nu_s$ gases are still dilute, by many orders of magnitude, relative to a degenerate Fermi gas with Fermi energy $\Lambda$.  Thus the occupation number $f_{\nu_s}(p)\ll 1$, even for this extreme case.

\subsection{Massive $\nu_s$ case}
\label{sect:massive_nu}
In section \ref{sec:TypeIaSNe}, we made a preliminary study of the effects of producing massive neutrinos plus massless ghosts, comparing to supernova data but not to the other cosmological data sets.  Here we undertake the more complete analysis,
implementing the energy-density evolution of $\phi$ and $\nu_s$ at the background level, given by Eq.~\eqref{eq:Ofits}, in a modified version of the \texttt{CAMB} code. 
It is  embedded in \texttt{CosmoMC} to scan the full parameter space.

In addition to the six parameters of $\Lambda$CDM, we considered $\Omega_g \equiv \Omega_{\nu_s} (a_0) + \Omega_{\phi} (a_0)$, the net contribution of
the ghost plus $\nu_s$ densities today,
and the equation of state for sterile neutrinos $w_{\nu_s}$ as the independent new-physics parameters, even though they are related to each other via 
Eqs.~(\ref{eq:Gamma},\,\ref{eq:Gamma_rho_approx}-\ref{eq:eos_nus_m}).
This allows for simultaneous study of the scalar and vector mediator models, which differ only  by how $\Omega_g$
and $w_{\nu_s}$ are related to the microphysical parameters $m_{\nu_s}$,  $\Lambda$, $M_i$.

The linear perturbations for the two new dark species were included in an approximate form, given by Eq.~\eqref{eq:perts_massive_approx} in Appendix~\ref{appB}. 
A more exact form for the perturbations for massive sterile neutrinos is technically challenging to derive,
and is left for a future study. We expect that the bounds on
$\Omega_g$
and $w_{\nu_s}$ derived here could only be 
moderately 
strengthened by such a study, since the
main effect is through 
the background evolution rather than the perturbations, as discussed in appendix~\ref{appB}.

Figure~\ref{fig:Dl_TT} shows the impact of massless ghosts and massive sterile neutrinos on the CMB temperature anisotropies and on the matter power spectrum, for $\Omega_g = 0.6$ and $w_{\nu_s} =0.02$.
To highlight the effects of the new parameters, those of $\Lambda$CDM  are fixed at their Planck 2018 best-fit values.  (Note that
 $\Omega_{\Lambda}$ is derived from the flatness assumption and is not a fundamental parameter in $\Lambda$CDM.) 
The features in Fig.~\ref{fig:Dl_TT} differ qualitatively from those 
where $m_{\nu_s}=0$, Fig.\ \ref{fig:Dl_TT_m0}.
In the CMB spectrum, there are two main effects: an overall shift of the CMB acoustic peaks toward smaller angular scales (larger $\ell$) and an increase of the late ISW contribution at large scales.

\begin{figure}[t]
\begin{center}
\includegraphics[scale=0.345]{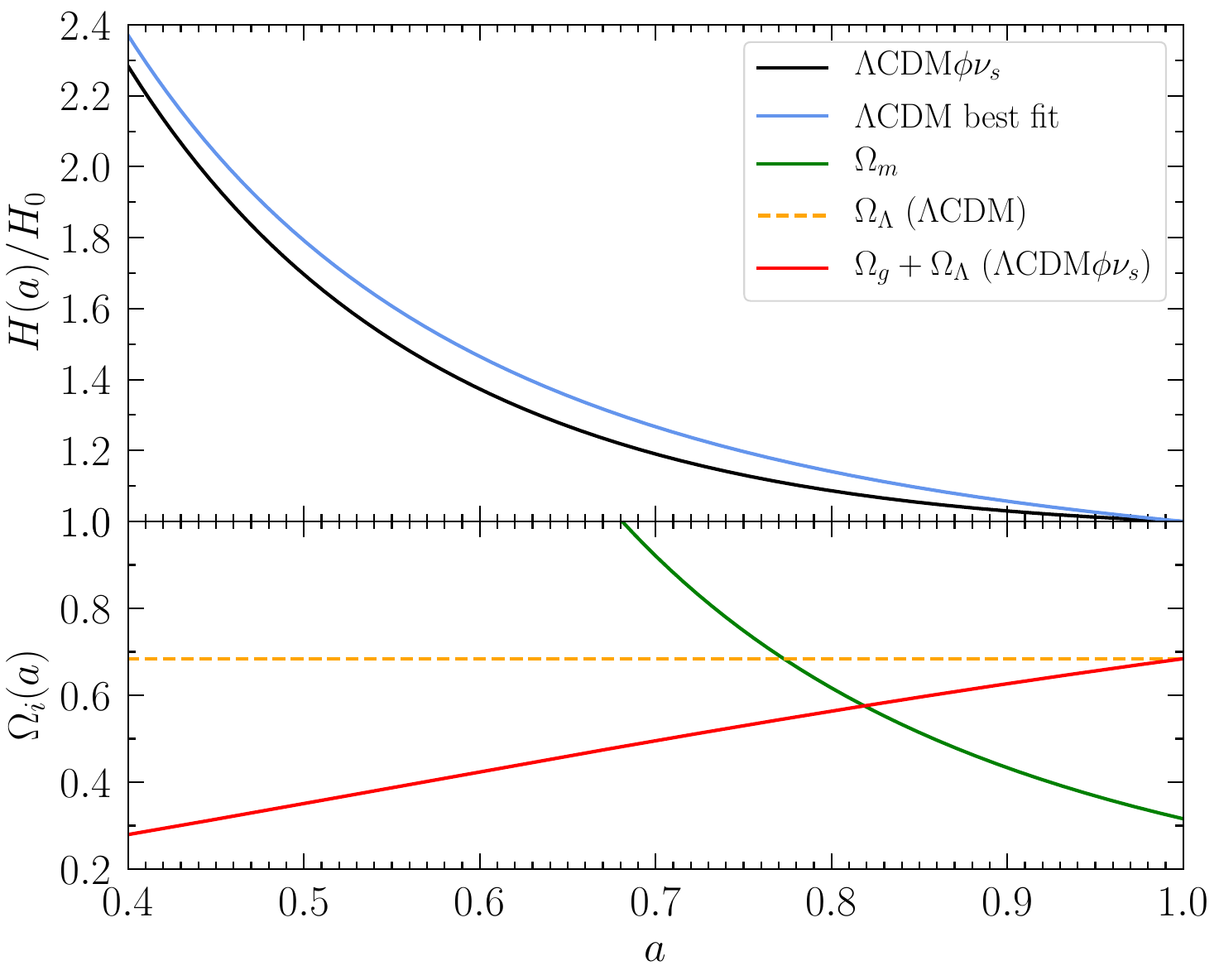}
\caption{Late-time evolution of (\textit{Top}) the Hubble parameter $H(a)$ normalized to its present-day value $H_0$, and (\textit{Bottom}) the energy densities of the main species normalized to the critical density of the Universe today. 
Parameter values are as in Fig.\ \ref{fig:Dl_TT}.
Red curve  is the total dark energy density,
$\Omega_\Lambda + \Omega_g$.
}
\label{fig:HoH0_cmp_densities}
\end{center} 
\end{figure}

\subsubsection{Acoustic peak shift}

The peak shift comes from the decrease of the dark-energy abundance $\Omega_{\Lambda}$~\cite{Hu:2001bc} (a derived parameter in $\Lambda$CDM), which  is required to preserve spatial flatness. If one simultaneously
keeps the other $\Lambda$CDM parameters fixed to their Planck best-fit values,~\footnote{In our model, the preferred values of $H_0$ and $\Omega_m$ change by relatively smaller amounts than $\Omega_{\Lambda}$ and have a subdominant effect on the peak shift.} the Hubble rate $H(a)$ in Eq.~\eqref{eq:Friedmann} decreases relative to the standard scenario because the density of the new species becomes significant only at late times,
as shown in  Fig.~\ref{fig:HoH0_cmp_densities} (top).
Since the age of the Universe is computed as
$t = \int d\ln a/H(a)$, reducing $H(a)$ results in an older universe, and the thermal history, including photon decoupling, has to occur earlier in the past relative to $\Lambda$CDM. This shifts the acoustic CMB peaks to smaller scales, as seen in Fig.~\ref{fig:Dl_TT} (left). 

In detail, the location of the first acoustic peak is given by $\ell_s \simeq \pi / \theta_{s}^{\ast}$, where~\cite{Hu:2001bc}
\be
    \theta_s^{\ast} = \frac{r_s^{\ast}}{D_A^{\ast}} 
\ee
is the angular size of the sound horizon at the time of last scattering,
\be
    r_s^{\ast} = r_s (z_{\ast}) = \int_{z_{\ast}}^{\infty} \frac{dz}{H(z)}\,c_s(z)
\ee
is the sound horizon with $c_s(z)$ the sound speed of the photon-baryon fluid, and
\be
    D_A^{\ast} \equiv D_A(z_{\ast}) = \int_0^{z_{\ast}} \frac{dz}{H(z)}
\ee
is the comoving angular diameter distance between the last-scattering surface at redshift $z_{\ast}$ and today.
A decrease of $\Omega_{\Lambda}$ does not affect the sound horizon $r_s^{\ast}$, since it depends only on the physics before the time of photon decoupling, when the dark energy and the new species were negligible, but it does affect $D_A^{\ast}$ by reducing $H(z)$. Therefore, the comoving angular diameter distance increases with respect to $\Lambda$CDM, reducing the value of $\theta_{s}^{\ast}$ and  shifting the location of the first CMB peak to larger $\ell$. 

\begin{figure*}[t]
\begin{center}
\includegraphics[scale=0.35]{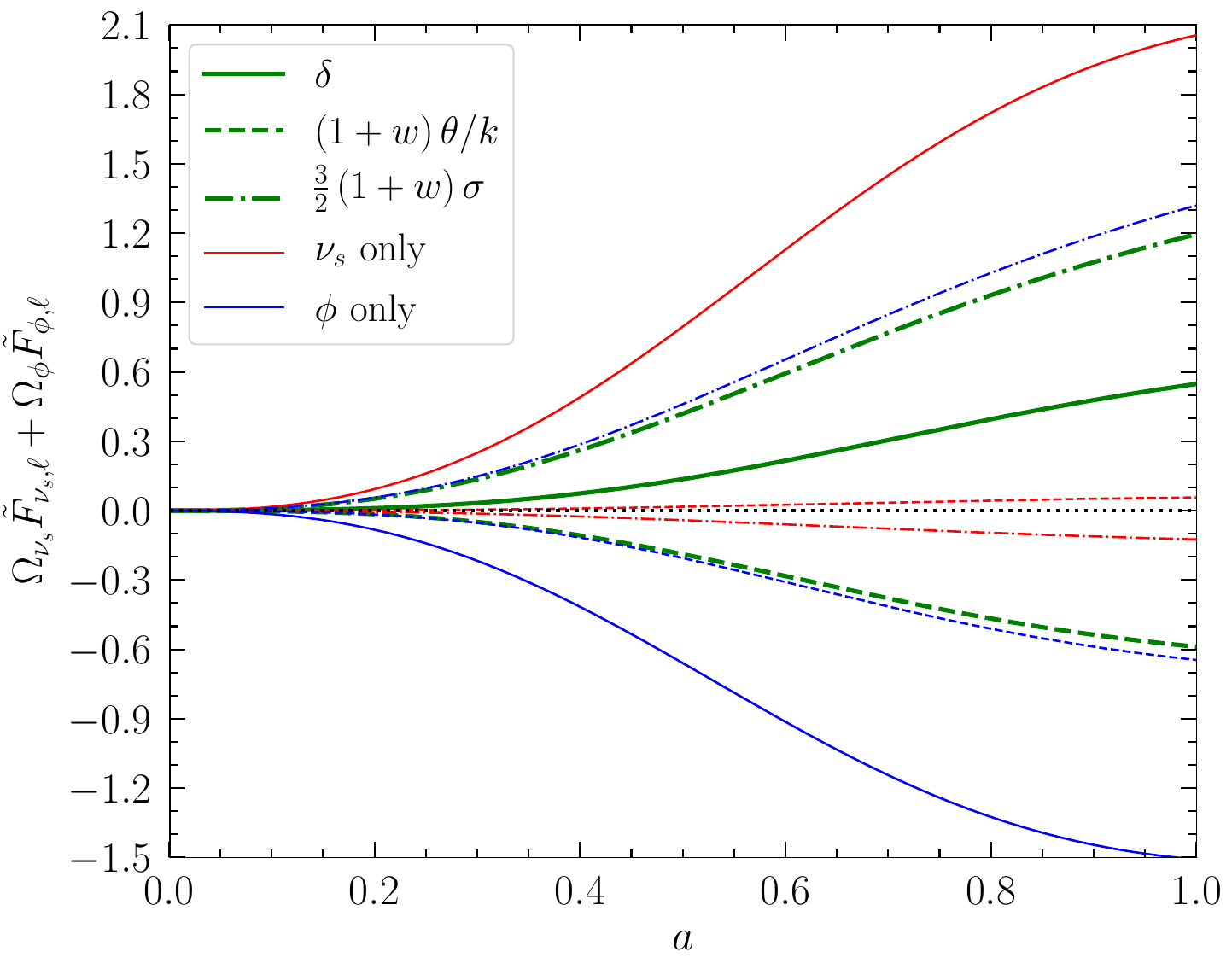}
\includegraphics[scale=0.35]{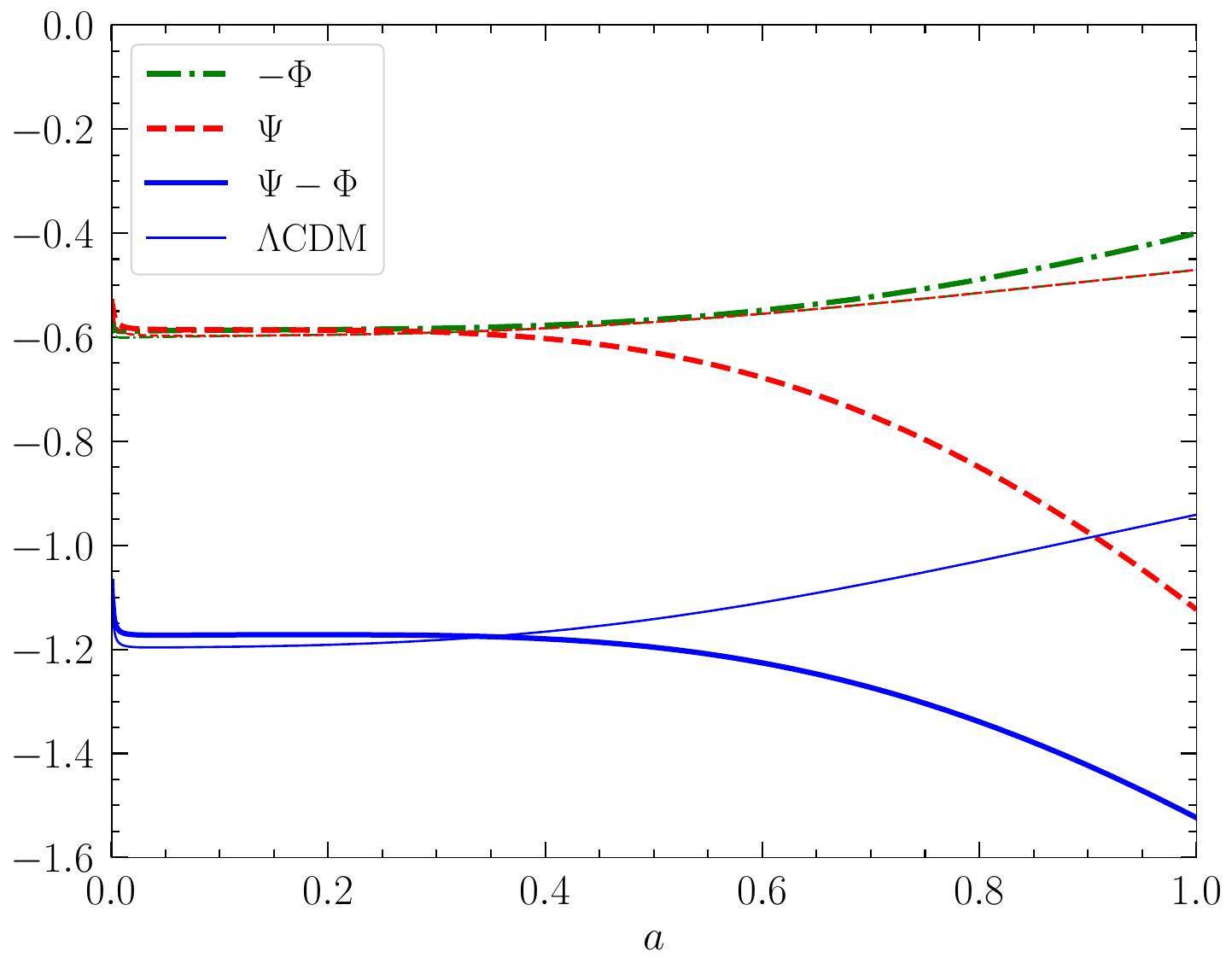}
\caption{Like Fig.\ \ref{fig:ISW_m0}, but for the massive $\nu_s$ case, with 
 $w_{\nu_s} \simeq 0.02$ and $\Omega_g \simeq 0.6$.
}
\label{fig:ISW}
\end{center} 
\end{figure*}

\subsubsection{Late ISW enhancement}

The second feature in the CMB temperature anisotropy spectrum, when nonrelativistic $\nu_s$ are included, is an increase in the late ISW contribution, which can be explained by the interplay of two competing phenomena: the behavior of the cosmological perturbations of $\nu_s$ and $\phi$, and the decrease of $\Omega_{\Lambda}$.

Fig.~\ref{fig:ISW} (left) shows the large-scale evolution of the synchronous-gauge physical perturbations for the ghosts and sterile neutrinos. 
Similarly to the massless $\nu_s$ case, the  physical shear perturbation $\rho_{\phi} \sigma_{\phi}$ for ghosts (blue dot-dashed line) dominates over that for sterile neutrinos (red dot-dashed curve), leading to an overall positive contribution to the total shear (green dot-dashed line) and a consequent decrease of the Newtonian potential $\Psi$, according to Eq.~\eqref{eq:Phi_Psi_potentials}.
The impact on the Newtonian curvature $\Phi$ from the perturbations of the new species is small, because the positive contribution of the density contrast $\delta$, which is dominated by $\nu_s$, is largely canceled by the negative contribution of the velocity perturbation $\theta$, dominated by $\phi$, as shown in Fig.~\ref{fig:ISW} (left).
The net effect on $(\Psi - \Phi)$, whose conformal time derivative enters the late ISW contribution in Eq.~\eqref{eq:Cl_ISW}, is to drive it to more negative values on a shorter timescale, leading to an increase in the late ISW effect, as shown in Fig.~\ref{fig:ISW} (right). 

The situation is complicated by the decrease of $\Omega_{\Lambda}$ in the background density evolution, which tends to counteract the late ISW effect. The dark energy impacts the decay of the gravitational wells which photons traverse. 
In $\Lambda$CDM, the Newtonian curvature $\Phi$ and potential $\Psi$, which parameterize the depth of the gravitational wells, and remain constant during matter domination, but they start to decay when dark energy takes over, illustrated by the thin curves in Fig.~\ref{fig:ISW} (right). 
This is due to the gravitational wells becoming shallower when the Universe starts accelerating, enhancing the late ISW effect.   Conversely,
the effect is diminished if $\Omega_{\Lambda}$ is reduced \cite{Hu:2001bc}, attenuating the increase of the late ISW contribution arising from the $\phi$ and $\nu_s$ perturbations. 

Which of these competing effects wins is determined
by  the relative values of $\Omega_{\Lambda}$ and $\Omega_g$, due to the ${\nu_s}$ contribution to the latter.  If $\Omega_{\Lambda}\gg \Omega_g$, the $\phi+\nu_s$ contribution becomes negligible and the late ISW effect is reduced.
In general, we see that for relativistic sterile neutrinos, with $w_{\nu_s} \gtrsim 0.1$, the late ISW effect in the CMB temperature anisotropy spectrum decreases, similarly to the massless $\nu_s$ case.

\subsubsection{Spatial curvature}

The previous discussion assumes that the Universe is spatially flat, which forces $\Omega_{\Lambda}$ to decrease with respect to the Planck prediction, whenever massive sterile neutrinos are produced. 
The presence of a positive curvature ($\Omega_k < 0$) might reduce the impact of the new species on the CMB spectrum since a nonzero curvature provides opposite effects to those seen here~\cite{Hu:2001bc}. 
A closed universe could also reduce some discrepancies between the $\Lambda$CDM model and the Planck 2018 data~\cite{DiValentino:2019qzk,Handley:2019tkm}, although it is still unclear whether they might be due to internal systematics~\cite{Vagnozzi:2020rcz}.
We leave the study of phantom fluids in a non-flat Universe for a future work.

\subsubsection{Matter power spectrum}\label{subsec:matter_power_spectrum}

As concerns the matter power spectrum, shown in  Fig.~\ref{fig:Dl_TT} (right), the vacuum decay into massless $\phi$ and nonrelativistic massive $\nu_s$ leads to a suppression at all scales. 
The dominant effect is the free streaming of the new species, which occurs immediately after they are produced, causing a damping of their density fluctuations compared to standard nonrelativistic species, such as CDM.

This is particularly important for sterile neutrinos, since they contribute to the total nonrelativistic matter and hence to the matter power spectrum. The density contrast for $\nu_s$ is smaller than that for CDM, $\delta_{\nu_s} < \delta_{\rm cdm}$, due to the collision term (see Appendix~\ref{appB}), whereas its energy density $\rho_{\nu_s}$ can be of the same order as $\rho_m$, but in any case it leads to $\delta \rho_{\nu_s} \lesssim \delta \rho_{\rm cdm}$.
%for the physical perturbations. 
Since the matter power spectrum is defined in terms of $\delta \rho_m / \rho_m$ to which both $\nu_s$ and CDM contribute, the net effect is a decrease of $P(k)$ relative to $\Lambda$CDM.  
This differs from massive SM neutrinos, where the suppression occurs only at scales smaller than the free-streaming length set by the time of relativistic-to-nonrelativistic transition~\cite{Lesgourgues:2006nd}.  Unlike SM neutrinos, the equation-of-state $w_{\nu_s}$ for $\nu_s$ from vacuum decay is approximately time-independent, depending only upon the value of $m_{\nu_s}$.

However, the free streaming of $\nu_s$ is not the only effect on $P(k)$.
Sterile neutrinos and ghosts contribute to the Friedmann equation and their abundance increases with time. 
If the $\Lambda$CDM parameters and $H_0$ are fixed at their Planck values, increasing $\Omega_g$ requires reducing $\Omega_{\Lambda}$ to keep the total dark energy unchanged. The net effect of the time evolution of $\Omega_g (a) + \Omega_{\Lambda}$ on the Friedmann equation is to lower the value of $H(a)$ (Fig.~\ref{fig:HoH0_cmp_densities}, upper plot) and to delay the transition from matter domination to the epoch when dark energy, including the $\phi + \nu_s$ fluid, started to dominate (Fig.~\ref{fig:HoH0_cmp_densities}, bottom plot).
This allowed more matter to fall into gravitational wells and thereby enhance the matter power spectrum.

Similarly, the perturbations of the new species, and in particular the nonzero shear for $\nu_s$ and $\phi$, tend to reduce the decay of the Newtonian gravitational potential $\Psi$ and curvature $\Phi$, leading to an increase of the matter clustering. This is a direct gravitational back-reaction effect at the level of perturbations, which adds to the background effect discussed above, leading to an increase of $P(k)$ primarily affecting  large angular scales, as in the case of massless $\nu_s$.

For nonrelativistic massive $\nu_s$, the free-streaming effect dominates over the background and back-reaction ones, as shown in Fig.~\ref{fig:Dl_TT} (right), producing less matter clustering. 
On the other hand, the latter two effects dominate in the case of relativistic or mildly relativistic massive $\nu_s$, because such sterile neutrinos do not contribute to the total nonrelativistic matter and hence do not directly affect the matter power spectrum.

\begin{figure*}[t]
\begin{center}
\includegraphics[scale=0.65]{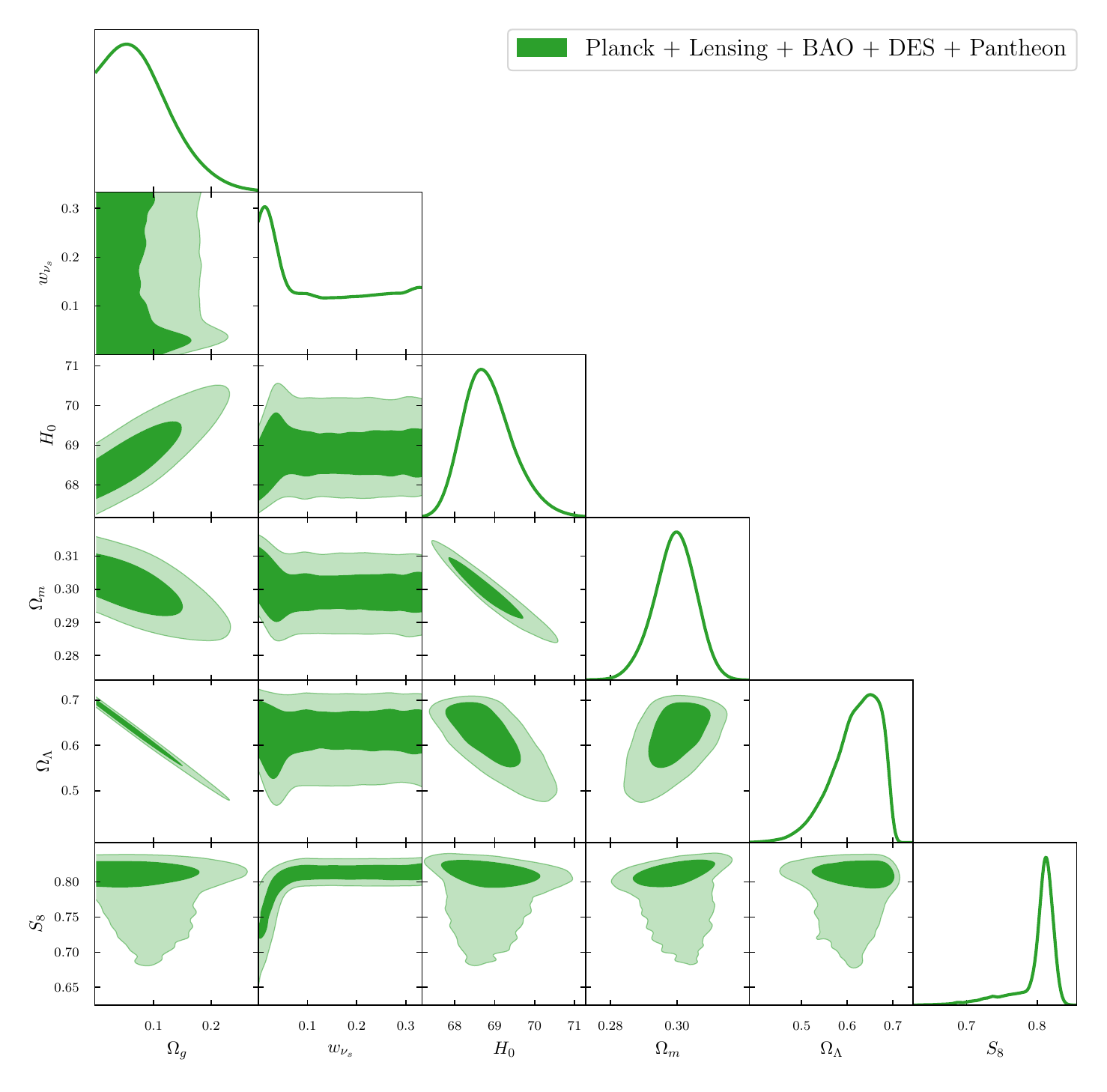}
\caption{Correlations between the $\Lambda$CDM$\phi\nu_s$ model parameters, as inferred from \texttt{CosmoMC} for massive $\nu_s$, valid independently of the mediator model used. The data sets Planck + Lensing + BAO + DES + Pantheon are the same as described in subsection~\ref{subsec:results_m0}. The darker and lighter green shaded regions correspond to the $68\%$ and $95\%$ C.L. intervals, respectively. 
}
\label{fig:cosmomc_m_red}
\end{center} 
\end{figure*}

\subsubsection{Results}

Similarly to the massless $\nu_s$ case, the MCMC code \texttt{CosmoMC} was used to constrain the two new parameters $\Omega_g$ and $w_{\nu_s}$, and to study their correlations with the $\Lambda$CDM ones. We adopted the same convergence requirement and data sets as used for the $m_{\nu_s}=0$ case, as described in  subsection~\ref{subsec:results_m0}.

Figure~\ref{fig:cosmomc_m_red} shows the correlations between the new parameters and those derived from $\Lambda$CDM that are significantly
correlated.
The latter include the total matter abundance $\Omega_m$, which is separate from the $\nu_s$ contribution, the dark energy abundance $\Omega_{\Lambda}$, the Hubble constant $H_0$, and the clustering amplitude parameter $S_8 = \sigma_8\sqrt{\Omega_m/0.3}$.
The extended version of the correlation parameter plot shown in Fig.~\ref{fig:cosmomc_m_red} is discussed in Appendix~\ref{appC}, including a table with the $68\%$ C.L. limits.

Increasing $\Omega_g$, defined as the present-day energy fractions of $\phi$ plus $\nu_s$, leads to an increase in $H_0$, which can be intuitively understood by the phantom nature of the effective $\phi+\nu_s$ fluid.
Its equation of state $w_{\rm eff}$  was found  to be $< -1.3$ (Fig.~\ref{fig:SN_bounds}), more negative than that of the cosmological constant.
Therefore, increasing the relative abundance of the phantom fluid causes the Universe to expand faster today than in the $\Lambda$CDM model, leading to a larger value of $H_0$.

Such accelerated expansion entails weaker growth of large scale structure, which is parameterized by $S_8 = \sigma_8 \sqrt{\Omega_m / 0.3}$, where $\sigma_8$ is the linear-theory standard deviation of matter density fluctuations in a sphere of radius $8\,h^{-1}\,\,\text{Mpc}$. Namely, Fig.~\ref{fig:cosmomc_m_red} shows a slow decrease in the parameter $S_8$ with increasing $H_0$, which is most pronounced 
 at small values of $w_{\nu_s}$. This confirms the observation made in subsection~\ref{subsec:matter_power_spectrum}, where the decrease of the amplitude of the matter power spectrum (and hence $\sigma_8$ and $S_8$) was observed for nonrelativistic massive $\nu_s$.
We will  quantitatively discuss the implications of those findings for $H_0$ and $S_8$ in the next subsection, devoted to possible resolutions of known cosmological tensions.

\begin{figure*}[t]
\begin{center}
\includegraphics[scale=0.35]{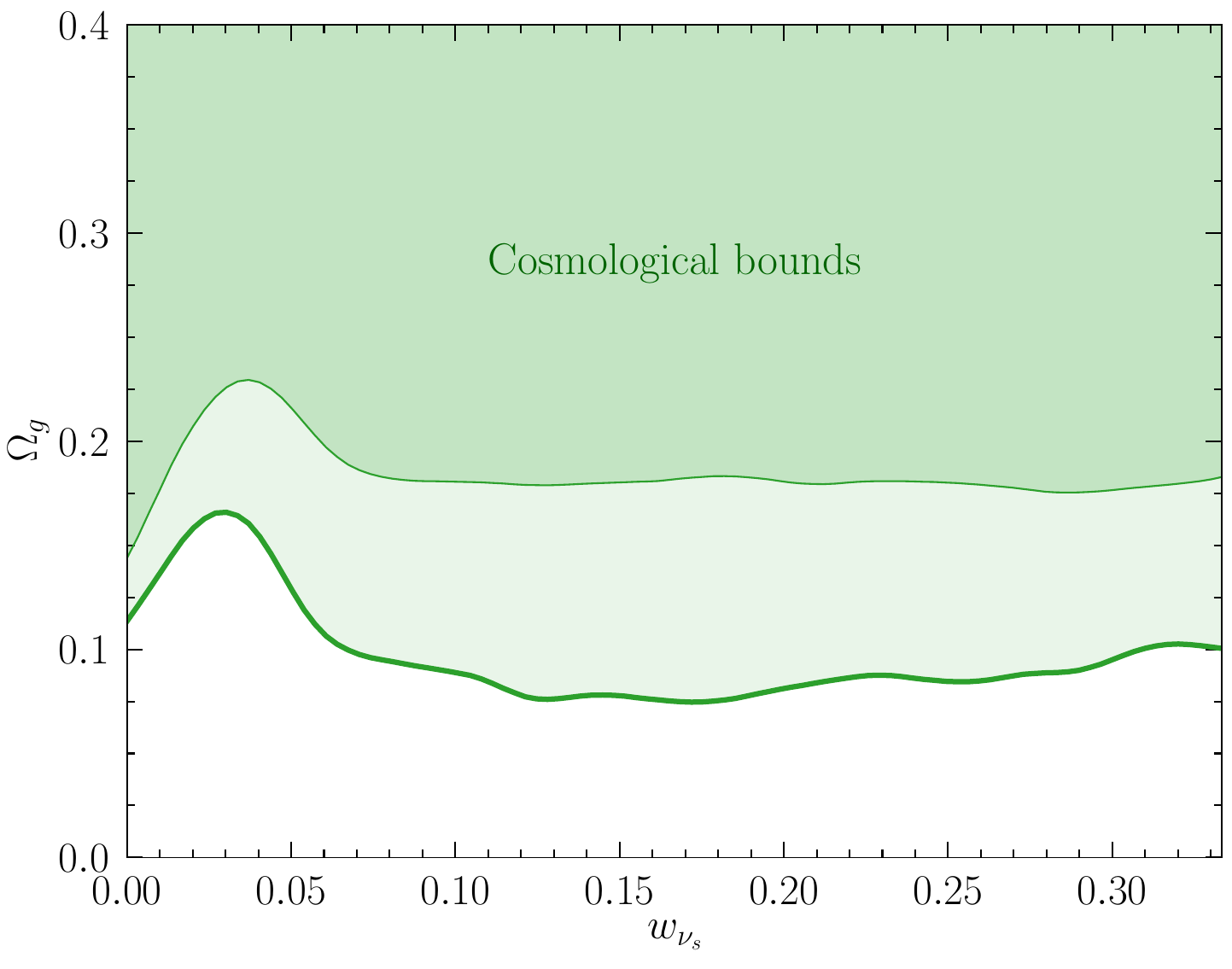}
\includegraphics[scale=0.345]{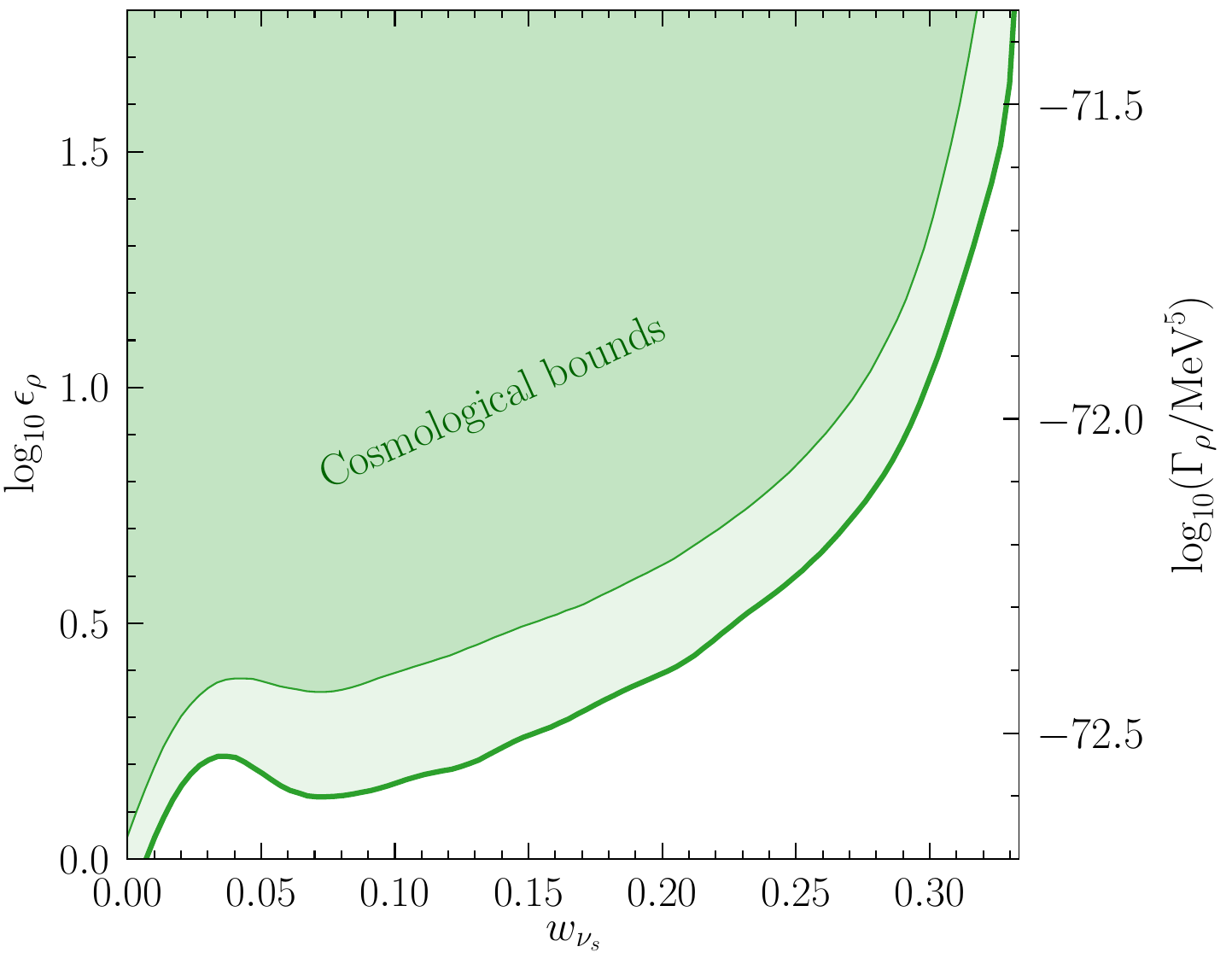}
\caption{\textit{Left}: Constraints on $\Omega_g = (\rho_\phi + \rho_{\nu_s})/\rho_{\rm crit}$ versus $w_{\nu_s} = P_{\nu_s}/\rho_{\nu_s}$ for the massive $\nu_s$ scenario, derived from Fig.~\ref{fig:cosmomc_m_red}.
\textit{Right}: Corresponding excluded regions in the plane of  $\epsilon_{\rho} = \Gamma_{\rho} / (3 H_0 \rho_{\rm{crit}})$, 
the dimensionless vacuum decay rate in the Boltzmann equations (\ref{eq:Boltzmann_y}), and $w_{\nu_s}$. (Right axis shows $\log_{10}\Gamma_\rho$ in MeV$^5$ units.)
In both plots, the thick and thin curves show the respective $68\%$ and $95\%$ C.L. exclusion limits. 
}
\label{fig:bound_Omegag_nus_m}
\end{center} 
\end{figure*}

\begin{figure*}[t]
\begin{center}
\includegraphics[scale=0.35]{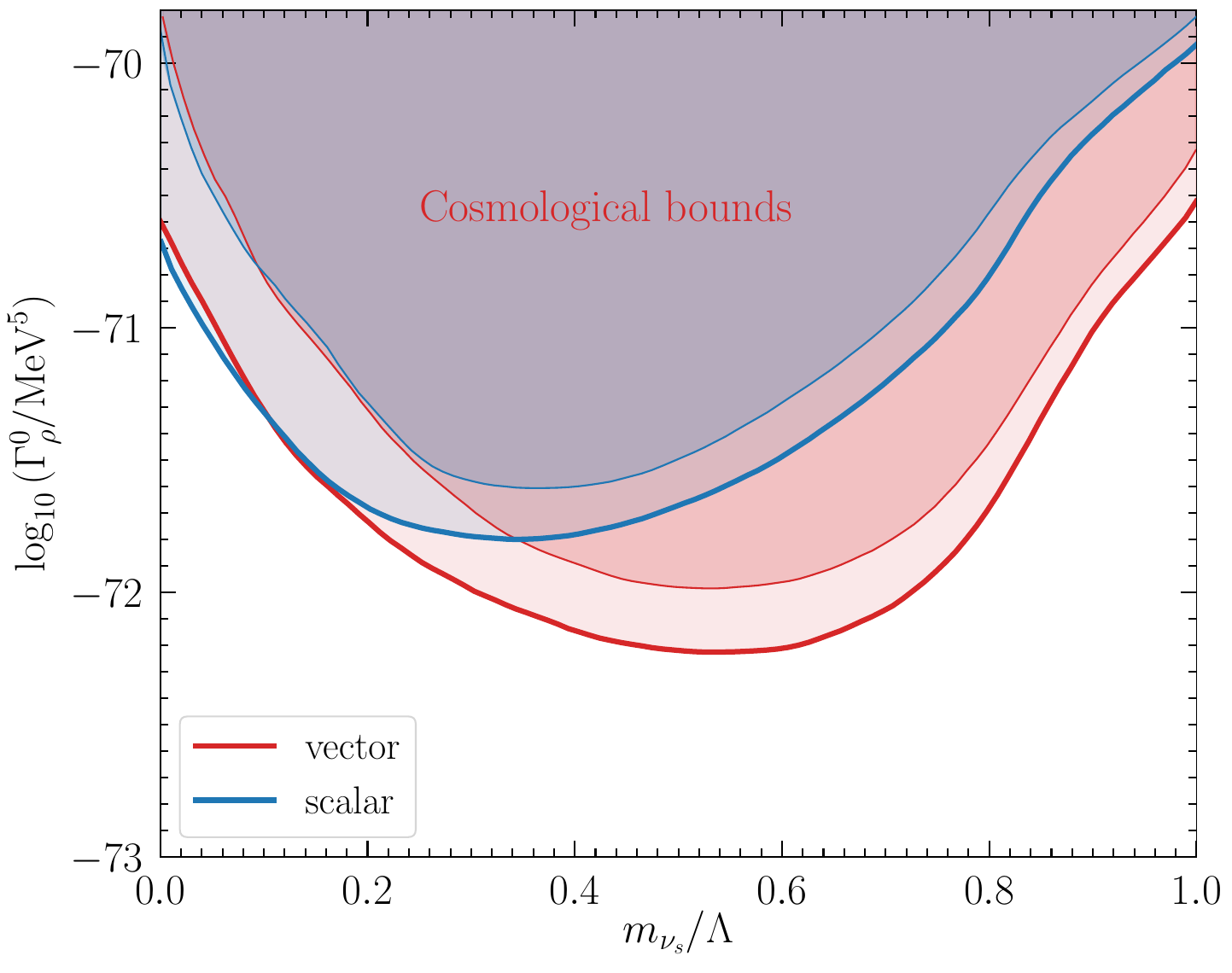}
\includegraphics[scale=0.35]{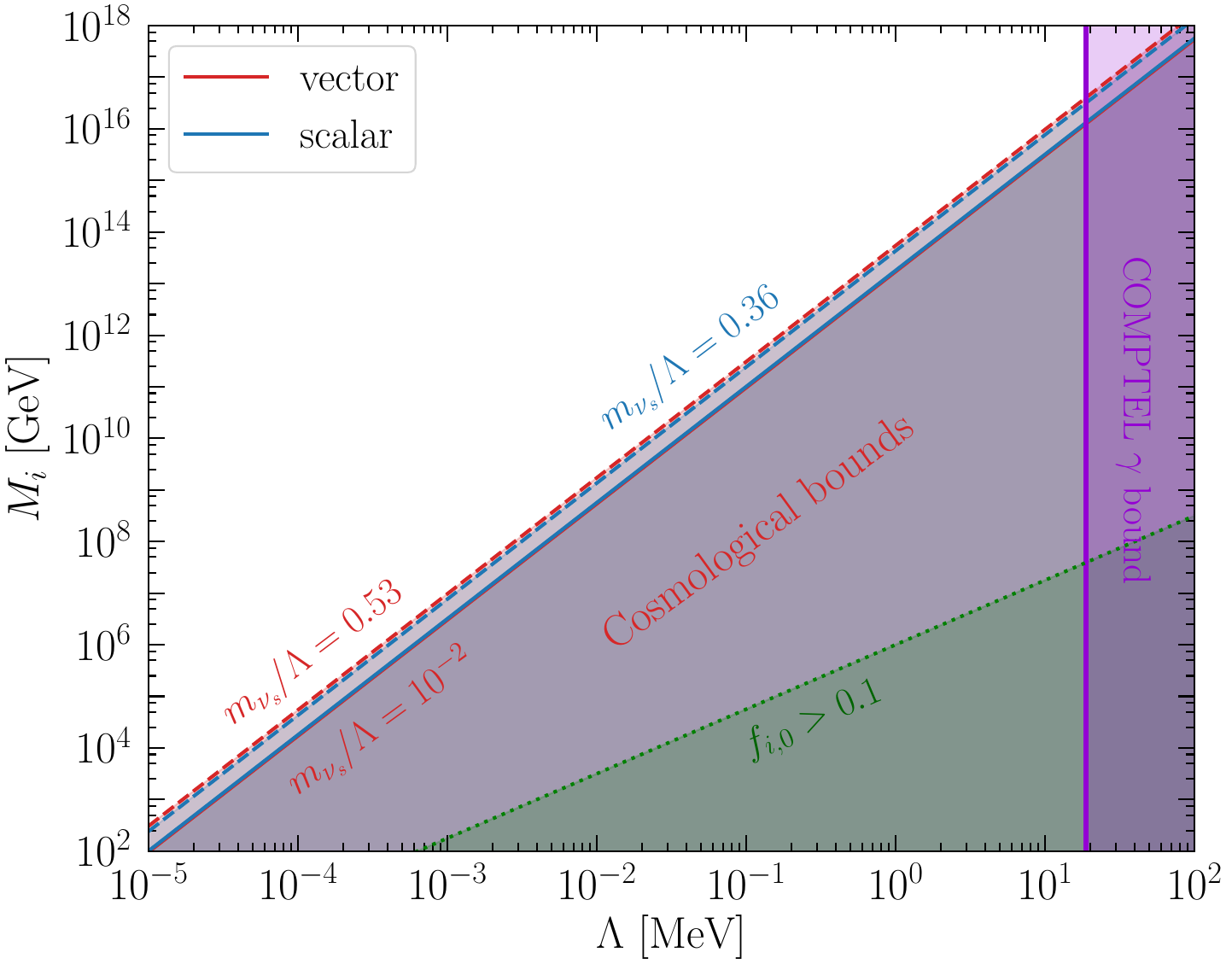}
\caption{\textit{Left}: Mapping of excluded regions of Fig.~\ref{fig:bound_Omegag_nus_m} (left) to the 
microphysics quantities $\Gamma_{\rho}^0 \equiv \Gamma_{\rho}|_{m_{\nu_s} = 0}$ (the vacuum decay rate) and $m_{\nu_s} / \Lambda$, for vector (red) and scalar (blue) mediator models. 
Thick (thin) curves show the $68\%$ ($95\%$) C.L.\ exclusion limits.
\textit{Right}: $95\%$ C.L. bounds on the parameters $M_i$ (the mediator mass
scale, with $i = s, v$), versus the ghost momentum cutoff $\Lambda$, corresponding to the left plot. Solid lines correspond to $m_{\nu_s} / \Lambda = 10^{-2}$, and dashed to the values that give the strongest constraints, namely $m_{\nu_s} / \Lambda \simeq 0.53\,\,(0.36)$ for the vector (scalar) mediator case.
}
\label{fig:bound_LogGammarho0_moL_m}
\end{center} 
\end{figure*}

Another striking correlation is the decrease of both $\Omega_{\Lambda}$ and $\Omega_{m}$ as the present-day abundance of the new species increases. The former effect derives from the flatness condition $\Omega_{\Lambda} + \Omega_m + \Omega_g = 1$.
Although $\Omega_m$ is also a derived parameter in $\Lambda$CDM, it is determined by  $\Omega_b h^2$ and $\Omega_c h^2$, which are constrained by the relative heights of the CMB peaks~\cite{Planck:2013pxb}.
Since our model affects only the late-time evolution and hence does not alter photon decoupling, whose physics impacts the peak height, the value of $(\Omega_b + \Omega_c) h^2$  remains nearly unchanged relative to $\Lambda$CDM. This is verified in the extended correlation plot in Appendix~\ref{appC}.
Consequently, the decrease in $\Omega_m$ with 
increasing $\Omega_g$ comes mostly from the growth of $H_0$, as explained above.

For constraining $\Omega_g$ and $w_{\nu_s}$, which are taken as independent input parameters, the second-from-top left plot 
of Fig.\ \ref{fig:cosmomc_m_red}
displays their allowed values.  It shows that $\Omega_g$ is roughly independent of $w_{\nu_s}$, which can be understood from the definition of $\Omega_g$ as the net amount of $\rho_{\nu_s}+ \rho_\phi$ that has been produced, whose value already incorporates the effect of $w_{\nu_s}$.  In the limit of $m_{\nu_s}\to 0$, $w_{\nu_s}\to 1/3$, with $\Gamma_\rho$ fixed, it is true that $\Omega_g$ would be driven to zero, but in the $\Omega_g$-$w_{\nu_s}$ parametrization, $\Gamma_\rho$ is not fixed and the former are independent of each other.
In fact, $w_{\nu_s}$ does not correlate strongly with the other $\Lambda$CDM parameters either. 
Its main effect is on the linear perturbations.  At the more important level of the homogeneous energy densities, $w_{\nu_s}$ has a relatively small impact.

The $68\%$ and $95\%$ C.L. exclusion plots for  $\Omega_g$ and $w_{\nu_s}$ corresponding to Fig.~\ref{fig:cosmomc_m_red} are shown in Fig.~\ref{fig:bound_Omegag_nus_m}
(left).  We also map them onto the plane of
 $w_{\nu_s}$ and 
$\epsilon_{\rho} = \Gamma_{\rho} / (3 H_0 \rho_{\rm crit})$, or directly $\Gamma_{\rho}$, using Eq.~\eqref{eq:Ofits} and Fig.~\ref{fig:cosmomc_m_red}. The resulting limits, shown in Fig.~\ref{fig:bound_Omegag_nus_m} (right), resemble those obtained from the supernova analysis of section~\ref{sec:TypeIaSNe}, Fig.~\ref{fig:SN_bounds} (left).
As expected, large values of  $\Gamma_{\rho}$ are allowed as $w_{\nu_s}\to 1/3$, since $\Omega_g$ can remain fixed in that limit.

Using Eqs.~\eqref{eq:Gamma_rate_int}, ~\eqref{eq:Gamma_rho_approx}, and~\eqref{eq:eos_nus_m}, one can constrain  $\Gamma_{\rho}^0 \equiv \Gamma_{\rho}|_{m_{\nu_s} = 0}$  versus $m_{\nu_s}/\Lambda$. This is shown in Fig.~\ref{fig:bound_LogGammarho0_moL_m} (left) for the vector and scalar mediator models.
It does not correspond to independent bounds on the microphysical parameters $\Lambda$ and $M_i$,
since cosmological observables are sensitive to
the derived combination $\Gamma_{\rho}^0$. 
However we can use Eqs.~\eqref{eq:Gamma_rho_approx} and~\eqref{eq:Gamma_rate_int}, together with the bounds on $\Gamma_{\rho}^0$, to limit $M_i$ as a function of $\Lambda$, for fixed values of $m_{\nu_s} / \Lambda$, 
as shown in 
Fig.~\ref{fig:bound_LogGammarho0_moL_m} (right).   

Analytically, the new physics
mass scales are bounded from below as
\be
    M_i \gtrsim (1-5)\times 10^{13}\left(\Lambda\over{\rm MeV}\right)^{2.25}{\rm GeV}
\ee
for neutrino masses $m_{\nu_s}/\Lambda \gtrsim 0.01$.  The strongest limits occur for particular values of  $m_{\nu_s} / \Lambda \simeq 0.53\,\,(0.36)$ for the vector (scalar) mediator respectively. 
As expected, these bounds are much stronger than those in the massless $\nu_s$ scenario, (section~\ref{subsec:results_m0}),  resulting mainly from modifications at the background level, as opposed to perturbations.
They depend weakly on $m_{\nu_s} / \Lambda$, since
$\Gamma_{\rho}^0$ in  Fig.~\ref{fig:bound_LogGammarho0_moL_m} (left) changes by at most few orders of magnitude from varying $m_{\nu_s}$.

\begin{figure*}[t]
\begin{center}
\includegraphics[scale=0.65]{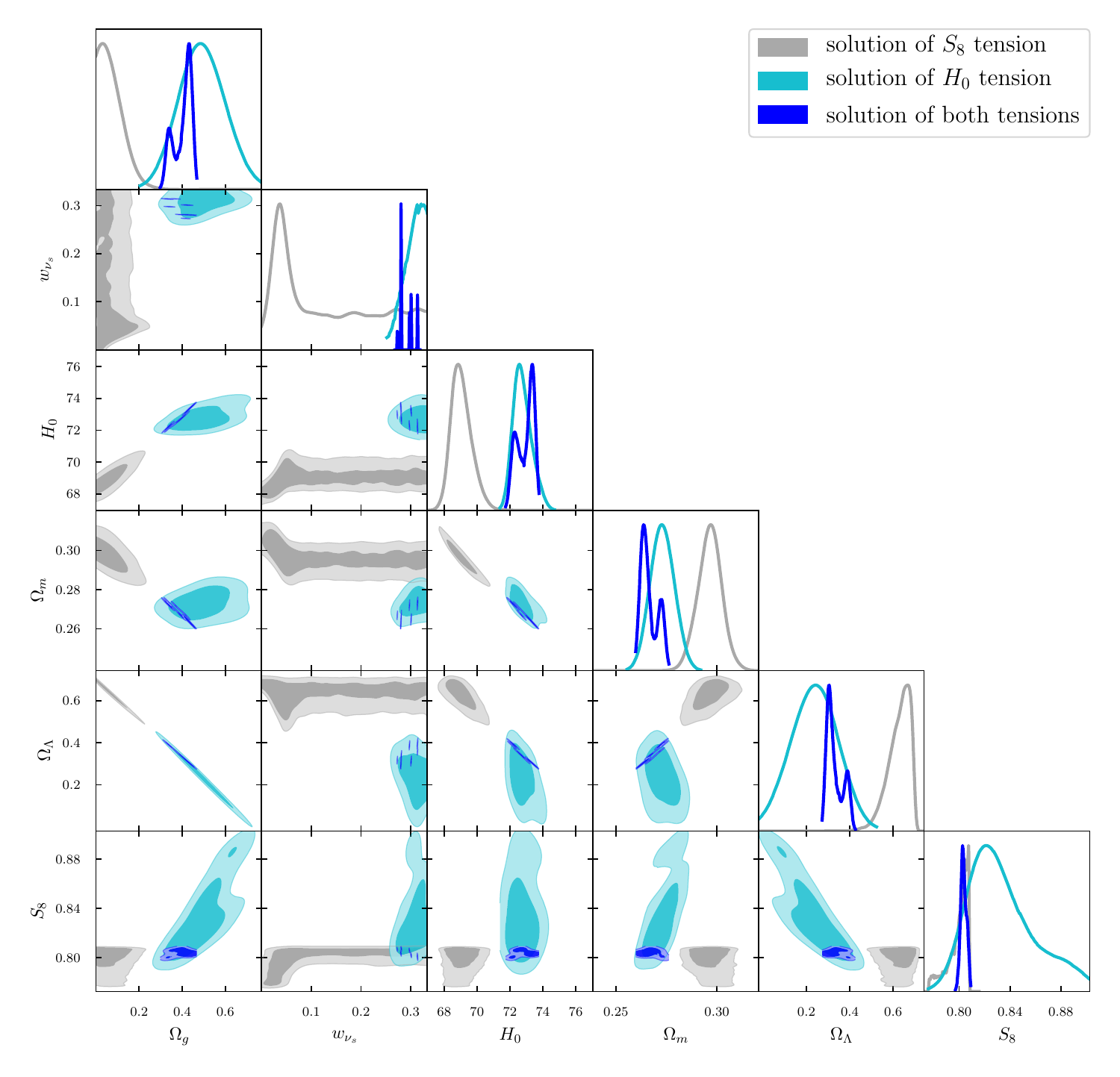}
\caption{Correlations between the $\Lambda$CDM$\phi\nu_s$ parameters for the models addressing the $H_0$ and $S_8$ tensions, selected from the full MCMC scan shown in Fig.~\ref{fig:cosmomc_m_red}.
Gray regions solve only the $S_8$ tension, cyan addresses only the $H_0$ tension, and blue indicates models ameliorating both.
Darker and lighter regions correspond to the $68\%$ and $95\%$ C.L.\ intervals,
respectively.
}
\label{fig:H0tension}
\end{center} 
\end{figure*}

\subsection{Cosmological tensions}
\label{sect:tensions}

In recent years, discrepancies have emerged among various observational probes regarding the values of certain cosmological parameters. The most notable one is the Hubble tension, which refers to a mismatch between early-time and late-time estimations of the Hubble constant $H_0$~\cite{Freedman:2017yms,Verde:2019ivm,DiValentino:2020zio,Abdalla:2022yfr,Dainotti:2021pqg,Dainotti:2022bzg,Lenart:2022nip,Bargiacchi:2023jse,Dainotti:2023ebr,Bargiacchi:2023rfd,Colgain:2022nlb,Malekjani:2023dky,Wong:2019kwg}.
This tension has crossed the critical $5\sigma$ level between the two most precise measurements: Planck finds $H_0 = 67.66 \pm 0.42$ km/s/Mpc~\cite{Planck:2018vyg} from fitting the CMB data within the $\Lambda$CDM framework, whereas SH0ES finds $H_0 = 73.04 \pm 1.04$ km/s/Mpc~\cite{Riess:2021jrx} based on the distance-ladder approach.~\footnote{A recent determination of $H_0$ from late-time measurements can be found in Ref.~\cite{Uddin:2023iob}.} 
A milder tension also exists  concerning the large-scale structure of the Universe~\cite{DES:2017myr,Hildebrandt:2018yau,KiDS:2020suj,Heymans:2020gsg,DES:2021wwk,HSC:2018mrq,Nunes:2021ipq,DiValentino:2020vvd}. Estimations of the amplitude of late-time matter fluctuations $S_8$ from galaxy redshift surveys are consistently lower than those inferred from CMB data, which points to $S_8 = 0.825 \pm 0.011$~\cite{Planck:2018vyg}, assuming the $\Lambda$CDM model.
The most recent late-time measurement $S_8 = 0.790^{+0.018}_{-0.014}$,
from the DES Y3$+$KIDS-1000 surveys~\cite{Kilo-DegreeSurvey:2023gfr}, agrees with the Planck value within the $\sim 2\sigma$ level.

This motivates us to reconsider our main results for the parameter distributions,
Fig.~\ref{fig:cosmomc_m_red}, with respect 
to the cosmological tensions.  In particular, 
we want to determine 
whether there exist models within the $\Lambda\text{CDM}\phi\nu_s$ scenario that can ameliorate the tensions, while remaining compatible with all the cosmological data. 
Figure~\ref{fig:H0tension} shows the same correlation plot between the parameters as in Fig.~\ref{fig:cosmomc_m_red}, restricted to those models that resolve the $H_0$ and $S_8$ tensions within the $1\sigma$ level.
Interestingly, both tensions can be resolved in a narrow region of the parameter space with $w_{\nu_s} \sim 0.3$ and $\Omega_g \sim 0.3 - 0.5$, forcing both $\Omega_{\Lambda}$ and $\Omega_m$ to decrease from their Planck 2018 best-fit values. 
In our MCMC chains, only about $0.015\%$ of the models can address both tensions, while keeping a total $\chi^2$ comparable to that of the $\Lambda$CDM model.

\begin{table}[t!]
\begin{tabular} {| c | c | c | c |}
\hline
 Parameter & $\Lambda$CDM & \multicolumn{2}{c|}{$\Lambda$CDM$\phi\nu_s$}  \\
  & & solve tensions & best fit  \\
\hline
\hline
{\boldmath$\Omega_b h^2   $} & $0.02246$ & $0.02222$ & $0.02250$ \\
{\boldmath$\Omega_c h^2   $} & $0.1192$ & $0.1182$ & $0.1185$ \\
{\boldmath$100\,\theta_{\rm MC} $} & $1.04093$ & $1.04087$ & $1.04119$ \\
{\boldmath$\tau           $} & $0.0530$ & $0.0609$ & $0.05817$ \\
{\boldmath${\rm{ln}}(10^{10} A_s)$} & $3.040$ & $3.0593$ & $3.0468$ \\
{\boldmath$n_s            $} & $0.9636$ & $0.9639$ & $0.9667$ \\
\hline
{\boldmath$\Omega_g$} & $0.0001$ & $0.3934$ & $0.1430$ \\
{\boldmath$w_{\nu_s}$} & $0.2480$ & $0.2811$ & $0.0548$ \\
\hline
$H_0 / \rm{[km/s/Mpc]}$ & $67.70$ & $72.71$ & $69.76$ \\
$\Omega_{\Lambda}           $ & $0.6893$ & $0.3397$ & $0.5659$ \\
$\Omega_m                  $ & $0.3106$ & $0.2669$ & $0.2911$ \\
$\sigma_8                  $ & $0.8062$ & $0.8556$ & $0.8118$ \\
$S_8                       $ & $0.820$ & $0.807$ & $0.800$ \\
\hline
$\log_{10}{(\Gamma_{\rho}^0 / \rm{MeV}^5)}$ & $-75.26\, (v)$ & $ -71.51\,(v)$ & $ -71.81\,(v)$ \\
 & $-75.02\, (s)$ & $ -71.38\,(s)$ & $ -70.87\,(s)$ \\
$m_{\nu_s} / \Lambda                $ & $0.331 \,(v)$ & $ 0.249\,(v)$ & $0.805\, (v)$ \\
 & $0.320 \,(s)$ & $ 0.235\,(s)$ & $0.820 \,(s)$ \\
$w_{\rm eff} (z=0)$ & $-1.2925$ & $-1.3462$ & $-1.3128$ \\
\hline
\hline
$\chi^2_{\rm Planck}$ & $2780.5$ & $2790.0$ & $2780.3$ \\
$\chi^2_{\rm Lensing}$ & $9.2$ & $8.4$ & $8.8$ \\
$\chi^2_{\rm BAO}$ & $5.7$ & $17.9$ & $7.4$ \\
$\chi^2_{\rm DES}$ & $523.4$ & $510.8$ & $512.0$ \\
$\chi^2_{\rm Pantheon}$ & $1035.0$ & $1053.1$ & $1036.5$ \\
$\chi^2_{\rm tot}$ & $4353.8$ & $4380.2$ & $4345.0$ \\
%$\chi^2_{\rm tot}$ & $4366.7$ & $4398.8$ & 4345.9 \\
{$\Delta \chi^2$} & $0$ & $+26.4$ & $-8.8$ \\
\hline
\end{tabular}
\caption{Cosmological parameters for 
$\Lambda$CDM (second column), for the preferred $\Lambda$CDM$\phi\nu_s$ model solving both the $H_0$ and $S_8$ tensions (third column), and for the best-fit $\Lambda$CDM$\phi\nu_s$ model found in the full MCMC scan, Fig.~\ref{fig:cosmomc_m_red} (last column). The first six  rows are the $\Lambda$CDM parameters, and the following two are the new-physics parameters. The remaining ones are derived from the cosmological model. 
$(v)$, $(s)$ indicate vector and scalar mediator
models, respectively. Bottom rows give the
 $\chi^2$ contribution from each data set, the total $\chi^2_{\rm tot}$, and $\Delta \chi^2 \equiv \chi^2_{\rm tot} - \chi^2_{\Lambda\text{CDM}}$.
}
\label{tab:best_params_m}
\end{table}

However, this does not constitute a successful resolution of the $H_0+S_8$ tensions, because it comes at the expense of creating new ones.  
In Table \ref{tab:best_params_m} we compare the
goodness of the fit for these models (third column) to $\Lambda$CDM (second column).  Although the fit to Dark Energy Survey (DES) data improves,
that to Planck, BAO and Pantheon degrades, resulting in a significantly worse global fit, with  $\Delta\chi^2 \simeq 26$.

Nevertheless, we find that the $\Lambda$CDM$\phi\nu_s$ model can ameliorate the $H_0$ tension to the $\sim 3\sigma$ level, with $H_0 \simeq 69.8\,\,\text{km/s/Mpc}$, while simultaneously addressing the $S_8$ tension. 
This is shown in the third column of Table~\ref{tab:best_params_m}, for the best-fit $\Lambda$CDM$\phi\nu_s$ model.
% found in the full MCMC scan is displayed. 
Coincidentally, this model predicts $\Omega_g \sim 0.15$, which is the same value that was
preferred by the Pantheon type Ia supernovae
in Fig.~\ref{fig:SN_likelihood}.
%%There is also a mild preference for this model over the vanilla $\Lambda$CDM one, which we think should deserve further study in the future.
The net improvement of $\Delta\chi^2 = -8.8$ relative to $\Lambda$CDM provides a mild preference of the ghost model, based on the Akaike and Bayesian Information Criteria~\cite{Cavanaugh:2019olr,CavanaughJosephE2012TBic}.
%The net improvement of $\Delta\chi^2 = -8.8$ relative to $\Lambda$CDM is statistically significant, given that $\Lambda$CDM$\phi\nu_s$ has just two additional parameters, and one of them $(w_{\nu_s})$ is essentially a nuisance parameter since it has little impact on the fits.
%
Similar conclusions were drawn in Ref.~\cite{Gangopadhyay:2023nli} applied to generic phantom dark energy models.

\begin{comment}
\mpuel{Perhaps discuss a bit how we can evade the existing bounds on $\sigma_{e, \rm{DM}}$. They all assume that the totality of nonrelativistic DM is interacting with electrons (not our case because we do have a CDM component, plus an additional one, which can be relativistic). 
The strongest bounds comes from CMB and Lyman-$\alpha$ data (see our paper on neutrino-DM constraints from blazars for literature), which however does not apply to our scenario because such bound is derived by modifications of the dynamics of the photon-baryon fluid during photon decoupling and the densities of $\nu_s$ in our model is negligible at that time.
Similarly, our model does not suffer from constraints on dark acoustic oscillations (Francis Cyr Racine and Kris Sigurdson had a paper on it a long time ago), which would constrain the abundance of interacting DM to be less than 5\%.
}

\end{comment}

\section{Conclusions}
\label{sec:conc}

In this work, we have reconsidered the cosmological consequences of a phantom field,
a scalar with the wrong-sign kinetic term.  Such a theory must be a low-momentum effective description, valid
below some cutoff $\Lambda$.  We refined
a previous estimated upper limit to show that
$\Lambda\lesssim 19\,$MeV, from observational constraints on diffuse
gamma rays.  At higher scales, phantom models must revert to a theory with normal-sign kinetic terms.

Unlike the great majority of phantom dark energy studies, we assumed the vacuum expectation value of the field vanishes, so that its potential makes no contribution to the Hubble expansion.  Instead, the spontaneous breakdown of the vacuum into phantom particles plus normal ones produces a fluid  that behaves as dark energy, as long as there is a mismatch between the masses of the two kinds of particles.  We showed that the equation of state of this exotic ``ghost fluid''
always violates the null energy condition (NEC), since it lies in the interval $-1.3 \lesssim w_{\rm eff} < -1.5$ at all times (see Fig.\ \ref{fig:SN_bounds}, right).  Therefore, even if at the present time the net contribution of the ghost fluid to the energy budget of the Universe is small, $\Omega_g \ll 1$, eventually it will come to dominate over the cosmological constant.  However since the phantom theory is only an 
effective description valid below the scale $\Lambda$, the usual ``big rip'' singularity that would occur in the future \cite{Caldwell:2003vq} is avoided.  Nevertheless, the accelerated expansion that occurs before the cutoff is reached will be sufficient to disrupt any object whose density falls below $\Lambda^4$.  For example,
the sun would be ripped apart if $\Lambda \gtrsim 10\,$keV.

We found that the NEC-violating
equation of state is at odds with combined cosmological data if one tries to make the ghost fluid account for all of the dark energy.  Instead, there is an upper bound $\Omega_g \lesssim 0.2$,
requiring it to be subdominant to the contribution from cosmological constant,
$0.5 \lesssim \Omega_\Lambda \le 0.7$.

Such large values of $\Omega_g$ cannot be attained if the phantoms couple with only gravitational strength to other particles.  Instead, we allowed for stronger interactions, mediating the production of phantoms plus light dark matter (with mass $< \Lambda$).  It leads to a larger Hubble rate at late times, in the case of massless phantoms, since the negative ghost energy redshifts faster than the positive dark matter density.  This constant rate of creation of energy density is reminiscent of the old steady-state 
cosmology \cite{1948MNRAS.108..252B,1948MNRAS.108..372H}, with the difference that the present mechanism is rooted in quantum field theory, rather than an {\it ad hoc} assumption.  Naively, one might 
think it could resolve the Hubble tension, but we find that this exacerbates the $S_8$ tension, as expected on general grounds for late-time mechanisms \cite{Cai:2021weh,Cai:2022dkh} (see however
Refs. \cite{Gariazzo:2021qtg,Krishnan:2020vaf,Krishnan:2021dyb}).  The best-fit
phantom model does soften the $H_0$ tension,
reducing it to $\sim 3\sigma$, and it provides a mild improvement relative to $\Lambda$CDM in fitting the data sets.

A special case is where massless phantoms couple to massless dark radiation with strength exceeding that of gravity.  Then their energy densities cancel exactly, and the magnitude of each can exceed usual bounds on a new light species by many orders of magnitude, since there is no impact on the Hubble expansion.  Nevertheless, linear-order cosmological perturbations grow in this scenario, and provide constraints via the ISW effect.  In the extreme case where the cutoff $\Lambda$ saturates the gamma-ray bound, and the new-physics scale coupling ghosts to dark radiation is $M\sim 10^{10}$\,GeV, the dark radiation could contribute as much as
$\Omega_{\nu_s}\sim 10^{26}$ (see Fig.\ \ref{fig:bounds_m0_gen}), being canceled by the negative contribution of the ghosts.

In a companion paper, we considered a possible observable effect of such a large density of 
dark radiation, whose energy is of order the cutoff $\Lambda$: in the presence of an additional interaction coupling it to electrons, it could be seen in direct detection experiments \cite{Cline:2023hfw}.  
It would be interesting to identify other potentially observable consequences of such a large population of energetic particles in the late Universe.  For example, there could be a mismatch between the preferred reference frame of the ghost fluid and that of the CMB, which for simplicity we neglected in the present study.  

\bigskip
{\bf Acknowledgments.}  We thank G. Alonso-Alvarez, J.\ Ambj\"orn, H.\ Bernardo, R.\ Brandenberger, S.\ Caron-Huot, B.\ Gavela, K. Rogers, K.\ Schutz, and M.\ Toomey for discussions, and T.\ Brinckmann,   A.\ Lewis, L.\ Lopez-Honorez, A.\ Singal, and S. Vagnozzi for very helpful correspondence.
We acknowledge Calcul Québec (\url{https://www.calculquebec.ca}) and Digital Research Alliance of Canada (\url{https://alliancecan.ca}) for supercomputing resources.
JC and MP are supported by NSERC (Natural Sciences and Engineering Research Council, Canada). MP was also partially supported by the Arthur B. McDonald Institute for Canadian astroparticle physics during the initial stages of the project.
TT is supported by JSPS Grant-in-Aid for Scientific Research KAKENHI Grant No. JP23H04004.

\begin{appendix}
\renewcommand{\setthesubsection}{\Alph{section}.\arabic{subsection}}

\begin{widetext}

\section{Computation of vacuum decay rate per unit volume}
\label{appA}
In this appendix the integral in Eq.~\eqref{eq:decay_rate} is evaluated, which generalizes the computation of Section~\ref{LambdaSect} when the produced particles (apart from the ghosts) are massive.
Eliminating the integral over $\vec{p}_4$ through the three-dimensional delta function, the vacuum decay rate into massless ghosts and massive sterile neutrinos is
\bea
\Gamma &=& \frac{1}{16\,(2\pi)^8} \int \frac{d^3 p_1}{p_1} \int \frac{d^3 p_2}{p_2} \int \frac{d^3 p_3}{E_3} \bigg(\frac{|\mathcal{M}|^2}{E_4}\bigg)_{\vec{p}_4 \to \vec{p}_1 + \vec{p}_2 - \vec{p}_3}\, \delta \Big(p_1 + p_2 - E_3 - \sqrt{(\vec{p}_1 + \vec{p}_2 - \vec{p}_3)^2 + m_{\nu_s}^2}\Big) \nn\\
& \times & \Theta (\Lambda - p_1)\, \Theta (\Lambda - p_2)\,,
\eea
where $p_i = |\vec{p}_i|$.
The two Heaviside functions restrict the ghost momenta to lie below the cutoff. The sterile neutrinos (with momenta $p_3$, $p_4$) are assumed to be Dirac fermions and the ghosts (momenta $p_1$, $p_2$) are complex scalars. For real scalar ghosts, an extra factor of $1/2$ should be included.

Using rotational invariance to make $\vec{p}_1$ parallel to $\hat{z}$ and $\vec{p}_2$ to lie in the $\hat{x}$-$\hat{z}$ plane,  the measure becomes
\be
\frac{d^3 p_1}{p_1} \frac{d^3 p_2}{p_2} \frac{d^3 p_3}{E_3} = 4 \pi\, p_1\, dp_1\, 2 \pi\, p_2\, dp_2\, d c_{1 2}\, p_3\, d E_3\, d c_{1 3}\, d\varphi_3\,,
\ee
where $c_{j k} \equiv \cos{\theta_{jk}}$ for $j, k = 1, 2, 3$, and we used $p_i dp_i = E_i dE_i$. Hence
\bea
\Gamma &=& \frac{1}{8 (2\pi)^6} \int_0^{\Lambda} dp_1 p_1 \int_0^{\Lambda} dp_2 \int_{-1}^1 d c_{1 2} \int_{-1}^1 d c_{1 3} \int d E_3 \int d\varphi_3\, \frac{p_2 p_3}{E_4}\, |\mathcal{M}|^2_{\vec{p}_4 \to \vec{p}_1 + \vec{p}_2 - \vec{p}_3} \nn\\ 
&\times & \delta \Big(p_1 + p_2 - E_3 - \sqrt{(\vec{p}_1 + \vec{p}_2 - \vec{p}_3)^2 + m_{\nu_s}^2}\Big)\,.
\eea
Using the remaining delta function to solve for $\varphi_3$ gives
\be
\delta (\dots) = \delta \Big(p_1 + p_2 - E_3 - \sqrt{p_1^2 + p_2^2 + E_3^2 + 2 p_1 p_2 c_{1 2} - 2 p_1 p_3 c_{1 3} - 2 p_2 p_3 c_{2 3} } \Big) \equiv \delta (g (\cos{\varphi_3}))\,,
\ee
where $c_{2 3} = c_{1 2} c_{1 3} + s_{1 2} s_{1 3} \cos{\varphi_3}$,
with $s_{j k} \equiv \sin{\theta_{jk}}$.  Here
$\delta (g (x)) = \sum_{x_s} \delta (x - x_s)/|g' (x_s)|$,
where $x_s$ are the roots of $g(x)$. In this case the single root is
\be
\label{eq:cos_phi3_0}
(\cos{\varphi_3})_{0} = - \frac{p_1 p_2 (1 - c_{1 2}) - p_1 (E_3 - p_3 c_{1 3}) - p_2 (E_3 - p_3 c_{1 2} c_{1 3})}{p_2 p_3 s_{1 2} s_{1 3}}\,,
\ee
with $g' (\cos{\varphi_3}) = ({p_2 p_3}/{E_4}) s_{1 2} s_{1 3}$.
Using $d c_{j k} = - s_{j k}\,d\theta_{j k}$, the decay rate integrals reduce to
\bea
\label{eq:Gamma_rate_int}
\Gamma &=& \frac{1}{8 (2\pi)^6} \int_0^{\Lambda} dp_1\, p_1 \int_0^{\Lambda} dp_2 \int_{0}^{\pi} d \theta_{1 2} \int_{0}^{\pi} d \theta_{1 3} \int_{m_{\nu_s}}^{2 \Lambda - m_{\nu_s}} d E_3 \int d\varphi_3\,\, |\mathcal{M}|^2 \,\delta (\cos{\varphi_3} - (\cos{\varphi_3})_0) \nn\\
& = & \frac{\Lambda^4}{4 (2\pi)^6} \int_0^{1} d x_1\, x_1 \int_0^{1} d x_2 \int_{0}^{\pi} d \theta_{1 2} \int_{0}^{\pi} d \theta_{1 3} \int_{\frac{m_{\nu_s}}{\Lambda}}^{2 - \frac{m_{\nu_s}}{\Lambda}} d y_3\,\, \frac{\Theta(1 - |(\cos{\varphi_3})_0|)}{|(\sin{\varphi_3})_0|}\,\, |\mathcal{M}|^2\,,
\eea
where  $x_j \equiv p_i / \Lambda$, $y_j \equiv E_j / \Lambda$ and there is a factor of $2$ for the two values of $\varphi_3$ satisfying Eq.~\eqref{eq:cos_phi3_0}. The upper limit on $E_3$ comes from energy conservation with $p_4 = 0$, the maximum energy allowed by kinematics.

If vacuum decay is mediated by gravitons, the matrix element \eqref{eq:M2_gravity} is
\be
|\mathcal{M}_g|^2 = -\frac{\Lambda^4}{32\,m_p^4} \bigg\{\frac{[\hat{t} - (m_{\nu_s} / \Lambda)^2]\, [\hat{s} + \hat{t} - (m_{\nu_s} / \Lambda)^2]\, [\hat{s} + 2\, \hat{t} - (m_{\nu_s} / \Lambda)^2]^2}{\hat{s}^2} \bigg\}\,,
\ee
with $\hat{s} = 2 x_1 x_2 (1 - c_{12})$, $\hat{t} = (m_{\nu_s} / \Lambda)^2 - 2 x_1 (y_3 - x_3 c_{13})$ are the usual Mandelstam variables divided by $\Lambda^2$ and $x_3 \equiv \sqrt{y_3^2 - (m_{\nu_s} / \Lambda)^2}$.
Numerically solving the integrals in Eq.~\eqref{eq:Gamma_rate_int} with $|\mathcal{M}_g|^2$ using \texttt{Vegas} \citep{Lepage:1977sw,Lepage:2020tgj}, we find 
the analytic fit
\be
\Gamma_g \approx 8.9 \times 10^{-9} \,\frac{\Lambda^8}{m_P^4}\,\bigg[1 - \bigg( \frac{m_{\nu_s}}{\Lambda}\bigg)^2\bigg]\,\exp{\bigg[-5.3\, \bigg(\frac{m_{\nu_s}}{\Lambda}\bigg)^{4.2} \bigg]}\,.
\ee
for $m_{\nu_s}\le\Lambda$.

For the scalar mediator, $|\mathcal{M}_n|^2$ in Eq.~\eqref{eq:M2} simplifies to
\be
|\mathcal{M}_s|^2 = 4\, {\Lambda^4 \over M_s^4}\,
    \Big[x_1 x_2\, (1 - c_{12}) - 2\, (m_{\nu_s} / \Lambda)^2 \Big]\,,
\ee
and the result for the corresponding decay rate is
\be
\Gamma_s \approx 1.1 \times 10^{-5}\,\frac{\Lambda^8}{M_s^4}\,\exp{\bigg[- 6.7\, \bigg(\frac{m_{\nu_s}}{\Lambda}\bigg)^{2.1} \bigg]}\,.
\ee
Similarly, for the vector mediator, the matrix element \eqref{eq:Gamma_rate_int} is
\be
\label{eq:M2_vector}
|\mathcal{M}_v|^2 = 32\, {\Lambda^4 \over M_v^4}\,x_1^2\,
    \Big[y_3\, (x_2 - y_3 ) + x_3\, (2 y_3 - x_2) c_{13} - x_3^2 \,c_{13}^2 - x_2\, c_{12}\, (y_3 - x_3 c_{13}) \Big]\,,
\ee
leading to a vacuum decay rate
\be
\Gamma_v \approx 1.1 \times 10^{-5}\,\frac{\Lambda^8}{M_v^4}\,\exp{\bigg[- 5.7\, \bigg(\frac{m_{\nu_s}}{\Lambda}\bigg)^{4.2} \bigg]}\,.
\ee

%\tt{In fact, I'm not sure whether one can take the integration intervals of $|p_1|$ and $|p_2|$ independently as $0\leq|p_1|\leq\Lambda$ and $0\leq|p_2|\leq\Lambda$. 
%At least for normal 4-body decays, the interval of $|p_2|$ seems to depend on $|p_1|$~\cite{Kumar:1969jjy}. But the situation may be different for vacuum decay as in our case.}
%\jc{Yes, I think nothing limits us from putting the upper limits independently, except that $p_3$ and $p_4$ are also supposed to be below $\Lambda$, which may have been neglected here.}
%\tt{OK, I see.}

\section{Derivation of Boltzmann equations and hierarchies}
\label{appB}
In this appendix, we derive the Boltzmann equations describing the evolution of the background densities for ghosts and sterile neutrinos and their perturbations at linear order. 
Following Refs.~\citep{Ma:1995ey,Piattella:2018hvi},  the phase-space distribution for the particle species $i$ is a background homogeneous contribution $f_{i}^0$ plus a perturbation, 
\be
\label{eq:fi_expansion}
f_i (\vec{x}, \vec{p}_i, \tau) =
    f_{i}^0 (q_i, \tau)\, [1 + \Psi_i (\vec{x}, \vec{q}_i, \tau)]\,,
\ee
where $\tau$ is the conformal time, related to the physical time $t$ by $d\tau = dt /a(t)$, with $a$ the scale factor, $\vec{q}_i \equiv a \vec{p}_i$ is the comoving three-momentum with magnitude $|\vec{q}_i| = q_i$ and unit-vector direction $\hat{q}_i$, and $\Psi_i (\vec{x}, \vec{q}_i, \tau)$ is the phase-space density contrast. 
The phase-space distribution evolves according to the Boltzmann equation, which is generically given in terms of the comoving momentum by
\be
\label{eq:Boltzmann_general}
\frac{d f_i}{d\tau} = \frac{\partial f_i}{\partial \tau} + \frac{q_i}{\varepsilon_i}\, \hat{q}_i \cdot \nabla_{\vec{x}} f_i + \frac{d q_i}{\partial \tau} \frac{\partial f_i}{\partial q_i} = \bigg(\frac{d f_i}{d\tau}\bigg)_{\mathcal{C}} \,,
\ee
where $\varepsilon_i = \sqrt{q_i^{\,2} + a^2 m_i^2} = a E_i$ is the comoving energy and we neglected the term proportional to the derivative of $f_i$ with respect to the direction $\hat{q}_i$, it being a second-order quantity.
Putting Eq.~\eqref{eq:fi_expansion} into Eq.~\eqref{eq:Boltzmann_general} gives at the background level
\begin{equation}
\label{eq:Boltzmann_bkg_q}
    \frac{\partial f_{i}^0}{\partial \tau} (q_i, \tau) = \bigg(\frac{d f_i}{d\tau}\bigg)_{\mathcal{C}}^{(0)} (q_i, \tau)\,,
\end{equation}
and at the linear-perturbation level
\begin{equation}
\label{eq:time_dep_Boltz_fi}
    \frac{\partial \Psi_i}{\partial \tau} (\vec{x}, \vec{q}_i, \tau) + \frac{q_i}{\varepsilon_i}\, \hat{q}_i \cdot \nabla_{\vec{x}} \Psi_i (\vec{x}, \vec{q}, \tau) + \frac{d q_i}{d\tau} \frac{\partial \ln{f_{i}^0}}{\partial q_i} =
    \frac{1}{f_{i}^0}\bigg(\frac{d f_i}{d\tau}\bigg)_{\mathcal{C}}^{(1)} -  \frac{1}{f_{i}^0}\bigg(\frac{d f_i}{d\tau}\bigg)_{\mathcal{C}}^{(0)}\, \Psi_i (\vec{x}, \vec{q}_i, \tau)\,,
\end{equation}
where $\big(\frac{d f_i}{d\tau}\big)_{\mathcal{C}}^{(1)}$ is the perturbed collision term. On the right-hand side of Eq.~\eqref{eq:time_dep_Boltz_fi} we used Eq.~\eqref{eq:Boltzmann_bkg_q} to replace the time-derivative of $f_{i}^0$ with the zeroth-order collision term. 
Equation~\eqref{eq:Boltzmann_bkg_q} reduces to Eq.~\eqref{eq:Botlzmann_eq_f} in terms of the physical time $t$ and momentum $p_i$, and identifying $S_i \equiv \big(\frac{d f_i}{d\tau}\big)_{\mathcal{C}}^{(0)}$.

It is common practice to decompose $\Psi_i$ into Legendre polynomials $P_{\ell} (\hat{k} \cdot \hat{q}_i)$, motivated by left-hand side of Eq.~\eqref{eq:time_dep_Boltz_fi} depending only on $q_i$, $k = |\vec{k}|$ and $\hat{k} \cdot \hat{q}_i$ in Fourier space. Hence \cite{Ma:1995ey}
\begin{equation}
\label{eq:Legendre-polynomials-decomposition}
\begin{split}
    \Psi_i (\vec{k}, \vec{q}_i, \tau) &= \sum_{\ell = 0}^{\infty} (-i)^{\ell} (2 \ell + 1) \Psi_{i, \ell} (k, q_i, \tau) P_{\ell} (\hat{k} \cdot \hat{q}_i)\,, \\
    \Psi_{i, \ell} (k, q_i, \tau) &= \frac{i^{\ell}}{2} \int_{-1}^{1} d (\hat{k} \cdot \hat{q}_i)\, \Psi_i (\vec{k}, \vec{q}_i, \tau) P_{\ell} (\hat{k} \cdot \hat{q}_i)\,.
\end{split}
\end{equation}
This allows the Fourier version of Eq.~\eqref{eq:time_dep_Boltz_fi} to be expressed as a hierarchy of equations for the multipole moments $\Psi_{i, \ell} (k, q_i, \tau)$~\cite{Ma:1995ey,Barenboim:2020vrr}
\begin{equation}
\label{eq:pert_eq_Psi_gen}
\begin{split}
    \Psi_{i, 0}' (k, q_i, \tau) &= - \frac{q_i k}{\varepsilon_i} \Psi_{i, 1} (k, q_i, \tau) + \frac{1}{6} \frac{\partial \ln{f_{i}^0}}{\partial \ln{q_i}}\, h' + \mathcal{C}_0^{(1)} [\Psi_i (k, q_i, \tau)] \,, \\
    \Psi_{i, 1}' (k, q_i, \tau) &= \frac{q_i k}{3 \varepsilon_i} \Big(\Psi_{i, 0} (k, q_i, \tau) - 2 \Psi_{i, 2} (k, q_i, \tau)\Big) + \mathcal{C}_1^{(1)} [\Psi_i (k, q_i, \tau)] \,, \\
    \Psi_{i, 2}' (k, q_i, \tau) &= \frac{q_i k}{5 \varepsilon_i} \Big(2 \Psi_{i, 1} (k, q_i, \tau) - 3 \Psi_{i, 2} (k, q_i, \tau) \Big) - \frac{\partial \ln{f_{i}^0}}{\partial \ln{q_i}} \frac{h' + 6 \eta'}{15} + \mathcal{C}_2^{(1)} [\Psi_i (k, q_i, \tau)] \,, \\
    \Psi_{i, \ell > 2}' (k, q_i, \tau) &= \frac{k}{2\ell +1} \frac{q_i}{\varepsilon_i} [\ell \Psi_{i, \ell -1} (k, q_i, \tau) - (\ell + 1) \Psi_{i, \ell+1} (k, q_i, \tau)] + \mathcal{C}_{\ell}^{(1)} [\Psi_i (k, q_i, \tau)] \,,
\end{split}
\end{equation}
where the collision-term multipoles are~\cite{Barenboim:2020vrr}
\begin{equation}
\label{eq:C_term_nus_ell}
    \mathcal{C}_{\ell}^{(1)} [\Psi_i (k, q_i, \tau)] \equiv \frac{1}{f_{i}^0} \bigg(\frac{d f_i}{d\tau} \bigg)^{(1)}_{\mathcal{C}, \ell} (k, q_i, \tau) - \frac{1}{f_{i}^0} \bigg(\frac{d f_i}{d\tau} \bigg)^{(0)}_{\mathcal{C}} \Psi_{i, \ell} (k, q_i, \tau)\,, 
\end{equation}
and 
\begin{equation}
    \bigg(\frac{d f_i}{d\tau} \bigg)^{(1)}_{\mathcal{C}, \ell} (k, q_i, \tau) \equiv \frac{i^{\ell}}{4 \pi} \int d\Omega_{\vec{k}}\,P_{\ell} (\hat{k} \cdot \hat{q}_i) \bigg(\frac{d f_i}{d\tau} \bigg)^{(1)}_{\mathcal{C}} (\vec{k}, \vec{q}_i, \tau)\,,
\end{equation}
with $\Omega_{\vec{k}}$ the solid angle subtended by the wave vector $\vec{k}$.

The collision term for sterile neutrinos $\nu_s$ on the right-hand side of Eq.~\eqref{eq:Boltzmann_general} comes from the vacuum decay into $\phi\phi\nu_s\bar\nu_s$, which has the same matrix element as the scattering process
 $\phi (q_1, \vec{q}_1) \phi (q_2, \vec{q}_2) \leftrightarrow \nu_s (\varepsilon_3, \vec{q}_3) \nu_s (\varepsilon_4, \vec{q}_4)$  when the
 ghost energies are treated as being positive
\begin{equation}
\begin{split}
    \bigg(\frac{d f_{\nu_s}}{d\tau}\bigg)_{\mathcal{C}} (\vec{q}_3) =& \frac{a}{2 \varepsilon_3} \int \frac{d^3 \vec{q}_1}{(2 \pi)^3 \,2 q_1} \int \frac{d^3 \vec{q}_2}{(2 \pi)^3 \,2 q_2} \int \frac{d^3 \vec{q}_4}{(2 \pi)^3 \,2 \varepsilon_4}\, |\mathcal{M}|^2 \,\,\times \\
    &\times \,\,(2 \pi)^4 \delta (q_1 + q_2 - \varepsilon_3 - \varepsilon_4) \delta^{(3)} (\vec{q}_1 + \vec{q}_2 - \vec{q}_3 - \vec{q}_4) \,\,\times \\ 
    &\times\,\, \Big[(1 + f_{\phi} (\vec{q_1})) (1 + f_{\phi} (\vec{q_2})) (1 - f_{\nu_s} (\vec{q_3})) (1 - f_{\nu_s} (\vec{q_4})) - f_{\phi} (\vec{q}_1) f_{\phi} (\vec{q}_2) f_{\nu_s} (\vec{q}_3) f_{\nu_s} (\vec{q}_4)\Big] \,.
\end{split}
\end{equation}
%where the ghosts $\phi$ here are considered as positive-energy particles, which is allowed by the energy-conserving delta function.
Expanding $f_i$ as in Eq.~\eqref{eq:fi_expansion} and neglecting the Pauli-blocking and Bose-Einstein stimulated-emission factors,  the background contribution and the linear-order perturbation of the collision term are
respectively
\begin{equation}
\label{eq:dfiodtau-0}
\begin{split}
    \bigg(\frac{d f_{\nu_s}}{d\tau}\bigg)_{\mathcal{C}}^{(0)} (q_3) =&\,\, \frac{a}{2 \varepsilon_3} \int \frac{d^3 \vec{q}_1}{(2 \pi)^3 \,2 q_1} \int \frac{d^3 \vec{q}_2}{(2 \pi)^3 \,2 q_2} \int \frac{d^3 \vec{q}_4}{(2 \pi)^3 \,2 \varepsilon_4}\, |\mathcal{M}|^2 \,\,\times \\
    &\times \,\,(2 \pi)^4 \delta (q_1 + q_2 - \varepsilon_3 - \varepsilon_4) \delta^{(3)} (\vec{q}_1 + \vec{q}_2 - \vec{q}_3 - \vec{q}_4) \,\,\times \\
    &\times \,\,\Big[1 + f_{\phi}^0 (q_1) +  f_{\phi}^0 (q_2) - f_{\nu_s}^0 (q_3) - f_{\nu_s}^0 (q_4)\Big]\,,
\end{split}
\end{equation}
and
\begin{equation}
\label{eq:dfiodtau-1-ell}
\begin{split}
    \bigg(\frac{d f_{\nu_s}}{d\tau}\bigg)_{\mathcal{C, \ell}}^{(1)} (k, q_3) \approx&\,\, a \frac{i^{\ell}}{4\pi} \int d\Omega_{\vec{k}}\, P_{\ell} (\hat{k} \cdot \hat{q}_3)\,\frac{1}{2 \varepsilon_3} \int \frac{d^3 \vec{q}_1}{(2 \pi)^3 \,2 q_1} \int \frac{d^3 \vec{q}_2}{(2 \pi)^3 \,2 q_2} \int \frac{d^3 \vec{q}_4}{(2 \pi)^3 \,2 \varepsilon_4}\, |\mathcal{M}|^2 \,\,\times \\
    &\times \,\,(2 \pi)^4 \delta (q_1 + q_2 - \varepsilon_3 - \varepsilon_4) \delta^{(3)} (\vec{q}_1 + \vec{q}_2 - \vec{q}_3 - \vec{q}_4) \,\,\times \\ 
    &\times\,\, \Big[2 f_{\phi}^0 (q_1) \Psi_{\phi} (\vec{q_1}) - f_{\nu_s}^0 (q_3) \Psi_{\nu_s} (\vec{q_3}) - f_{\nu_s}^0 (q_4) \Psi_{\nu_s} (\vec{q_4})\Big]\,,
\end{split}
\end{equation}
where the dependence on $(\vec{k}, \tau)$ is suppressed and  the symmetry $q_1 \leftrightarrow q_2$ was used to combine two terms in the last line. 
Similar expressions can be derived for the ghosts $\phi$.
Subleading contributions were kept in Eq.~\eqref{eq:dfiodtau-0}, arising from the terms $f_{i}^0 \ll 1$, that will be relevant in the collision multipoles \eqref{eq:C_term_nus_ell}.

Integrating Eq.~\eqref{eq:dfiodtau-0}, weighted by the sterile neutrino energy $\varepsilon_3$, and neglecting subleading contributions gives the numerical fit~\footnote{See footnote \ref{fnl}}
\be
\label{eq:Gamma_rho}
\Gamma_{\rho} \equiv \frac{1}{a} \int \frac{d^3 q_3}{(2\pi)^3}\,\varepsilon_3\, \bigg(\frac{d f_{\nu_s}}{d\tau}\bigg)_{\mathcal{C}}^{(0)} \approx
\left\{\arraycolsep=1.4pt\def\arraystretch{2.2}
    \begin{array}{ll}
   8.4 \times 10^{-6}\,\frac{\Lambda^9}{M_s^4}\,\exp{\big[- 6.6\, \big(\frac{m_{\nu_s}}{\Lambda}\big)^{2.1} \big]}, & \hbox{\ scalar}\\
   9.0 \times 10^{-6}\,\frac{\Lambda^9}{M_v^4}\,\exp{\big[- 5.5\, \big(\frac{m_{\nu_s}}{\Lambda}\big)^{4.3} \big]},& \hbox{\ vector}
    \end{array}\right. \,,
\ee
which appears in Eq.~\eqref{eq:Boltzmann_eq_t}. 
The same result arises from the integral of $\big(\frac{d f_{\phi}}{d\tau}\big)_{\mathcal{C}}^{(0)}$ weighted by the ghost energy $q_1$.

The computation of the perturbed collision term in Eq.~\eqref{eq:dfiodtau-1-ell} in the case of massive sterile neutrinos is nontrivial, requiring not only the integral of  $\Psi_{i, \ell}$ (defined in Eq.~\eqref{eq:Legendre-polynomials-decomposition}) and solution of the multipole hierarchy~\eqref{eq:pert_eq_Psi_gen}, but also depending on two free parameters: the  $\Lambda$  and $a m_{\nu_s}$. 
It simplifies in the relativistic limit, $q \gg a m_{\nu_s}$, approximated by $m_{\nu_s} \sim 0$ in the preceding expressions. 
This is the most interesting case for the perturbation analysis, since then the background contributions to the energy density from ghosts and sterile neutrinos cancel and modifications of the CMB and matter power spectrum can arise only through the perturbations.
We consider this scenario in detail in the following subsection, \ref{appB}.\ref{subsec:perts_m0}, and derive an adequate approximate description for the more general case in subsection~\ref{appB}.\ref{subsec:perts_m}.
%leaving a better treatment to a future work. 

\subsection{Relativistic sterile neutrinos}~\label{subsec:perts_m0}
In the relativistic limit, it is a common simplification to integrate over $\vec{q}_i$
in~\eqref{eq:pert_eq_Psi_gen}, weighted by the energy and the background phase-space distribution~\cite{Ma:1995ey}.  The equations can
then be expressed in terms of the relative perturbation variables: the density contrast $\delta_{i}$, the divergence of the velocity field $\theta_{i}$ and the anisotropic stress $\sigma_{i}$, where $i=\phi,\,\nu_s$. They are defined in terms of
\begin{equation}
\label{eq:Fil_def}
    F_{i, \ell} \equiv \frac{\int dq\, q^3 f_{i}^0 (q) \Psi_{i, \ell} (q)}{\int dq\, q^3 f_{i}^0 (q)}\,,
\end{equation}
by $\delta_i = F_{i,0}$, $\theta_i = 3 k F_{i, 1} / 4$ and $\sigma_i = F_{i, 2} / 2$.
Then the perturbation equations~\eqref{eq:pert_eq_Psi_gen} reduce to~\cite{Ma:1995ey,Barenboim:2020vrr}
\begin{equation}
\label{eq:pert_eq_F_gen}
\begin{split}
    \delta_i' &= -\frac{4}{3} \theta_i + \frac{2}{3} h' + C_0^{(1)} [F_i]\,, \\
    \theta_i' &= k^2 \bigg(\frac{1}{4} \delta_i - \sigma_i \bigg) + C_1^{(1)} [F_i]\,, \\
    F_{i, 2}' &= 2 \sigma_i' = \frac{8}{15} \theta_i - \frac{3}{5} k F_{i, 3} + \frac{4}{15} h' + \frac{8}{5} \eta' + C_2^{(1)} [F_i]\,, \\
    F_{i, \ell}' &= \frac{k}{2\ell + 1} [\ell F_{i, \ell -1} - (\ell + 1) F_{i, \ell + 1}] + C_{\ell}^{(1)} [F_i]\,,
\end{split}
\end{equation}
with collision multipoles
\begin{equation}
\label{eq:C_term_nus_ell_m0}
    C_{\ell}^{(1)} [F_i] \equiv \frac{\int dq\,q^3 \Big(\frac{df_i}{d\tau}\Big)^{(1)}_{\mathcal{C}, \ell}}{\int dq\,q^3 f_{i}^0 (q)} - F_{i, \ell} \frac{\int dq\,q^3 \Big(\frac{df_i}{d\tau}\Big)^{(0)}_{\mathcal{C}}}{\int dq\,q^3 f_{i}^0 (q)}\,.
\end{equation}
The numerators can be evaluated by integrating Eqs.~\eqref{eq:dfiodtau-0} and~\eqref{eq:dfiodtau-1-ell}. In terms of $x_j \equiv q_j / \Lambda$, the collision terms for $\nu_s$ are
\begin{equation}
\label{eq:int_q3_C_bkg}
\begin{split}
    \int dq_3\, q_3^3\, \bigg(\frac{d f_{\nu_s}}{d\tau}\bigg)_{\mathcal{C}}^{(0)} =&\,\, \frac{a\,\Lambda^5}{8 (2\pi)^4}\,\bigg[\int_0^{\infty} dx_3\, \mathcal{K}_{3} (x_3) + 2 \int_0^1 dx_1\, f_{\phi}^0 (x_1)\, \mathcal{K}_{1, 0} (x_1) - \int_0^{\infty} dx\, f_{\nu_s}^0 (x)  \Big(\mathcal{K}_3 (x) + \mathcal{K}_{4, 0} (x) \Big) \bigg]\,,
\end{split}
\end{equation}
\begin{equation}
\label{eq:int_dfdtau_nus_1st}
    \int dq_3\, q_3^3\,\bigg(\frac{d f_{\nu_s}}{d\tau}\bigg)_{\mathcal{C, \ell}}^{(1)} =  \frac{a\,\Lambda^5}{8 (2\pi)^4}\,\bigg[2 \int_0^1 dx_1\,f_{\phi,0} (x_1)\, \Psi_{\phi, \ell} (x_1)\, \mathcal{K}_{1, \ell} (x_1) - \int_{0}^{\infty} dx\, f_{\nu_s}^0 (x)\, \Psi_{\nu_s, \ell} (x) \Big(\mathcal{K}_3 (x) + \mathcal{K}_{4, \ell} (x) \Big) \bigg]\,,
\end{equation}
depending on the dimensionless kernels
\begin{equation}
\label{eq:K_kernels_def}
\begin{split}
    \mathcal{K}_{1, \ell} (x_1) &= \int_0^{\infty} dx_3\, x_3^2\, \int_{0}^1 dx_2 \int_{0}^{\pi} d\theta_{13}\,P_{\ell} (\cos{\theta_{13}})\, \int_{0}^{\pi} d\theta_{23}\, |\mathcal{M}|^2\, \frac{\Theta{(1 - |\cos{\varphi_2}|)}}{|\sin{\varphi_2}|}\,, \\
    \mathcal{K}_3 (x_3) &= x_3^2 \int_0^1 dx_1 \int_0^1 dx_2 \int_0^{\pi} d\theta_{13} \int_0^{\pi} d\theta_{23}\,|\mathcal{M}|^2\, \frac{\Theta{(1 - |\cos{\varphi_2}|)}}{|\sin{\varphi_2}|}\,, \\
    \mathcal{K}_{4, \ell} (x_4) &= \int_0^{\infty} dx_3\, x_3^2 \int_0^{1} dx_1\, \int_{0}^{\pi} d\theta_{3 4}\, P_{\ell} (\cos{\theta_{3 4}}) \int_{0}^{\pi} d\theta_{1 3}\, |\mathcal{M}|^2\, \frac{\Theta (1 - |\cos{\varphi_1}|)}{|\sin{\varphi_1}|}\,\Theta(1 + x_1 - x_3 - x_4)\,,
\end{split}
\end{equation}
with $\cos{\varphi_2}$ and $\cos{\varphi_1}$  given by
\begin{equation}
\begin{split}
    \cos{\varphi_2} &= \frac{x_1 x_2 (1 - c_{13} c_{23}) - x_1 x_3 (1 - c_{13}) - x_2 x_3 (1 - c_{23})}{x_1 x_2 s_{13} s_{23}} \,, \\
    \cos{\varphi_1} &= \frac{x_1 x_3 (1 - c_{13}) + x_1 x_4 (1 - c_{13} c_{34}) - x_3 x_4 (1 - c_{34})}{x_1 x_4 s_{13} s_{34}} \,,
\end{split}
\end{equation}
and $c_{jk} = \cos{\theta_{jk}}$,  $s_{jk} = \sin{\theta_{jk}}$.
Analogously, the collision terms for ghosts (with overall $-$ signs from their negative energy) are
\begin{equation}
\label{eq:int_q1_C_bkg}
    \int dq_1\, q_1^3\, \bigg(\frac{d f_{\phi}}{d\tau}\bigg)_{\mathcal{C}}^{(0)} = - \frac{a\,\Lambda^5}{8 (2\pi)^4}\,\bigg[\int_0^1 dx_1\, \mathcal{W}_1 (x_1) + \int_0^1 dx\,f_{\phi,0} (x) \Big(\mathcal{W}_{1} (x) + \mathcal{W}_{2, 0} (x) \Big) - 2 \int_0^{\infty} dx_3\,f_{\nu_s}^0 (x_3)\,\mathcal{W}_{3, 0} (x_3) \bigg]\,,
\end{equation}
\begin{equation}
\label{eq:int_dfdtau_phi_1st}
    \int dq_1\, q_1^3\,\bigg(\frac{d f_{\phi}}{d\tau}\bigg)_{\mathcal{C, \ell}}^{(1)} = - \frac{a\,\Lambda^5}{8 (2\pi)^4}\,\bigg[\int_0^1 dx\,f_{\phi,0} (x)\,\Psi_{\phi, \ell} (x) \Big(\mathcal{W}_1 (x) + \mathcal{W}_{2,\ell} (x) \Big) - 2 \int_0^{\infty} dx_3\,f_{\nu_s}^0 (x_3)\,\Psi_{\nu_s, \ell} (x_3)\,\mathcal{W}_{3,\ell} (x_3) \bigg]\,,
\end{equation}
where
\begin{equation}
\label{eq:W_kernels_def}
\begin{split}
    \mathcal{W}_1 (x_1) &= x_1^2 \, \int_0^1 dx_2 \int_0^{\infty} dx_3 \int_0^{\pi} d\theta_{12} \int_0^{\pi} d\theta_{13}\,|\mathcal{M}|^2 \frac{\Theta{(1 - |\cos{\varphi_3}|)}}{|\sin{\varphi_3}|}\,, \\
    \mathcal{W}_{2, \ell} (x_2) &= \int_0^{1} dx_1\, x_1^2\, \int_{0}^{\infty} dx_3 \int_{0}^{\pi} d\theta_{12}\,P_{\ell} (\cos{\theta_{12}})\, \int_{0}^{\pi} d\theta_{13}\, |\mathcal{M}|^2 \frac{\Theta{(1 - |\cos{\varphi_3}|)}}{|\sin{\varphi_3}|}\,, \\
    \mathcal{W}_{3,\ell} (x_3) &= \int_0^{1} dx_1\, x_1^2\, \int_{0}^1 dx_2 \int_{0}^{\pi} d\theta_{13}\,P_{\ell} (\cos{\theta_{13}})\, \int_{0}^{\pi} d\theta_{12}\, |\mathcal{M}|^2 \frac{\Theta{(1 - |\cos{\varphi_3}|)}}{|\sin{\varphi_3}|}\,,
\end{split}
\end{equation}
and 
\begin{equation}
    \cos{\varphi_3} = -\frac{x_1 x_2 (1 - c_{12}) - x_1 x_3 (1 - c_{13}) - x_2 x_3 (1 - c_{12} c_{13})}{x_2 x_3 s_{12} s_{13}}\,.
\end{equation}
The kernels in Eqs.~\eqref{eq:K_kernels_def} and~\eqref{eq:W_kernels_def} are fitted by polynomials
\begin{equation}
\label{eq:Kernels_poly}
    (\mathcal{K}, \mathcal{W})\, (x) = x^3\,(a_0 + a_1\,x + a_2\,x^2 + a_3\,x^3 + a_4\,x^4 + a_5\,x^5 + a_6\,x^6)\,,
\end{equation}
with coefficients $a_j$ given in Table~\ref{tab:kernels_coeffiecients} for the $\mathcal{K}$ and $\mathcal{W}$ kernels.
%, where $\chi^2 / \rm{d.o.f.} < 1$ for all of them and we checked explicitly that the other terms with higher $\ell$ are completely negligible. 

\begin{table}[t!]
%\centering
    \begin{tabular}{| c | c c c c c c c | c c c c c c c |}
    \hline
    & \multicolumn{7}{c|}{Vector mediator} & \multicolumn{7}{c|}{Scalar mediator} \\
    \hline \hline
    $\mathcal{K}_{i, \ell}$ & $a_0$ & $a_1$ & $a_2$ & $a_3$ & $a_4$ & $a_5$ & $a_6$  & $a_0$ & $a_1$ & $a_2$ & $a_3$ & $a_4$ & $a_5$ & $a_6$ \\
    \hline \hline
    $\mathcal{K}_{1,0}$ & $2.2$ & $2.8$ & $0$ & $0$ & $0$ & $0$ & $0$  & $6.1$ & $-2.5$ & $0$ & $0$ & $0$ & $0$ & $0$  \\
    $\mathcal{K}_{1,1}$ & $-1.1$ & $2.8$ & $0$ & $0$ & $0$ & $0$ & $0$  & $0.8$ & $0.8$ & $0$ & $0$ & $0$ & $0$ & $0$ \\
    $\mathcal{K}_{1,2}$ & $-0.9$ & $0.8$ & $0$ & $0$ & $0$ & $0$ & $0$  & $-0.3$ & $0.9$ & $0$ & $0$ & $0$ & $0$ & $0$ \\
    $\mathcal{K}_{1,3}$ & $-0.1$ & $-0.4$ & $0$ & $0$ & $0$ & $0$ & $0$  & $-0.3$ & $0.5$ & $0$ & $0$ & $0$ & $0$ & $0$ \\
    $\mathcal{K}_{1,4}$ & $0.2$ & $-0.5$ & $0$ & $0$ & $0$ & $0$ & $0$  & $-0.2$ & $0.3$ & $0$ & $0$ & $0$ & $0$ & $0$ \\
    $\mathcal{K}_{1,5}$ & $0.2$ & $-0.3$ & $0$ & $0$ & $0$ & $0$ & $0$  & $-0.1$ & $0.1$ & $0$ & $0$ & $0$ & $0$ & $0$ \\
    $\mathcal{K}_{3}$ & $-6.1$ & $75.8$ & $-164.8$ & $156.3$ & $-75.2$ & $18.1$ & $-1.7$  & $3.7$ & $33.1$ & $-97.1$ & $100.9$ & $-50.2$ & $12.0$ & $-1.1$ \\
    $\mathcal{K}_{4,0}$ & $42.0$ & $-94.5$ & $62.7$ & $8.4$ & $-28.2$ & $12.2$ & $-1.7$  & $80.3$ & $-288.3$ & $447.1$ & $-381.1$ & $186.8$ & $-49.5$ & $5.5$ \\
    $\mathcal{K}_{4,1}$ & $-18.2$ & $22.1$ & $27.7$ & $-67.4$ & $49.2$ & $-16.0$ & $2.0$  & $-25.1$ & $78.0$ & $-104.8$ & $79.3$ & $-35.9$ & $9.2$ & $-1.0$ \\
    $\mathcal{K}_{4,2}$ & $-1.3$ & $30.9$ & $-83.0$ & $93.1$ & $-52.6$ & $14.9$ & $-1.7$  & $-3.1$ & $23.2$ & $-50.7$ & $51.6$ & $-27.3$ & $7.3$ & $-0.8$ \\
    $\mathcal{K}_{4,3}$ & $5.1$ & $-31.2$ & $62.9$ & $-60.5$ & $30.6$ & $-7.9$ & $0.8$  & $2.5$ & $-13.0$ & $23.8$ & $-20.7$ & $9.3$ & $-2.1$ & $0.2$ \\
    \hline
    \end{tabular}\\
    %\hspace{0.5cm}
    \vspace{0.5cm}
    \begin{tabular}{| c | c c c c c c c | c c c c c c c |}
    \hline
    & \multicolumn{7}{c|}{Vector mediator} & \multicolumn{7}{c|}{Scalar mediator} \\
    \hline \hline
    $\mathcal{W}_{i, \ell}$ & $a_0$ & $a_1$ & $a_2$ & $a_3$ & $a_4$ & $a_5$ & $a_6$  & $a_0$ & $a_1$ & $a_2$ & $a_3$ & $a_4$ & $a_5$ & $a_6$ \\
    \hline \hline
    $\mathcal{W}_{1}$ & $0.1$ & $5.5$ & $0$ & $0$ & $0$ & $0$ & $0$  & $4.2$ & $0$ & $0$ & $0$ & $0$ & $0$ & $0$ \\
    $\mathcal{W}_{2,0}$ & $4.4$ & $0.1$ & $0$ & $0$ & $0$ & $0$ & $0$  & $8.1$ & $-5.1$ & $0$ & $0$ & $0$ & $0$ & $0$\\
    $\mathcal{W}_{2,1}$ & $-2.2$ & $0$ & $0$ & $0$ & $0$ & $0$ & $0$  & $-2.7$ & $1.7$ & $0$ & $0$ & $0$ & $0$ & $0$ \\
    $\mathcal{W}_{2,2}$ & $0.4$ & $0$ & $0$ & $0$ & $0$ & $0$ & $0$  & $0$ & $0$ & $0$ & $0$ & $0$ & $0$ & $0$ \\
    $\mathcal{W}_{3,0}$ & $17.9$ & $-11.5$ & $-46.1$ & $76.8$ & $-48.5$ & $14.1$ & $-1.6$  & $41.9$ & $-127.2$ & $173.7$ & $-138.5$ & $67.3$ & $-18.4$ & $2.2$ \\
    $\mathcal{W}_{3,1}$ & $-12.0$ & $45.8$ & $-61.5$ & $36.6$ & $-8.4$ & $-0.4$ & $0.3$  & $-10.9$ & $56.2$ & $-102.5$ & $91.8$ & $-44.0$ & $10.9$ & $-1.1$ \\
    $\mathcal{W}_{3,2}$ & $-0.2$ & $-7.7$ & $21.5$ & $-23.0$ & $11.9$ & $-3.0$ & $0.3$  & $-8.1$ & $41.1$ & $-79.2$ & $77.4$ & $-41.1$ & $11.3$ & $-1.3$ \\
    $\mathcal{W}_{3,3}$ & $-0.8$ & $4.3$ & $-11.6$ & $15.1$ & $-9.8$ & $3.1$ & $-0.4$  & $-4.9$ & $23.7$ & $-42.6$ & $37.9$ & $-17.8$ & $4.2$ & $-0.4$ \\
    $\mathcal{W}_{3,4}$ & $0.8$ & $-5.0$ & $11.7$ & $-14.0$ & $9.0$ & $-3.0$ & $0.4$  & $-2.1$ & $8.0$ & $-10.0$ & $3.8$ & $1.4$ & $-1.4$ & $0.3$ \\
    \hline
    \end{tabular}
    \caption{Numerical coefficients for the polynomials in Eqs.~\eqref{eq:Kernels_poly} used to fit the kernels in Eq.~\eqref{eq:K_kernels_def} and~\eqref{eq:W_kernels_def}. Kernels with higher $\ell$ than those displayed are negligible. Kernels with subscripts $i = 1, 2$ refer to ghost quantities, whereas those with subscripts $i = 3, 4$ refer to sterile neutrinos. $\mathcal{K}_3$ and $\mathcal{W}_1$ do not have the $\ell$ subscript because they refer to the background contribution.}
    \label{tab:kernels_coeffiecients}
\end{table}

The kernels associated with ghost particles (subscripts $i = 1, 2$) have only the two
leading coefficients nonvanishing;
therefore, the integrals over ghost momenta  in Eqs.~\eqref{eq:int_dfdtau_nus_1st} and~\eqref{eq:int_dfdtau_phi_1st} take the form
\begin{equation}
%\label{eq:int_poly_kernels}
    \int dx\,x^3\,f_{\phi}^0 (x) \Psi_{\phi, \ell} (x)\,(\mathcal{K}, \mathcal{W}) (x) \approx \int dx\,x^3\,f_{\phi}^0 (x) \Psi_{\phi, \ell} (x)\,(a_0 + a_1 \,x)\,,
\end{equation}
the first term of which is proportional to $F_{\phi, \ell}$, while the second one is a higher-order correction. To estimate the latter, we assume $\Psi_{\phi, \ell} (x)$ depends weakly on $x$ so that
\begin{equation}
\label{eq:int_fPsi_phi_approx}
\begin{split}
    \int dx\,x^3\,f_{\phi}^0 (x) \Psi_{\phi, \ell} (x)\,(\mathcal{K}, \mathcal{W}) (x) &\sim \int dx\,x^3\,f_{\phi}^0 (x) \Psi_{\phi, \ell} (x)\,\bigg[a_0 + a_1 \frac{\int dx\,x^4\,f_{\phi}^0 (x) }{\int dx\,x^3\,f_{\phi}^0 (x)} \bigg] \\
    &\sim F_{\phi, \ell}\,\frac{\rho_{\phi}}{\Lambda^4}%\,[a_0 + 0.69\,a_1]
    \left\{\begin{array}{ll}
    &[a_0 + 0.66\,a_1], \quad \hbox{scalar}\\
    &[a_0 + 0.69\,a_1], \quad \hbox{vector}
    \end{array}\right. \,,
\end{split}
\end{equation}
using $f_{\phi}^0 (x)$ defined in Eq.~\eqref{eq:fi0_gen_sol}; see Fig.~\ref{fig:spectra}.
%and it can be inferred from the $m_{\nu_s} = 0$ result shown in Fig.~\ref{fig:spectra} once it is divided by $4\pi\, x^2 / (2\pi)^3$. 
The factor $1/\Lambda^4$ comes from the normalization of the background phase-space distribution according to the number density, so that $f_{i}^0 (x) \propto \Gamma_{\rho} / (\Lambda^4 H_0)$, while $\rho_{i} (a) \propto \Gamma_{\rho} / H_0$ as in Eq.~\eqref{eq:energy_densities_approx}.
For $\nu_s$, the analogous procedure leads to
\begin{equation}
\label{eq:int_fPsi_nus_approx}
\begin{split}
    \int dx\, f_{\nu_s}^0 (x) \Psi_{\nu_s, \ell} (x)\,(\mathcal{K}, \mathcal{W}) &\approx \int dx\,x^3\, f_{\nu_s}^0 (x) \Psi_{\nu_s, \ell} (x)\, (a_0 + a_1\,x + a_2\,x^2 + a_3\,x^3 + a_4\,x^4 + a_5\,x^5 + a_6\,x^6) \\
    &\sim \int dx\,x^3\,f_{\nu_s}^0 (x) \Psi_{\nu_s, \ell} (x)\,\bigg[a_0 + a_1\,\frac{\int dx\,x^4\,f_{\nu_s}^0 (x) }{\int dx\,x^3\,f_{\nu_s}^0 (x)} + a_2\,\frac{\int dx\,x^5\,f_{\nu_s}^0 (x) }{\int dx\,x^3\,f_{\nu_s}^0 (x)} + \\
    &\qquad \qquad \qquad \qquad \qquad \qquad \quad\, + a_3\,\frac{\int dx\,x^6\,f_{\nu_s}^0 (x) }{\int dx\,x^3\,f_{\nu_s}^0 (x)} + a_4\,\frac{\int dx\,x^7\,f_{\nu_s}^0 (x) }{\int dx\,x^3\,f_{\nu_s}^0 (x)} + \\
    &\qquad \qquad \qquad \qquad \qquad \qquad \quad\, + a_5\,\frac{\int dx\,x^8\,f_{\nu_s}^0 (x) }{\int dx\,x^3\,f_{\nu_s}^0 (x)} + a_6\,\frac{\int dx\,x^9\,f_{\nu_s}^0 (x) }{\int dx\,x^3\,f_{\nu_s}^0 (x)} \bigg] \\
    &\sim F_{\nu_s, \ell}\,\frac{\rho_{\nu_s}}{\Lambda^4}%\,[a_0 + 0.75\,a_1 + 0.63\,a_2 + 0.58\,a_3 + 0.58\,a_4 + 0.61\,a_5 + 0.68\,a_6]
    \left\{\begin{array}{ll}
    &[a_0 + 0.72\,a_1 + 0.59\,a_2 + 0.54\,a_3 + 0.53\,a_4 + 0.56\,a_5 + 0.63\,a_6], \quad \hbox{scalar}\\
    &[a_0 + 0.75\,a_1 + 0.63\,a_2 + 0.58\,a_3 + 0.58\,a_4 + 0.61\,a_5 + 0.68\,a_6], \quad \hbox{vector}
    \end{array}\right. \,.
\end{split}
\end{equation}
The series $[a_0 + \dots]$ in Eqs. (\ref{eq:int_fPsi_phi_approx},\ref{eq:int_fPsi_nus_approx}) differ from each other because of the larger phase space available for
$\nu_s$ relative to $\phi$.

Equations~\eqref{eq:int_q3_C_bkg} and~\eqref{eq:int_q1_C_bkg} can be similarly estimated, but more simply since all the functions are now known.
Therefore, Eqs.~\eqref{eq:int_q3_C_bkg},~\eqref{eq:int_dfdtau_nus_1st},~\eqref{eq:int_q1_C_bkg} and~\eqref{eq:int_dfdtau_phi_1st} reduce to
\begin{equation}
\label{eq:numerators_term_C}
\begin{split}
     \int dq_3\,q_3^3\,\bigg(\frac{d f_{\nu_s}}{d\tau} \bigg)^{(0)}_{\mathcal{C}} &\approx a \Gamma_{\rho} + a \frac{\Gamma_{\rho}}{\Lambda^4}\, \bigg[a_{\nu_s, 0}\,\rho_{\phi} + b_{\nu_s, 0}\,\rho_{\nu_s} \bigg]\,, \\
     \int dq_1\,q_1^3\,\bigg(\frac{d f_{\phi}}{d\tau} \bigg)^{(0)}_{\mathcal{C}} &\approx -a \Gamma_{\rho} + a \frac{\Gamma_{\rho}}{\Lambda^4}\, \bigg[a_{\phi, 0}\,\rho_{\phi} + b_{\phi,0}\,\rho_{\nu_s} \bigg]\,, \\
     \int dq_{\nu_s}\,q_{\nu_s}^3\,\bigg(\frac{df_{\nu_s}}{d\tau}\bigg)^{(1)}_{\mathcal{C}, \ell} &\approx a\,\frac{\Gamma_{\rho}}{\Lambda^4}\,\bigg[a_{\nu_s, \ell}\,\rho_{\phi}\,F_{\phi, \ell} + b_{\nu_s, \ell}\,\rho_{\nu_s}\,F_{\nu_s, \ell} \bigg]\,, \\
     \int dq_{\phi}\,q_{\phi}^3\,\bigg(\frac{df_{\phi}}{d\tau}\bigg)^{(1)}_{\mathcal{C}, \ell} &\approx a\,\frac{\Gamma_{\rho}}{\Lambda^4}\,\bigg[a_{\phi, \ell}\,\rho_{\phi}\,F_{\phi, \ell} + b_{\phi, \ell}\,\rho_{\nu_s}\,F_{\nu_s, \ell} \bigg]\,,
\end{split}
\end{equation}
where $a_{i, \ell}$ and $b_{i, \ell}$ are given in Table~\ref{tab:first_term_C_coeff}. There one notices that the $\ell = 0, 1$ terms of Eq.~\eqref{eq:numerators_term_C} for $\nu_s$ are equal and opposite in sign to those for $\phi$, as well as the background terms. This follows from energy and momentum conservation both at the zeroth and linear order in the perturbations~\cite{Barenboim:2020vrr}. 
Moreover, the second term in the first two equations in~\eqref{eq:numerators_term_C} is subleading compared to $\Gamma_{\rho}$, scaling as $ \Gamma_{\rho}^2 / (H_0\,\Lambda^4)$. Although it can be neglected in the background Boltzmann equations, it is needed in the perturbation equations since it is of the same order as the perturbed collision term (see Eqs.~\eqref{eq:C_term_nus_ell_m0} and~\eqref{eq:numerators_term_C}).

\begin{table}[t!]
\centering
    \begin{tabular}{| c | c  c | c  c | c  c | c  c |}
    \hline
    & \multicolumn{4}{c|}{Vector mediator} & \multicolumn{4}{c|}{Scalar mediator} \\
    \hline \hline
    & \multicolumn{2}{c|}{$\nu_s$} & \multicolumn{2}{c|}{$\phi$}  & \multicolumn{2}{c|}{$\nu_s$} & \multicolumn{2}{c|}{$\phi$} \\
    \hline
    $\ell$ & $a_{\nu_s, \ell}$ & $b_{\nu_s, \ell}$ & $a_{\phi, \ell}$ & $b_{\phi, \ell}$  & $a_{\nu_s, \ell}$ & $b_{\nu_s, \ell}$ & $a_{\phi, \ell}$ & $b_{\phi, \ell}$ \\
    \hline \hline
    $0$ & $7.5$ & $-8.1$    & $-7.5$ & $8.1$  & $8.1$ & $-9.7$ & $-8.1$ & $9.7$ \\
    $1$ & $1.5$ & $0.2$     & $-1.5$ & $-0.2$  & $2.3$ & $-0.8$ & $-2.3$ & $0.8$ \\
    $2$ & $-0.7$ & $-43.3$  & $-3.9$ & $32.0$  & $0.5$ & $-79.3$ & $-3.8$ & $90.7$ \\
    $3$ & $-0.6$ & $-2.6$   & $-3.5$ & $-0.3$  & $-0.1$ & $-3.4$ & $-3.8$ & $-0.1$ \\
    $4$ & $-0.3$ & $-3.0$   & $-3.5$ & $-0.1$  & $0$ & $-3.4$ & $-3.8$ & $0$ \\
    $5$ & $-0.1$ & $-2.9$   & $-3.5$ & $0$  & $0$ & $-3.4$ & $-3.8$ & $0$ \\
    $\geq 6$ & $0$ & $-2.9$ & $-3.5$ & $0$  & $0$ & $-3.4$ & $-3.8$ & $0$ \\
    \hline
    \end{tabular}
    \caption{Numerical value of the coefficients $a_{i, \ell}$ and $b_{i, \ell}$ entering Eq.~\eqref{eq:numerators_term_C}.}
    \label{tab:first_term_C_coeff}
\end{table}

Substituting Eq.~\eqref{eq:numerators_term_C} into Eq.~\eqref{eq:C_term_nus_ell_m0}, the full perturbation equations~\eqref{eq:pert_eq_F_gen} for $\nu_s$ are
\begin{equation}
\label{eq:nus_pert_equations_final}
\begin{split}
    \delta_{\nu_s}' &\approx -\frac{4}{3} \theta_{\nu_s} - \frac{2}{3} h'+ a\,\frac{\Gamma_{\rho}}{\Lambda^4}\,\bigg[%7.5
    a_{\nu_s,0}\,\frac{\rho_{\phi}}{\rho_{\nu_s}}\,(\delta_{\phi} - \delta_{\nu_s}) \bigg] - a\, \frac{\Gamma_{\rho}}{\rho_{\nu_s}}\,\delta_{\nu_s}\,, \\
    \theta_{\nu_s}' &\approx k^2 \bigg(\frac{1}{4} \delta_{\nu_s} - \sigma_{\nu_s} \bigg) + a\,\frac{\Gamma_{\rho}}{\Lambda^4}\,\bigg[%1.5
    a_{\nu_s,1}\,\frac{\rho_{\phi}}{\rho_{\nu_s}}\,\theta_{\phi} + \bigg(%8.3 - 7.5
    B_{\nu_s, 1} -a_{\nu_s, 0}\,\frac{\rho_{\phi}}{\rho_{\nu_s}}\bigg)\,\theta_{\nu_s} \bigg] - a\,\frac{\Gamma_{\rho}}{\rho_{\nu_s}}\,\theta_{\nu_s}\,, \\
    F_{\nu_s, 2}' &= 2 \sigma_{\nu_s}' \approx \frac{8}{15} \theta_{\nu_s} - \frac{3}{5} k F_{\nu_s, 3} + \frac{4}{15} h' + \frac{8}{5} \eta' + 2\, a\,\frac{\Gamma_{\rho}}{\Lambda^4}\,\bigg[%-0.7
    a_{\nu_s, 2}\,\frac{\rho_{\phi}}{\rho_{\nu_s}}\,\sigma_{\phi} + \bigg(%-35.2 - 7.5
    B_{\nu_s, 2} - a_{\nu_s, 0}\,\frac{\rho_{\phi}}{\rho_{\nu_s}}\bigg)\,\sigma_{\nu_s} \bigg] - 2\,a\,\frac{\Gamma_{\rho}}{\rho_{\nu_s}}\,\sigma_{\nu_s}\,, \\
    F_{\nu_s, \ell \geq 2}' &\approx \frac{k}{2\ell + 1} [\ell F_{\nu_s, \ell -1} - (\ell + 1) F_{\nu_s, \ell + 1}] + a\,\frac{\Gamma_{\rho}}{\Lambda^4}\,\bigg[\bigg(%5.2 - 7.5
    B_{\nu_s, \ell} - a_{\nu_s, 0}\,\frac{\rho_{\phi}}{\rho_{\nu_s}}\bigg)\,F_{\nu_s, \ell} \bigg] - a\,\frac{\Gamma_{\rho}}{\rho_{\nu_s}}\,F_{\nu_s, \ell}\,,
\end{split}
\end{equation}
where $B_{\nu_s, \ell} \equiv (b_{\nu_s, \ell} - b_{\nu_s, 0})$, and for the ghosts they are
\begin{equation}
\label{eq:ghost_pert_equations_final}
\begin{split}
    \delta_{\phi}' &\approx -\frac{4}{3} \theta_{\phi} - \frac{2}{3} h' + a\,\frac{\Gamma_{\rho}}{\Lambda^4}\,\bigg[%-8.1
    -b_{\phi,0}\,\frac{\rho_{\nu_s}}{\rho_{\phi}}\,(\delta_{\phi} - \delta_{\nu_s}) \bigg] + a\,\frac{\Gamma_{\rho}}{\rho_{\phi}}\,\delta_{\phi}\,, \\
    \theta_{\phi}' &\approx k^2 \bigg(\frac{1}{4} \delta_{\phi} - \sigma_{\phi} \bigg) + a\,\frac{\Gamma_{\rho}}{\Lambda^4}\,\bigg[\bigg(%6.0 - 8.1
    A_{\phi,1} - b_{\phi, 0}\,\frac{\rho_{\nu_s}}{\rho_{\phi}}\bigg)\,\theta_{\phi} %- 0.2
    + b_{\phi,1}\,\frac{\rho_{\nu_s}}{\rho_{\phi}}\,\theta_{\nu_s} \bigg] + a\,\frac{\Gamma_{\rho}}{\rho_{\phi}}\,\theta_{\phi}\,, \\
    F_{\phi, 2}' &= 2 \sigma_{\phi}' \approx \frac{8}{15} \theta_{\phi} - \frac{3}{5} k F_{\phi, 3} + \frac{4}{15} h' + \frac{8}{5} \eta' + 2\,a\,\frac{\Gamma_{\rho}}{\Lambda^4}\,\bigg[\bigg(%3.6 - 8.1
    A_{\phi,2} - b_{\phi,0}\,\frac{\rho_{\nu_s}}{\rho_{\phi}}\bigg)\,\sigma_{\phi} %+ 32.0
    + b_{\phi,2}\,\frac{\rho_{\nu_s}}{\rho_{\phi}}\,\sigma_{\nu_s} \bigg] + 2\,a\,\frac{\Gamma_{\rho}}{\rho_{\phi}}\,\sigma_{\phi}\,, \\
    F_{\phi, \ell \geq 2}' &\approx \frac{k}{2\ell + 1} [\ell F_{\phi, \ell -1} - (\ell + 1) F_{\phi, \ell + 1}] + a\,\frac{\Gamma_{\rho}}{\Lambda^4}\,\bigg[\bigg(%4.0 - 8.1
    A_{\phi, \ell} - b_{\phi,0}\,\frac{\rho_{\nu_s}}{\rho_{\phi}}\bigg)\,F_{\phi, \ell} \bigg] + a\,\frac{\Gamma_{\rho}}{\rho_{\phi}}\,F_{\phi, \ell}\,,
\end{split}
\end{equation}
where $A_{\phi, \ell} \equiv (a_{\phi, \ell} - a_{\phi, 0})$.
For both species, we solved the perturbation equations with \texttt{CAMB} up to the hierarchy multipole $\ell_{\rm max} = 45$, as done for massless neutrinos and photons when no approximation scheme is applied.

\subsubsection{Qualitative features of the perturbed massless equations}
\label{subsec:considerations_perts}

Eqs.~\eqref{eq:nus_pert_equations_final} and~\eqref{eq:ghost_pert_equations_final} can be further simplified since $\rho_{\nu_s} (a) = - \rho_{\phi} (a)$ for massless sterile neutrinos.
Then the last term on the right-hand sides, which arise from the time dependence of the background phase-space distributions $f_{i}^0$, are negative assuming $F_{i, \ell} > 0$ and therefore provide damping of the perturbations for all values of $\ell$. Such damping is typical of particle decay processes ({\it e.g.}, Ref.~\cite{Ichiki:2004vi}), hence we will refer to it as the ``decay term'' below.  Its  effect is larger at early times because $\rho_i (a)$ evolves faster than $a$.
The damping is independent of  $\Gamma_{\rho}$ since $\rho_{i} \propto \Gamma_{\rho} / H_0$, according to Eq.~\eqref{eq:energy_densities_approx}. 
%This does not harm the physical perturbations because the latter are defined in terms of the relative perturbations $F_{i, \ell}$ as $\rho_{i}\,F_{i, \ell}$.

On the other hand, the second-to-last term of the right-hand sides of Eqs.~\eqref{eq:nus_pert_equations_final} and~\eqref{eq:ghost_pert_equations_final}, which we will refer to as the ``collision term,''
scales as $a$, so that it grows at late times. 
Being  proportional to $\Gamma_{\rho} / \Lambda^4 \sim \Lambda\, (\Lambda / M_i)^{4}$, it is generally smaller than the last term. %depending on the choice of $\Lambda$ and $M_i$.
We expect that deviations in the CMB or matter power spectra with respect to $\Lambda$CDM  will be apparent only when the collision term starts to dominate over the decay term, otherwise Eqs.~\eqref{eq:nus_pert_equations_final} and~\eqref{eq:ghost_pert_equations_final} become identical, giving $F_{\nu_s, \ell} \sim F_{\phi, \ell}$, implying $\rho_{\nu_s} F_{\nu_s, \ell} + \rho_{\phi} F_{\phi, \ell} \sim 0$.
However, the collision term %not only generally leads to oscillations for the low $\ell$ multipole equations since it couples ghost and sterile neutrino variables with different strength (i.e. different numerical coefficients), but 
in the high-$\ell$ equations leads to exponential instabilities because of the positive sign in front of the relative perturbation $F_{i, \ell}$.~\footnote{The sign follows from the numerical values of the coefficients in Table \ref{tab:first_term_C_coeff}.}\ \  
Hence, it cannot be too much larger than the decay term while remaining consistent with the CMB and matter power spectrum. 
This criterion gives the estimate 
$\Gamma_{\rho} / \Lambda^4 \lesssim H_0 \sim 10^{-39}$ MeV, in rough agreement with the
quantitative result of
Section~\ref{sect:cmb}.

This reasoning also explains why $\Gamma_\rho$ 
becomes unconstrained at sufficiently small values of $\Gamma_\rho / \Lambda^4$ (see Fig.\ \ref{fig:bounds_m0_gen}).  In that regime the collision term is small compared to the decay term, and the perturbations become irrelevant.
Moreover, deviations between $\Lambda$CDM$\phi\nu_s$ and $\Lambda$CDM are more pronounced at small $k$ (large scales), because the Liouville term (the first term) of the perturbation equations above becomes negligible, being proportional to $k$.
This is borne out in Fig.~\ref{fig:Dl_TT_m0}.

It is noteworthy that $F_{\nu_s, \ell}$ is less suppressed than  $F_{\phi, \ell}$, at $\ell>2$, where the $\phi$ and $\nu_s$ equations decouple. This is because $(a_{\phi,\ell} - a_{\phi,0}) + b_{\phi,0} < (b_{\nu_s,\ell} - b_{\nu_s,0}) + a_{\nu_s,0}$ in the comparison of their collision terms;
hence $F_{\nu_s, \ell} \gtrsim F_{\phi, \ell}$ for $\ell \gtrsim 2$.~\footnote{Since the relative perturbation $F_{i, \ell}$ for $\ell \gtrsim 2$ is negative for relativistic particles, such as massless SM neutrinos and photons, the condition $F_{\nu_s, \ell} \gtrsim F_{\phi, \ell}$ implies that $F_{\phi, \ell}$ is more negative than $F_{\nu_s, \ell}$. Therefore $ 0 \lesssim |\rho_{\nu_s} F_{\nu_s, \ell}| \lesssim \rho_{\phi} F_{\phi, \ell}$ for the physical perturbations, since $\rho_{\nu_s} = - \rho_{\phi} \gtrsim 0$.
This is confirmed by Fig.~\ref{fig:ISW_m0} (left).}
The physical origin for this can be traced to the larger phase space available for sterile neutrinos than for ghosts, which increases its collision term.
For the  $\ell = 0, 1$ perturbations, the situation is more complicated since the $\nu_s$ and $\phi$ equations are coupled. 

The perturbations for the scalar mediator have a modestly larger effect on  cosmological observables relative to the vector mediator.
This is ultimately due to the kinematical dependences of the matrix elements (\ref{eq:M2}),
which result in the coefficients for the
scalar model in Table~\ref{tab:first_term_C_coeff} being larger than the vector ones.
This trend is confirmed by the \texttt{CosmoMC} results in Fig.~\ref{fig:bounds_m0_gen}.

\subsection{Massive sterile neutrinos}~\label{subsec:perts_m}
A detailed analysis of the cosmological perturbations for the most general case
where $\nu_s$ can be nonrelativistic, is beyond the scope of the present paper, since 
in that case the perturbations play a subdominant role compared to the background
contributions to the energy density.  
Nevertheless, we develop an approximate treatment in this appendix.

The perturbation equation hierarchy for massive $\nu_s$ is given in Eq.~\eqref{eq:pert_eq_Psi_gen}, from which the perturbed energy density, pressure, energy flux and shear stress in $k$-space are~\cite{Ma:1995ey}
\bea
\label{eq:physical_pert_variables_massive}
    \delta \rho_{\nu_s} &=& \frac{1}{a^4} \int dq\,q^2\,\varepsilon\, f_{{\nu_s},0}(q) \Psi_{{\nu_s},0} \,, \nn\\
    \delta P_{\nu_s} &=& \frac{1}{3\,a^4} \int dq\,q^2\,\frac{q^2}{\varepsilon}\, f_{{\nu_s},0}(q) \Psi_{{\nu_s},0} \,, \nn\\
    (\rho_{\nu_s} + P_{\nu_s})\, \theta_{\nu_s} &=& \frac{k}{a^4} \int dq\,q^3\, f_{{\nu_s},0}(q) \Psi_{{\nu_s},1} \,, \nn\\
    (\rho_{\nu_s} + P_{\nu_s})\, \sigma_{\nu_s} &=& \frac{2}{3\,a^4} \int dq\,q^2\,\frac{q^2}{\varepsilon}\, f_{{\nu_s},0}(q) \Psi_{{\nu_s},2}\,,
\eea
where $\varepsilon = \sqrt{q^2 + a^2 m_{\nu_s}^2}$ is the comoving energy of $\nu_s$.
Proceeding analogously to the massless case, one can integrate Eq.~\eqref{eq:pert_eq_Psi_gen} over $q$, weighting by $\varepsilon$ or $q^2 / \varepsilon$, to obtain  Eq.~\eqref{eq:physical_pert_variables_massive}. 

We take as an example  the $\ell=0$ equation, giving the density contrast $\delta_{\nu_s}$, since the same procedure can be straightforwardly applied to higher multipoles. 
From Eq.~\eqref{eq:physical_pert_variables_massive}, one obtains
\bea
\label{eq:int_Psi0_massive}
    \int dq\,q^2 \varepsilon f_{{\nu_s},0} (q) \Psi_{{\nu_s},0}' &=& - k \int dq\,q^3 f_{{\nu_s},0} (q) \Psi_{{\nu_s},1} + \frac{h'}{6} \int dq\,q^3 \varepsilon \frac{d f_{{\nu_s},0}}{dq} + \int dq\,q^2 \varepsilon f_{{\nu_s},0} (q)\, \mathcal{C}_0^{(1)}[\Psi_{{\nu_s},0}] \nn \\
    &=& - \frac{a^4}{4\pi} (\rho_{\nu_s} + P_{\nu_s})\, \theta_i - \frac{h'}{2} \frac{a^4}{4\pi} (\rho_{\nu_s} + P_{\nu_s}) + \int dq\,q^2 \varepsilon \bigg[\bigg(\frac{d f_{\nu_s}}{d\tau}\bigg)_{\mathcal{C}, 0}^{(1)} - \bigg(\frac{d f_{\nu_s}}{d\tau}\bigg)_{\mathcal{C}}^{(0)} \Psi_{{\nu_s}, 0} \bigg] \nn\\
    &=& - \frac{a^4 \rho_{\nu_s}}{4\pi}\, (1 + w_{\nu_s}) \,\bigg(\theta_{\nu_s} + \frac{h'}{2} \bigg) + \int dq\,q^2 \varepsilon \bigg(\frac{d f_{\nu_s}}{d\tau}\bigg)_{\mathcal{C}, 0}^{(1)} - \int dq\,q^2 \varepsilon \bigg(\frac{d f_{\nu_s}}{d\tau}\bigg)_{\mathcal{C}}^{(0)} \Psi_{{\nu_s}, 0}\,,
\eea
in the second step using the definition of $\mathcal{C}_0^{(1)} [\Psi_{{\nu_s}, 0}]$ in Eq.~\eqref{eq:C_term_nus_ell}, integrating the second integral on the right-hand side by parts, and using the definitions of the background density and pressure given by Eq.~\eqref{eq:rho_P_omega_def} expressed in comoving variables. 
The left-hand side of Eq.~\eqref{eq:int_Psi0_massive} can be related to the time derivative of the density contrast $\delta_{\nu_s}$ via
\bea
\label{eq:delta_prime_expansion_massive}
    \delta_{\nu_s}' &=& \frac{\delta \rho_{\nu_s}'}{\rho_{\nu_s}} - \frac{\rho_{\nu_s}'}{\rho_{\nu_s}} \delta_{\nu_s} = \frac{\int dq\,q^2 (\varepsilon f_{{\nu_s},0} (q) \Psi_{{\nu_s},0})'}{\int dq\,q^2 \varepsilon f_{{\nu_s},0}(q)} - \frac{\int dq\,q^2 (\varepsilon f_{{\nu_s},0} (q))'}{\int dq\,q^2 \varepsilon f_{{\nu_s},0}(q)} \nn\\
    &=& \frac{4\pi}{a^4 \rho_{\nu_s}}\, a^2 m_{\nu_s}^2 \mathcal{H} \int dq\,\frac{q^2}{\varepsilon} f_{{\nu_s},0} (q) \Psi_{{\nu_s},0} + \frac{4\pi}{a^4 \rho_{\nu_s}}\, \int dq\,q^2 \varepsilon \bigg(\frac{d f_{{\nu_s},0}}{d\tau} \bigg)_{\mathcal{C}}^{(0)} \Psi_{{\nu_s},0} + \frac{4\pi}{a^4 \rho_{\nu_s}}\, \int dq\,q^2 \varepsilon f_{{\nu_s},0} (q_i) \Psi_{{\nu_s},0}' \nn\\
    &\quad& -\frac{4\pi}{a^4 \rho_{\nu_s}}\, \delta_{\nu_s}\, \bigg[a^2 m_{\nu_s}^2 \mathcal{H} \int dq\, \frac{q^2}{\varepsilon} f_{{\nu_s},0}(q) + \int dq \,q^2 \varepsilon \bigg(\frac{d f_{{\nu_s},0}}{d\tau} \bigg)_{\mathcal{C}}^{(0)} \bigg]\,,
\eea
where we used the background Boltzmann equation~\eqref{eq:Boltzmann_bkg_q} and $\varepsilon_{\nu_s}' = a^2 m_{\nu_s}^2 \mathcal{H} / \varepsilon_{\nu_s}$, with $\mathcal{H} = d \ln{a} / d\tau = a H$, the comoving Hubble rate.
Combining Eq.~\eqref{eq:delta_prime_expansion_massive} with~\eqref{eq:int_Psi0_massive}, 
\bea
\label{eq:delta_equation_massive}
    \delta_{\nu_s}' &=& - (1 + w_{\nu_s}) \bigg(\theta_{\nu_s} + \frac{h'}{2}\bigg) + \frac{4\pi}{a^4 \rho_{\nu_s}}\,\mathcal{H}\,a^2 m_{\nu_s}^2 \bigg[\int dq\, \frac{q^2}{\varepsilon} f_{{\nu_s},0} (q) \Psi_{{\nu_s},0} - \delta_{\nu_s} \int dq\,\frac{q^2}{\varepsilon} f_{{\nu_s},0} (q) \bigg] + \mathcal{C}_0^{(1)} [\delta_{\nu_s}]\,,
\eea
by defining a new collision term
\be
\label{eq:C01}
    \mathcal{C}_{0}^{(1)} [\delta_{\nu_s}] \equiv \frac{\int dq\,q^2 \varepsilon \Big(\frac{d f_{\nu_s}}{d\tau}\Big)_{\mathcal{C}, 0}^{(1)}}{\int dq\,q^2 \varepsilon f_{{\nu_s},0}(q)} - \delta_{\nu_s}\,\frac{\int dq\,q^2 \varepsilon \Big(\frac{d f_{\nu_s}}{d\tau}\Big)_{\mathcal{C}}^{(0)}}{\int dq\,q^2 \varepsilon f_{{\nu_s},0}(q)}\,,
\ee
in analogy with Eq.~\eqref{eq:C_term_nus_ell_m0} for the massless sterile neutrino case.
Notice that Eq.~\eqref{eq:delta_equation_massive} reduces to the first equation of~\eqref{eq:pert_eq_F_gen} in the massless case, $m_{\nu_s} \to 0$ and $w_{\nu_s} \to 1/3$.

Although Eq.\ (\ref{eq:C01}) is exact, it can be simplified in the nonrelativistic limit, for which $q \ll a m_{\nu_s}$. The terms in the square parenthesis reduce to
\bea
\label{eq:am_term_cancel}
    \mathcal{H} \frac{a^2 m_{\nu_s}^2}{\rho_{\nu_s}} \bigg[\int dq\, \frac{q^2}{\varepsilon} f_{{\nu_s},0} (q) \Psi_{{\nu_s},0} - \delta_{\nu_s} \int dq\,\frac{q^2}{\varepsilon} f_{{\nu_s},0} (q) \bigg] &\simeq& \frac{\mathcal{H}}{\rho_{\nu_s}} \int dq\, a m_{\nu_s} \,q^2 \bigg(1 - \frac{1}{2} \frac{q^2}{a^2 m_{\nu_s}^2} + \mathcal{O}\bigg(\frac{q^4}{a^4 m_{\nu_s}^4}\bigg) \bigg) f_{{\nu_s},0} (q) \Psi_{{\nu_s},0} \nn\\
    &\quad& - \mathcal{H} \frac{\delta_{\nu_s}}{\rho_{\nu_s}} \,\int dq\,a m_{\nu_s}\,q^2 \bigg(1 - \frac{1}{2} \frac{q^2}{a^2 m_{\nu_s}^2} + \mathcal{O}\bigg(\frac{q^4}{a^4 m_{\nu_s}^4}\bigg) \bigg) f_{{\nu_s},0} (q) \nn\\
    &\simeq& \mathcal{H}\bigg(\frac{\delta \rho_{\nu_s}}{\rho_{\nu_s}} - 3 \frac{\delta P_{\nu_s}}{\rho_{\nu_s}}\bigg) - \mathcal{H}\bigg(1 - 3 \frac{P_{\nu_s}}{\rho_{\nu_s}} \bigg) \delta_{\nu_s} \nn \\
    &\simeq& - 3 \mathcal{H}\, (c_{s, {\nu_s}}^2 - w_{\nu_s} )\, \delta_{\nu_s}\,,
\eea
by introducing the fluid sound speed $c_{s, {\nu_s}}^2 \equiv \delta P_{\nu_s} / \delta \rho_{\nu_s}$ and using  $\varepsilon \simeq a m_{\nu_s}\, [1 + q^2 / (2 a^2 m_{\nu_s}^2) + \mathcal{O}(q^4 / (a^4 m_{\nu_s}^4))]$. 
For a barotropic fluid, applicable since the equation of state for the new species is constant in time, $c_{s, {\nu_s}}^2 \simeq w_{\nu_s} \approx \text{constant}$ and therefore Eq.~\eqref{eq:am_term_cancel} vanishes.  Since it is  
negligible both in the relativistic and nonrelativistic limits,
we neglect it for general $m_{\nu_s}$.  Then Eq.~\eqref{eq:delta_equation_massive} 
simplifies to
\be
    \delta_{\nu_s}' \simeq - (1 + w_{\nu_s}) \bigg(\theta_{\nu_s} + \frac{h'}{2}\bigg) + \mathcal{C}_0^{(1)} [\delta_{\nu_s}]\,,
\ee
which agrees with the derivation based on   energy-momentum conservation~\cite{Ma:1995ey}.

The new collision term given by Eq.~\eqref{eq:C01} is complicated in the general $m_{\nu_s}>0$ case.  However, based on the results we obtained for $m_{\nu_s}=0$, some
inferences can be made.
In particular, the first term of Eq.~\eqref{eq:C01} is proportional to $\Gamma_{\rho} / \Lambda^4$, since it depends linearly on the background phase-space distribution $f_{i}^0 (q)$ with $i = \phi, \nu_s$, whereas the second term is exactly given by $\Gamma_{\rho}\, \delta_{\nu_s} / \rho_{\nu_s}$ and is thus proportional to $H_0$.
In order to be consistent with type Ia supernovae data,  Fig.~\ref{fig:SN_bounds} 
implies that
 $\Gamma_{\rho}\lesssim \mathcal{O}(10^{-70})\,\,\text{MeV}^5$.
For such small values, the constraints on  $\Gamma_{\rho}$ and $\Gamma_{\rho} / \Lambda^4$ for $m_{\nu_s}=0$,  Fig.~\ref{fig:bounds_m0_gen}, become almost independent of $\Gamma_{\rho} / \Lambda^4$,
suggesting that the first term of Eq.~\eqref{eq:C01} is negligible compared to the second. 
Physically, this means that redshifting of the new species, described by the Liouville terms and the second part of the collision term, is more important than the growth from particle production, when $m_{\nu_s}>0$.
In this case the perturbations of the new species decouple from each other.

By neglecting the first piece of the new collision term in Eq.~\eqref{eq:C01}, we expect to find conservative
bounds for the parameter $\Gamma_{\rho}$, or quantities derived from it such as $\Omega_g$, that might be moderately strengthened by a more exact treatment.
The bounds derived for $m_{\nu_s}>0$ come primarily from the background contributions of the new species, rather their perturbations. 

Carrying out
 the same procedure as for $\delta_{\nu_s}$ to higher $\ell$, the  perturbation equations for  $\phi$ and $\nu_s$ are 
\begin{equation}
\label{eq:perts_massive_approx}
    \begin{split}
        \delta_{\nu_s}' &\sim - (1 + w_{\nu_s}) \bigg(\theta_{\nu_s} + \frac{h'}{2} \bigg) - a \frac{\Gamma_{\rho}}{\rho_{\nu_s}} \delta_{\nu_s} \,, \\
        \theta_{\nu_s}' &\sim - \mathcal{H} (1 - 3 w_{\nu_s})\, \theta_{\nu_s} + \frac{w_{\nu_s}}{1 + w_{\nu_s}} k^2 \delta_{\nu_s} - k^2 \sigma_{\nu_s} - a \frac{\Gamma_{\rho}}{\rho_{\nu_s}} \theta_{\nu_s} \,, \\
        \sigma_{\nu_s}' &\sim - 3 \mathcal{H} \sigma_{\nu_s} + \frac{8}{3} \frac{w_{\nu_s}}{1 + w_{\nu_s}} \bigg(\theta_{\nu_s} + \frac{h'}{2} + 3 \eta' \bigg) - a \frac{\Gamma_{\rho}}{\rho_{\nu_s}} \sigma_{\nu_s} \,, \\
        \delta_{\phi}' &\sim - \frac{4}{3} \theta_{\phi} + \frac{2}{3} h' + a \frac{\Gamma_{\rho}}{\rho_{\phi}} \delta_{\phi} \,, \\
        \theta_{\phi}' &\sim k^2 \bigg(\frac{1}{4} \delta_{\phi} - \sigma_{\phi} \bigg) + a \frac{\Gamma_{\rho}}{\rho_{\phi}} \theta_{\phi} \,, \\
        \sigma_{\phi}' &\sim - 3 \mathcal{H} \sigma_{\phi} + \frac{2}{3} \bigg(\theta_{\phi} + \frac{h'}{2} + 3 \eta' \bigg) + a \frac{\Gamma_{\rho}}{\rho_{\phi}} \sigma_{\phi}\,,
    \end{split}
\end{equation}
assuming a barotropic fluid, where we neglected the equations for the multipoles with $\ell > 2$ since they will become negligible  for nonrelativistic $\nu_s$. 
Similar equations for the anisotropic stress $\sigma_i$ were derived  in Refs.~\cite{Hu:1995fqa,Hu:1998kj,Koivisto:2005mm} to approximate the closing of the $\ell$ hierarchy at the quadrupole, valid for both massless and massive particles. We take the viscosity parameter in Ref.\ \cite{Hu:1995fqa} to be $c_{i, \rm vis}^2 \simeq w_i$. These equations encapsulate the free-streaming effect of the new species, which cannot be neglected; Ref.~\cite{Hu:1995fqa} showed that doing so in the SM with massless neutrinos induces an error of $\sim 10\%$ in the CMB power spectrum.
Different prescriptions for the shear stress equation can be found in Ref.~\cite{Lesgourgues:2011}, which provide similar numerical accuracy.

The perturbation equations  derived above become the same for $\phi$ and $\nu_s$  in the limit $w_{\nu_s} \to 1/3$, since the part of the collision term sensitive to their phase space differences has been neglected.  
Hence the 
physical perturbations of the new species  cancel each other in that limit, $(\rho_{\nu_s} \delta_{\nu_s} + \rho_{\phi} \delta_{\phi}) \to 0$.
Similarly to the $m_{\nu_s}=0$ treatment, the equations
 are solved within \texttt{CAMB} up to the $\ell_{\rm max} = 45$.

\section{Correlations between parameters}
\label{appC}

In this appendix, we present the full correlation plots including all  parameters in the $\Lambda$CDM$\phi\nu_s$ model, as inferred from \texttt{CosmoMC}.

\subsection{Massless $\nu_s$ case}
For $m_{\nu_s}=0$, the input parameters are the standard six of $\Lambda$CDM, plus  $\Gamma_{\rho}$, which controls the amplitude of $\phi$ and $\nu_s$ energy densities, and $\Gamma_{\rho} / \Lambda^4$, which sets the strength of the perturbed collision term in the linear perturbations. 
Table~\ref{tab:params_m0} shows the best-fit values for these parameters and their associated $68\%$ C.L.\ intervals.
We verified consistency with the Planck 2018 best-fit parameters~\cite{Planck:2018vyg}, which is expected
since for $m_{\nu_s}=0$ the homogeneous components of the Universe remain the same as in $\Lambda$CDM.

The new parameters, $\Gamma_{\rho}$ and $\Gamma_{\rho} / \Lambda^4$, are degenerate with several $\Lambda$CDM parameters, as shown in Fig.~\ref{fig:cosmomc_m0}. 
This occurs because the linear-order $\phi$ and $\nu_s$ perturbations affect the CMB and matter power spectra through their energy densities, whose magnitude is of order $\Gamma_{\rho} / (H_0\, \sqrt{\Omega_{\Lambda}})$, from Eq.~\eqref{eq:energy_densities_approx}, with $\Omega_{\Lambda} \approx 1 - \Omega_m$.

The choice of the vector versus scalar mediator has a small impact on the results of Table~\ref{tab:params_m0} and Fig.~\ref{fig:cosmomc_m0}, and only the correlation between $\Gamma_{\rho}$ and $\Gamma_{\rho} / \Lambda$  differs slightly between the two cases. 
This can be traced back to the numerical coefficients in the perturbation equations \eqref{eq:nus_pert_equations_final} and~\eqref{eq:ghost_pert_equations_final}; see Table~\ref{tab:first_term_C_coeff}.

\begin{table}[t!]
\begin{tabular} {| c | c | c |}
\hline
\multicolumn{3}{|c|}{Massless $\nu_s$} \\
\hline
 Parameter &  vector & scalar \\
\hline
\hline
{\boldmath$\Omega_b h^2   $} & $0.02235\pm 0.00015$ & $0.02235\pm 0.00015$ \\
{\boldmath$\Omega_c h^2   $} & $0.1202\pm 0.0012$ & $0.1202\pm 0.0012$ \\
{\boldmath$100\,\theta_{\rm MC} $} & $1.04089\pm 0.00031$ & $1.04090\pm 0.00031$ \\
{\boldmath$\tau           $} & $0.0549\pm 0.0075$ & $0.0550\pm 0.0075$ \\
{\boldmath${\rm{ln}}(10^{10} A_s)$} & $3.046\pm 0.015$ & $3.046\pm 0.015$ \\
{\boldmath$n_s            $} & $0.9642\pm 0.0042$ & $0.9643\pm 0.0042$ \\
\hline
{\boldmath$\log_{10}{(\Gamma_{\rho} / \rm{MeV}^5})$} & --- & --- \\
{\boldmath$\log_{10}{(\Gamma_{\rho}/\Lambda^4 / \rm{MeV})}$} & $< -58.1 $ & $< -58.4 $ \\
\hline
$H_0 / \rm{[km/s/Mpc]}                       $ & $67.25\pm 0.54 $ & $67.26\pm 0.55$ \\
$\Omega_{\Lambda}           $ & $0.6832\pm 0.0075 $ & $0.6834\pm 0.0075$ \\
$\Omega_m                  $ & $0.3168\pm 0.0075 $ & $0.3166\pm 0.0075$ \\
$\sigma_8                  $ & $0.8122\pm 0.0061 $ & $0.8123\pm 0.0061$ \\
$S_8                       $ & $0.835\pm 0.013 $ & $0.834\pm 0.013$ \\
\hline
\end{tabular}
\caption{$68\%$ C.L. parameter intervals for the $\Lambda$CDM$\phi\nu_s$ model, as inferred from \texttt{CosmoMC} in the massless $\nu_s$ scenario, for the vector and scalar mediators. The first six parameters in bold are for the $\Lambda$CDM model, whereas the following two are associated to the new physics. The remaining ones are derived from the cosmological model.}
\label{tab:params_m0}
\end{table}

\begin{figure*}[t]
\begin{center}
\includegraphics[scale=0.7]{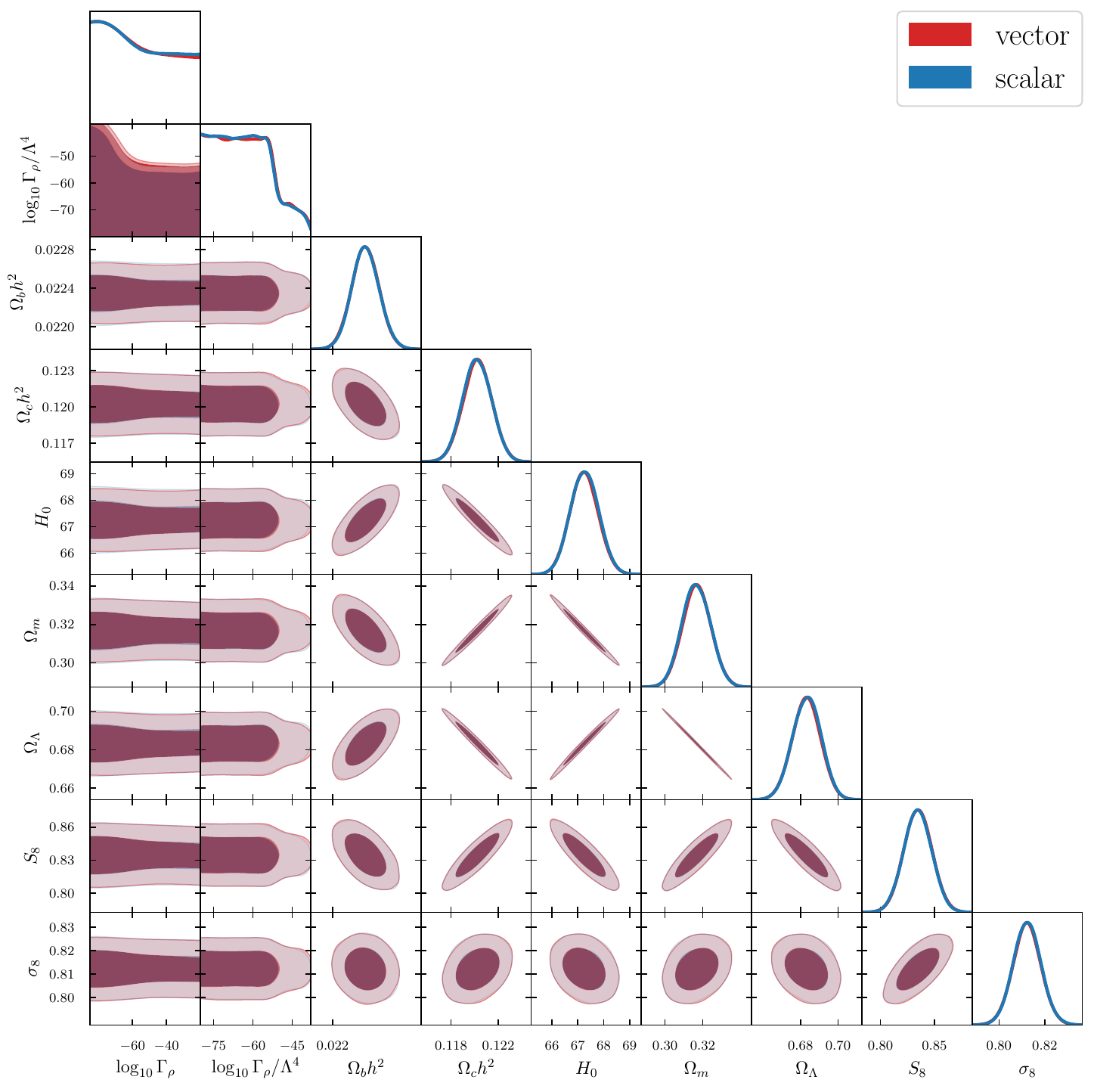}
\caption{Correlation between the main $\Lambda$CDM$\phi\nu_s$ model parameters, as inferred from \texttt{CosmoMC} for massless $\nu_s$, for the vector (red) and scalar (blue) mediator models.
The data sets Planck + Lensing + BAO + DES + Pantheon we consider here are described in the main text. The darker and lighter shaded regions correspond to the $68\%$ and $95\%$ C.L. intervals, respectively.
%\mpuel{Perhaps show all the parameters of Table~\ref{tab:params_m0}\dots}
}
\label{fig:cosmomc_m0}
\end{center} 
\end{figure*}

\subsection{Massive $\nu_s$ case}

\begin{table}[t!]
\begin{tabular} {| c | c |}
\hline
\multicolumn{2}{|c|}{Massive $\nu_s$} \\
\hline
 Parameter &  both mediators \\
\hline
\hline
{\boldmath$\Omega_b h^2   $} & $0.02246\pm 0.00014$ \\
{\boldmath$\Omega_c h^2   $} & $0.11878\pm 0.00094$ \\
{\boldmath$100\,\theta_{\rm MC} $} & $1.04102\pm 0.00029$ \\
{\boldmath$\tau           $} & $0.0556\pm 0.0074$ \\
{\boldmath${\rm{ln}}(10^{10} A_s)$} & $3.045\pm 0.014$ \\
{\boldmath$n_s            $} & $0.9670\pm 0.0037$ \\
\hline
{\boldmath$\Omega_g$} & $< 0.102$ \\
{\boldmath$w_{\nu_s}$} & --- \\
\hline
$H_0 / \rm{[km/s/Mpc]}$ & $68.88^{+0.48}_{-0.75}$ \\
$\Omega_{\Lambda}           $ & $0.617^{+0.069}_{-0.027}$ \\
$\Omega_m                  $ & $0.2992^{+0.0068}_{-0.0056}$ \\
$\sigma_8                  $ & $0.801^{+0.027}_{-0.081}$ \\
$S_8                       $ & $0.799^{+0.027}_{-0.077}$ \\
\hline
\end{tabular}
\caption{Similar to Table~\ref{tab:params_m0}, but for the massive $\nu_s$ scenario. The results are independent of the mediator model chosen.
}
\label{tab:params_m}
\end{table}

\begin{figure}[t]
\begin{center}
\includegraphics[scale=0.7]{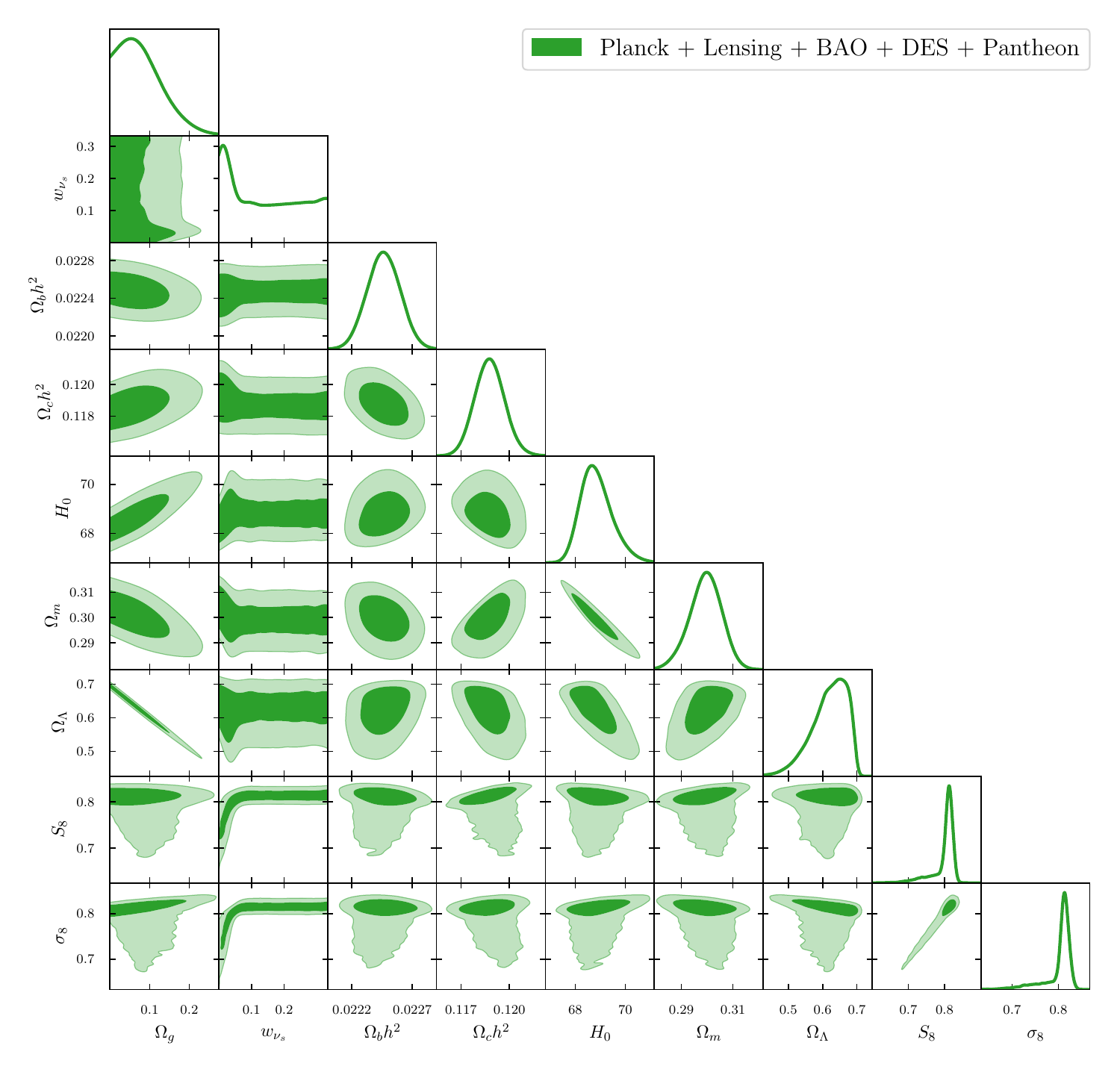}
\caption{Similar to Fig.~\ref{fig:cosmomc_m0}, but for the massive $\nu_s$ case. These results apply to both the vector and scalar mediator models. 
}
\label{fig:cosmomc_m}
\end{center} 
\end{figure}

For massive $\nu_s$, the new-physics parameters
are taken to be 
 $\Omega_g$ and $w_{\nu_s}$.
%Table~\ref{tab:best_params_m} displays the best-fit values of the $\Lambda$CDM$\phi\nu_s$ model parameters and their corresponding $68\%$ C.L.\ interval. 
%The six standard $\Lambda$CDM parameters are compatible with their Planck 2018 best-fit values within $1\sigma$ (see last column of Table 2 of Ref.~\cite{Planck:2018vyg}).
%On the other hand, the derived $\Lambda$CDM parameters, such as $\Omega_\Lambda$, differ from their $\Lambda$CDM values due to the
%preference for  $\Omega_g \neq 0$. They are however still compatible within $3\sigma$.
They show degeneracies with several $\Lambda$CDM quantities in Fig.~\ref{fig:cosmomc_m}. 
As noted previously, $w_{\nu_s}$ is largely completely unconstrained, having no effect
on most of the $\Lambda$CDM parameters. 
The exceptions are $\sigma_8$ and $S_8$, which are related to the amplitude of the matter perturbations. This occurs because $w_{\nu_s}$ impacts the linear perturbation equations.
Table~\ref{tab:params_m} shows the $68\%$ C.L. parameter intervals for the $\Lambda$CDM$\phi \nu_s$ model for the massive $\nu_s$ scenario, as derived from \texttt{CosmoMC}.

\end{widetext}

\end{appendix}

\bibliography{references}
\bibliographystyle{utphys}

\end{document}